\documentclass[a4paper,10pt]{article}
\pdfoutput=1 

\usepackage{jheppub} 

\usepackage[T1]{fontenc} 

\usepackage{graphicx}
\usepackage{verbatim,epsfig}
\usepackage{epstopdf}
\usepackage[utf8]{inputenc}
\usepackage[colorlinks=true,linkcolor=blue,citecolor=blue]{hyperref}
  \usepackage{bm}
   \usepackage{amsmath}
    \usepackage{amssymb}
     \usepackage{pifont}
      \usepackage{multirow,boldline}

\newcommand{\nn}{\nonumber}

\newcommand{\ot}{\leftarrow}

\renewcommand{\(}{\left(}
\renewcommand{\)}{\right)}
\renewcommand{\[}{\left[}
\renewcommand{\]}{\right]}

\renewcommand{\vec}[1]{\bm{#1}}

\newcommand{\specialcellleft}[2][l]{\begin{tabular}[#1]{@{}l@{}}#2\end{tabular}}
\newcommand{\specialcellcenter}[2][c]{\begin{tabular}[#1]{@{}c@{}}#2\end{tabular}}

\title{Non-perturbative structure of semi-inclusive deep-inelastic and Drell-Yan scattering  at small  transverse momentum}

\author[1]{Ignazio Scimemi}
\affiliation[1]{Departamento de F\' isica Te\'orica and IPARCOS \\ Universidad Complutense de Madrid, \\ Plaza de Ciencias 1, Ciudad Universitaria,
 28040 Madrid, Spain}
\author[2]{Alexey Vladimirov}
\affiliation[2]{Institut f\"ur Theoretische Physik, \\ Universit\"at Regensburg,\\
D-93040 Regensburg, Germany}

\emailAdd{ignazios@ucm.es}
\emailAdd{alexey.vladimirov@physik.uni-regensburg.de}

\abstract{
We consider semi-inclusive deep inelastic scattering (SIDIS) and Drell-Yan events within transverse momentum dependent (TMD) factorization. Based on the simultaneous fit of multiple data points, we extract the unpolarized TMD distributions and the non-perturbative evolution kernel. The high quality of the fit confirms a complete universality of TMD non-perturbative distributions. The extraction is supplemented by phenomenological analyses of various parts of the TMD factorization, such as sensitivity to non-perturbative parameterizations, perturbative orders, collinear distributions, correlations between parameters, and others. 
}

\begin{document}
\maketitle
\flushbottom

\section{Introduction}
The factorization theorem for the differential cross-sections of  boson production (Drell-Yan process or DY in this paper) and semi-inclusive deep inelastic scattering (SIDIS) identifies clearly the sources of non-perturbative QCD effects as the transverse momentum dependent (TMD) distributions and, separately, their evolution kernel \cite{Collins:1989gx,Bacchetta:2006tn,Bacchetta:2008xw,Becher:2010tm,Collins:2011zzd,GarciaEchevarria:2011rb,Echevarria:2012js,Echevarria:2014rua,Chiu:2012ir,Vladimirov:2017ksc,Scimemi:2018xaf}. The extraction of these non-perturbative (NP) elements from data is then a major challenge for modern phenomenology~\cite{Angeles-Martinez:2015sea}. 

In this article, we consider the unpolarized observables that have the simplest structure and are accessible in a relatively large number of experiments. 
They allow us to extract the quark unpolarized TMD distributions and the non-perturbative part of TMD evolution. In the literature one can find many extractions of these elements within various schemes \cite{Sun:2013hua,Anselmino:2013lza,Signori:2013mda,DAlesio:2014mrz,Aidala:2014hva,Bacchetta:2017gcc,Scimemi:2017etj,Bertone:2019nxa,Vladimirov:2019bfa,Bacchetta:2019sam}. The distinctive feature of this work is the simultaneous study of two kinds of reactions: DY and SIDIS. Previously, a global fit of both processes has been attempted only in ref.~\cite{Bacchetta:2017gcc}. \textit{We demonstrate that the global description of both processes is straightforward and does not meet any obstacle.} The description is based on the latest theory developments, such as next-to-next-to-leading order (NNLO) and N$^3$LO perturbative parts together with $\zeta$-prescription. In addition, we make a special emphasis on some topics not so often discussed in the literature, that is,   universality and  theory uncertainties of the TMD.


The factorization theorem declares that the TMD non-perturbative parts have a certain degree of universality, as explained in the following: a) the evolution kernel is the same for all processes where the TMD factorization theorem is valid; b) the TMD parton distribution functions (TMDPDF) are the same in DY and SIDIS experiments. Testing universality needs an analysis of different types of experiments at the same time. Although the universality is a cornerstone of the approach, we have not found any dedicated phenomenological study in the literature. In order to check and proof universality properties of the TMD approach, we perform an  analysis in three steps:
\begin{itemize}
\item[I.] Firstly, we consider only the DY measurements, and analyze TMDPDF $f_1(x,b)$ and rapidity anomalous dimension (RAD), $\mathcal{D}(\mu,b)$. The DY data sets have a vast span in $x$ and $Q$, therefore, it is possible to extract $f_1$ (that dictates the $x$-dependence of the cross-section) and $\mathcal{D}$ (that dictates the $Q$-dependence of the cross-section) without a significant correlation between these functions. This analysis is conceptually similar to the previous work \cite{Bertone:2019nxa}, albeit some improvements. 
\item[II.] Using the  outcome of the previous step ($\mathcal{D}$ and $f_1$), we consider the SIDIS measurements and extract the TMDFF, $D_1$. Assuming the universality of TMD distributions, one should be able to describe the SIDIS cross-section with a single extra function $D_1$. This is a non-trivial statement since the SIDIS cross-section has 4-degrees of freedom,  and only two of them are affected by $D_1$.  Additionally, the present SIDIS data are concentrated in a range of small-$Q$ that is unreachable for DY experiments.  

\item[III.] Finally, we perform a simultaneous fit of DY and SIDIS data. Given the excellent quality of the separate DY and SIDIS fits, this stage  provides only a fine-tune of non-perturbative parameters as well as a consistency check of previous step II.
\end{itemize}
These three independent analyses provide a consistent and congruent picture of the TMD factorization and allow the extraction of three non-perturbative functions (unpolarized quark TMDPDF, TMDFF and quark evolution kernel). We find that our  results are in full agreement with the depicted scenario, which gives a solid confirmation of the declared universality. 

On top of the described test of universality and the extraction of TMD distributions, in this work we perform many additional studies  of the TMD approach, some of which should be better addressed elsewhere: we test  the phenomenological limits of the TMD factorization for SIDIS; we check  the dependence of the TMD prediction on the collinear inputs; we perform an overall test of the impact of power suppressed contributions to the TMD factorization; we check the impact of experimental constraints on the final phase space configurations (like fiducial cross sections and lepton cuts at LHC, bin shapes in HERMES kinematics). Altogether the tests can form a comprehensive picture of TMD factorization and its accuracy. 
We have observed that the impact of some input uncertainties, f.i. the ones from collinear PDF, to the prediction is unlucky large. 
Still, we restrict ourself to the  indication of problematic issues, leaving it as an invitation for the further developments in the future.


The theoretical work done in recent years for the development of the elements of TMD factorization has been noticeable. Significant efforts have been committed in the perturbative calculations for TMD distributions at small-$b$ \cite{Gehrmann:2014yya,Echevarria:2015byo,Echevarria:2015usa,Echevarria:2016scs,Li:2016ctv,Vladimirov:2016dll,Luo:2019hmp}. Together with the N$^3$LO results for universal QCD anomalous dimensions \cite{Gehrmann:2010ue,Baikov:2016tgj,Moch:2017uml,Moch:2018wjh,Lee:2019zop}, it leads to an extremely accurate perturbative input. The consistent composition of all elements is made employing the $\zeta$-prescription \cite{Scimemi:2018xaf,Vladimirov:2019bfa}. The $\zeta$-prescription is essential for current and future TMD phenomenology because it grants a unified approach to observables irrespectively of the order of perturbative matching. So, the collinear matching procedure that is fundamental for resummation approaches (such as in refs.~\cite{Landry:2002ix,Qiu:2000hf,Bozzi:2008bb,Becher:2010tm,Catani:2012qa,Bizon:2018foh,Bizon:2019zgf}) or $b^*$-like prescriptions (such as in refs.\cite{Collins:1981va,Collins:2011zzd,Aybat:2011zv,Bacchetta:2017gcc}), is considered just as  part of the model for a TMD distribution in the $\zeta$-prescription. Therefore, \textit{unpolarized TMD distributions (extracted in this work with NNLO matching) and the TMD evolution (extracted in this work with NNLO and N$^3$LO matching) are entirely universal and could be used for the description of other processes}, where the matching is not known at such a high order.

Given the number of details needed for the presentation of this work, we split the discussions into almost independent parts. The first part, sec.~\ref{sec:2}, contains the description of the TMD factorization theorem for unpolarized DY and SIDIS cases. In this section, we articulate all relevant formulas, including a lot of small corrections and details that we have not found mentioned in previous literature.  This part provides a comprehensive collection of theory results, which can be useful for comparison with other works and future tests, and it can be seen as a theory review.  Some of the issues reported here are expected to be addressed in separate works.
Sec.~\ref{sec:data} is devoted to the review of the available SIDIS and DY data suitable for unpolarized TMD phenomenology.
Sec.~\ref{sec:fitprocedure} presents the details of the comparison of the theory expression with the experimental data. It contains the definition of $\chi^2$-test, the interpretation of the experimental environment, and some details of the numerical implementation that is made by \texttt{artemide} package \cite{web}. The following sections \ref{sec:DY-fit},~\ref{sec:SIDIS-fit} and \ref{sec:GLOBAL-fit} describe the fit program outlined earlier, and they are devoted to DY(only), SIDIS(only), and DY and SIDIS(together) fits.
 Each of these sections contains several subsections describing the specific impact of each process on TMD extraction. Finally, we collect the information on the resulting NP functions in sec.~\ref{sec:final}.

\section{Cross sections in TMD factorization}
\label{sec:2}

In this section, we present in detail  the cross sections of SIDIS and DY processes in TMD factorization, skipping their derivation that can be found in refs.~\cite{Collins:1989gx,Bacchetta:2006tn,Bacchetta:2008xw,Becher:2010tm,Collins:2011zzd,GarciaEchevarria:2011rb,Echevarria:2012js,Echevarria:2014rua,Chiu:2012ir,Vladimirov:2017ksc,Scimemi:2018xaf}. The main purpose of this section is to collect all pieces of information about theoretical approximations and models that are used in the fit procedure.

\subsection{SIDIS cross-section}

The (semi-inclusive) deep-inelastic scattering (SIDIS) is defined by the reaction
\begin{eqnarray}
\ell(l)+H(P)\to \ell(l')+h(p_h)+X,
\end{eqnarray}
where $\ell$ is a lepton, $H$ and $h$ are respectively the target and the fragmenting hadrons, and $X$ is the undetected final state. The vectors in brackets denote the momenta of each particle. The masses of the particles are
\begin{eqnarray}
P^2=M^2,\qquad p_h^2=m^2,\qquad l^2=l'^2=m_l^2\simeq0.
\end{eqnarray}
In the following, we neglect the lepton masses, but keep the effects of  the hadron masses.

Approximating the interaction of a lepton and a hadron by a single photon exchange, one obtains  the differential cross-section
\begin{eqnarray}\label{def:xSec-0}
d\sigma=\frac{2}{s-M^2} \frac{ \alpha^2_{\text{em}}}{(q^2)^2} L_{\mu\nu}W^{\mu\nu}\frac{d^3l'}{2E'}\frac{d^3p_h}{2E_{h}},
\end{eqnarray}
with $q=l-l'$ being the momentum of the intermediate photon. Here, the scattering flux-factor, $((s-(m_l+M)^2)(s-(m_l-M)^2))^{-1/2}$ is evaluated at vanishing lepton mass; the factors $q^2$ come from the photon propagators $\Delta^{\mu\nu}=g^{\mu\nu}/(q^2+i0)$ and  $\alpha_{\text{em}}=e^2/4\pi$ is QED coupling constant. The last factors in eq.~(\ref{def:xSec-0}) are the phase-space differentials  for the detected hadron and lepton, with $E'$($E_h$) being their energies. The leptonic and hadronic tensors ($L^{\mu\nu}$ and $W^{\mu\nu}$) are \begin{eqnarray}
L_{\mu\nu}&=&e^{-2}\langle l'|J_{\mu}(0)|l\rangle \langle l|J^\dagger_{\nu}(0)|l'\rangle,\nn
\\
W_{\mu\nu}&=&e^{-2}\int \frac{d^4x}{(2\pi)^4}e^{-i(x\cdot q)}\sum_X\langle P|J^\dagger_{\mu}(x)|p_h,X\rangle \langle p_h,X|J_{\nu}(0)|P\rangle,
\end{eqnarray}
where $e$ is the lepton charge, and $J^\mu$ is the electro-magnetic current.

\subsubsection{Kinematic variables for SIDIS}
The formulation of the factorization theorem in SIDIS is done in the hadronic Breit frame (alternatively, we can call it  "the factorization frame"), where the momenta of hadrons are almost light-like and back-to-back. The light-like direction to which the hadrons are aligned defines the decomposition of their momenta, 
\begin{eqnarray}
P^\mu&=&P^+ \bar n^\mu+\frac{M^2}{2P^+}n^\mu,\qquad p_h^\mu =p_h^- n^\mu+\frac{m^2}{2p^-_h}\bar n^\mu,
\end{eqnarray}
with
$n^2=\bar n^2=0,$ $ (n\bar n)=1$.  Here, we have also introduced the common notation of a vector decomposition
\begin{eqnarray}
v^\mu=v^+ \bar n^\mu+v^- n^\mu+v_T^\mu,\qquad v^+=(nv),\qquad v^-=(\bar nv),\qquad (nv_T)=(\bar nv_T)=0.
\end{eqnarray}
The transverse component of a vector is extracted with the projector
\begin{eqnarray}\label{def:gT}
v_T^\mu=g_T^{\mu\nu}v_\nu,\qquad g_T^{\mu\nu}=g^{\mu\nu}-n^\mu \bar n^\nu-\bar n^\mu n^\nu.
\end{eqnarray}
We also use the convention that the bold font denotes vectors that have only transverse components. So, they obey Euclidian scalar product: 
\begin{eqnarray}
\vec v_T^2=-v_T^2>0.
\end{eqnarray}

For the SIDIS cross-section one introduces the following scalar variables:
\begin{eqnarray}
Q^2=-q^2,\qquad x=\frac{Q^2}{2(Pq)},\qquad y=\frac{(Pq)}{(Pl)},\qquad z=\frac{(Pp_h)}{(Pq)}.
\end{eqnarray}
In the experimental environment one typically measures the transverse momentum defined as the one of the produced hadron with respect to the plane formed by vectors $q$ and $P$. The projector corresponding  to these transverse components  is given by the tensor $g_\perp^{\mu\nu}$ defined as
\begin{eqnarray}\label{def:gPerp-SIDIS}
g_\perp^{\mu\nu}&=&g^{\mu\nu}-\frac{1}{M^2Q^2+(Pq)^2}\[Q^2 P^\mu P^\nu+(Pq)(P^\mu q^\nu+q^\mu P^\nu)-M^2 q^\mu q^\nu\]
\\\nn
&=&g^{\mu\nu}-\frac{1}{Q^2(1+\gamma^2)}\[2x^2 P^\mu P^\nu+2x(P^\mu q^\nu+q^\mu P^\nu)-\gamma^2 q^\mu q^\nu\].
\end{eqnarray}
In what follows, we  denote the transverse components of $v^\mu$ in the factorization frame as $v_T^\mu$, see eq.~(\ref{def:gT}), while transverse components projected by $g_\perp$  are $v_\perp^\mu$. 

In order to describe the target- and produced-mass corrections, it is convenient to use the following combinations
\begin{eqnarray}\label{def:mass-var}
\gamma=\frac{2Mx}{Q},\qquad \varsigma=\gamma\frac{m}{zQ},\qquad \varsigma^2_\perp=\gamma^2\frac{m^2+\vec p^2_{h\perp}}{z^2Q^2}.
\end{eqnarray}
The definition of $\varsigma^2_\perp$ in eq.~(\ref{def:mass-var}) contains ${\vec p}^2_{h\perp}=p_{h \mu} p_{h \nu } g_\perp^{\mu\nu}$. 

The measured transverse momentum $p_\perp$ is different from the one defined in  TMD factorization. In fact, the transverse momentum used in the factorization theorem, $q_T$, is defined with respect to the hadron-hadron-plane and the corresponding transverse components are extracted by the tensor $g_T^{\mu\nu}$ in eq.~(\ref{def:gT}). In terms of hadron momenta the tensor $g_T$ reads
\begin{eqnarray}\label{def:gT-SIDIS}
g_T^{\mu\nu}&=&g^{\mu\nu}-\frac{1}{m^2M^2-(Pp_h)^2}\[m^2 P^\mu P^\nu-(Pp_h)(P^\mu p_h^\nu+p_h^\mu P^\nu)+M^2 p_h^\mu p_h^\nu\]
\\\nn
&=&g^{\mu\nu}+\frac{1}{Q^2(1-\varsigma^2)}\[4\frac{x^2}{\gamma^2}\varsigma^2 P^\mu P^\nu-\frac{2x}{z}(P^\mu p_h^\nu+p_h^\mu P^\nu)+\frac{\gamma^2}{z^2} p_h^\mu p_h^\nu\].
\end{eqnarray}
Using the projectors in eq.~(\ref{def:gPerp-SIDIS}) and eq.~(\ref{def:gT-SIDIS}), it is straightforward to derive the relation between $q_T^2=q_\mu q_\nu g_T^{\mu\nu}$ and $p_\perp^2=p_h^\mu p_h^\nu g_{\perp,\mu\nu}$:
\begin{eqnarray}\label{def:qT<->pT}
q_T^2=\frac{p_\perp^2}{z^2}\frac{1+\gamma^2}{1-\varsigma^2}.
\end{eqnarray}

Using these definition we can rewrite the elements of the SIDIS cross-section formula in terms of observable variables. The differential volumes of the phase space are
\begin{eqnarray}\label{th:phase-elem-1}
\frac{d^3l'}{2E'}=\frac{y}{4x}dQ^2dx d\psi,\qquad \frac{d^3p_h}{2E_h}=\frac{1}{\sqrt{1-\varsigma^2_\perp}}\frac{dz d^2p_\perp}{2z}
=\frac{1}{\sqrt{1-\varsigma^2_\perp }}\frac{dz d\vec p^2_\perp d \varphi}{4z},
\end{eqnarray}
where $\psi$ is the azimuthal angle of scattered lepton, and $\varphi$ is the azimuthal angle of the produced hadron. 

In the following we find important to introduce the variables $x_S$ and $z_S$, that are the collinear fractions of parton momentum which include kinematic power corrections,
\begin{eqnarray}\label{def:x1z1}
x_S=-\frac{q^+}{P^+},\qquad z_S=\frac{p_h^-}{q^-},
\end{eqnarray}
which are invariant under boosts along the direction of $n$, $\bar n$, but are not invariant for a generic Lorentz transformation.
The  variables $x_S$ and $z_S$ in eq.~(\ref{def:x1z1}) are
\begin{eqnarray}\label{def:SIDIS-x1}
x_S&=&-x\frac{2}{\gamma^2}\(1-\sqrt{1+\gamma^2\(1-\frac{\vec q_T^2}{Q^2}\)}\),
\\\label{def:SIDIS-z1}
z_S&=&-z\frac{1-\sqrt{1+\gamma^2\(1-\frac{\vec q_T^2}{Q^2}\)}}{\gamma^2}\frac{1+\sqrt{1-\varsigma^2}}{1-\frac{\vec q_T^2}{Q^2}}=z\frac{x_S}{x}\frac{1+\sqrt{1-\varsigma^2}}{2\(1-\frac{\vec q_T^2}{Q^2}\)},
\end{eqnarray}
where we have used the variable $\vec q_T^2$ (\ref{def:qT<->pT}) for simplicity.

The kinematic corrections presented above are usually small when  $Q\gg M,m$. In this case the relation between observed and factorization variables simplifies
\begin{eqnarray}
\label{eq:APPX}
\vec q_T^2\simeq \frac{\vec p_\perp^2}{z^2},\qquad x_S\simeq x\(1-\frac{\vec q_T^2}{Q^2}\),\qquad z_S\simeq z.
\end{eqnarray}
Notice that the data of SIDIS at our disposal are taken at energies comparable with hadron masses and thus target mass correction could be significant. The contributions dependent on hadron masses could in principle be classified as power corrections. 
However we consider more appropriate to distinguish these corrections from others of different origin. Thus we will not use the approximate formulas in eq.~(\ref{eq:APPX}). The phenomenological test of this assumption is given in sec.~\ref{sec:SIDIS-power-corr}.
 
 \subsubsection{Factorization for the hadronic tensor in SIDIS}
In this work we consider the  transverse momentum dependence of the cross section which is factorizable in terms of transverse momentum dependent (TMD) distributions in the limit of $q_T\ll Q$, where $q_T$ is defined in  eq.~(\ref{def:qT<->pT}) and  $Q$ is the di-lepton invariant mass. We refer to the literature about the proof of  factorization of  the processes related to this work~\cite{Collins:1989gx,Becher:2010tm,Collins:2011zzd,GarciaEchevarria:2011rb,Echevarria:2012js,Echevarria:2014rua,Chiu:2012ir,Vladimirov:2017ksc}. In order to specify the properties of the TMD distributions and the factorized hadronic tensor, we start fixing the basic notation.

For unpolarized hadrons, the factorized hadronic tensor  and in its complete  form reads
\begin{eqnarray}\label{SIDIS:Wmunu}
W^{\mu\nu}&=&-2z_S \sum_{f}e_f^2 |C_V(Q^2,\mu^2)|^2\int \frac{d^2b}{(2\pi)^2}e^{-i(bq_T)} 
\\\nn &&\times \Big[ g_T^{\mu\nu}
f_{1,f\ot H}\(x_{S},b;\mu,\zeta_1\)D_{1,f\to h}\(z_S,b;\mu,\zeta_2\)
\\\nn && +(g_T^{\mu\nu}b^2-2b^\mu b^\nu)\frac{m\,M}{4} h_{1,f\ot H}^\perp\(x_S,b;\mu,\zeta_1\)H_{1,f\to h}^\perp\(z_S,b;\mu,\zeta_2\)\Big]+
O\(\frac{q_T^2}{Q^2}\),
\end{eqnarray}
where the index $f$ in the sum  runs through all quark flavours (including anti-quarks), $e_f$ is a charge of a quark measured in units of $e$. The function $C_V$ is the matching coefficient for vector current to collinear/anti-collinear vector current and the factorization ($\mu$) and rapidity ($\zeta$) scales typical of the TMD factorization are shown explicitly.
 
The unpolarized TMDPDF and TMDFF from partons of flavor $f$ are defined as
\begin{eqnarray}\label{def:f1}
&&f_{1,f\ot h}(x,b;\mu,\zeta)=
\\\nn &&\qquad\int \frac{d\lambda}{2\pi}e^{-ix\lambda p^+}
\sum_X \langle h(p)|\bar q(n \lambda+b) W^\dagger_n(n\lambda+b)\frac{\gamma^+}{2}|X\rangle\langle X|W_n(0)q(0)|h(p)\rangle,
\\\label{def:D1}
&&D_{1,f\ot h}(z,b;\mu,\zeta)=
\\\nn &&\qquad\frac{1}{2zN_c}\int \frac{d\lambda}{2\pi}e^{i\lambda p^+/z}\sum_X\langle 0|\frac{\gamma^+}{2}W_n(n\lambda+b) q(n \lambda+b)|h(p),X\rangle\langle h(p),X|\bar  q(0)W^\dagger_n(0)|0\rangle.
\end{eqnarray}
 Here, $W_v(x)$ are Wilson lines rooted at $x$ and pointing along vector $v$ to infinity. In the case of SIDIS, the Wilson lines in TMDPDF(TMDFF) points to future (past) infinity. The functions $h_1^\perp$ and $H_1^\perp$ are Boer-Mulders and Collins functions respectively and they are defined as
\begin{eqnarray}\label{def:h1}
&&iM \epsilon_T^{\alpha\beta}b_\beta h^\perp_{1,f\ot h}(x,b;\mu,\zeta)=
\\\nn &&\qquad \int \frac{d\lambda}{2\pi}e^{-ix\lambda p^+}\sum_X\langle h(p)|\bar q(n \lambda+b)W^\dagger_n(n\lambda+b)\frac{i\sigma^{\alpha +}\gamma^5}{2}|X\rangle\langle X|W_n(0)q(0)|h(p)\rangle,
\\\label{def:H1}
&&iM \epsilon_T^{\alpha\beta}b_\beta H^\perp_{1,f\ot h}(z,b;\mu,\zeta)=
\\\nn&& \frac{1}{2zN_c}\int \frac{d\lambda}{2\pi}e^{i\lambda p^+/z}\langle 0|\frac{i\sigma^{\alpha +}\gamma^5}{2}W_n(n\lambda+b)q(n \lambda+b)|h(p),X\rangle\langle h(p),X|\bar  q(0)W^\dagger_n(0)|0\rangle,
\end{eqnarray}
where $\epsilon_T^{\mu\nu}=\epsilon^{+-\mu\nu}$. In formulas (\ref{def:f1}-\ref{def:H1}) we have omitted  for brevity the obvious details of operator definitions, such as $T$($\bar T$)-ordering, color and spinor indices, rapidity and ultraviolet renormalization factors.

Boer-Mulders and Collins functions in eq.~(\ref{def:h1},~\ref{def:H1}) do not contribute to the angle averaged cross-section, but they can appear when cuts on  phase space distributions of final particles are introduced by the experimental setup. 
 In this work we will not consider these effects, and leave their study for the future (see discussion in sec.~\ref{sec:power}).

The TMD distributions depend on $\vec b^2$ only. Therefore, the angular dependence can be integrated explicitly with the result
\begin{eqnarray}\label{SIDIS:Wmunu-J}
W^{\mu\nu}&=&\frac{z_S}{\pi} \sum_{f}e_f^2 \Big[(-g_T^{\mu\nu})W^f_{f_1D_1}(Q,|q_T|,x_S,z_S)
\\\nn && 
+\(g_T^{\mu\nu}-2\frac{q_T^\mu q_T^\nu}{q_T^2}\)W^f_{h_1^\perp H_1^\perp}(Q,|q_T|,x_S,z_S)\Big]+
O\(\frac{q_T^2}{Q^2}\),
\end{eqnarray}
where 
\begin{eqnarray}\label{def:WfD}
W^f_{f_1D_1}(Q,q_T,x_S,z_S)&=&|C_V(Q^2,\mu^2)|^2
\\\nn && \times\int_0^\infty db\,b J_0(bq_T)f_{1,f\ot H}\(x_S,b;\mu,\zeta_1\)D_{1,f\to h}\(z_S,b;\mu,\zeta_2\),
\\\label{def:WhH}
W^f_{h^\perp_1H^\perp_1}(Q,q_T,x_S,z_S)&=&\frac{mM}{4}|C_V(Q^2,\mu^2)|^2 
\\\nn &&  \times \int_0^\infty db\,b^3 J_2(bq_T)h_{1,f\ot H}^\perp\(x_S,b;\mu,\zeta_1\)H_{1,f\to h}^\perp\(z_S,b;\mu,\zeta_2\).
\end{eqnarray}
The functions $W^f_{ab}$ are dimensionless and scale-independent functions. The experimental configurations are not usually provided in the factorization frame, and the correspondence between the  measured quantities and the ones that appear in the factorization theorem is often non-trivial. It happens  in fact, that  a Lorentz transformation affects the power corrections to the  cross section presented here. We detail this in the next sections.

\subsubsection{Leptonic tensor in SIDIS}
The  leptonic tensor  for unpolarized SIDIS is
\begin{eqnarray}
L_{\mu\nu}=2(l_\mu l'_\nu+l'_\mu l_\nu- (ll') g_{\mu\nu} ).
\end{eqnarray}
In order to express the convolution of the leptonic tensor with a hadronic tensor we define the azimuthal angle of a produced hadron as \cite{Bacchetta:2006tn}:
\begin{eqnarray}
\cos\phi=\frac{-l_\mu p_{h\nu}g_\perp^{\mu\nu}}{\sqrt{-l_\alpha l_\beta g_\perp^{\alpha\beta} }\sqrt{-p_{h\alpha'} p_{h\beta'} g_\perp^{\alpha'\beta'} }}
\end{eqnarray}
and we  define
$$
\varepsilon=\frac{1-y-\frac{\gamma^2y^2}{4}}{1-y+\frac{y^2}{2}+\frac{\gamma^2 y^2}{4}}.
$$
As the result we obtain
\begin{eqnarray}\label{th:SIDIS-L1}
(-g_T^{\mu\nu}) L_{\mu\nu}&=&\frac{2Q^2}{1-\varepsilon}\Big[1+\frac{\vec p_\perp^2}{Q^2 z^2}\frac{\varepsilon-\frac{\gamma^2}{2}}{1-\varsigma^2}
\\\nn &&
 -\cos\phi \frac{\sqrt{2\varepsilon(1+\varepsilon)\vec p_\perp^2}}{zQ}\frac{\sqrt{1-\varsigma_\perp^2}}{1-\varsigma^2}
 -\cos(2\phi)\frac{\varepsilon\vec p_\perp^2 \gamma^2 }{2 z^2 Q^2(1-\varsigma^2)}
\Big].
\end{eqnarray}
The kinematical rearrangements of the variables produce the appearance of  the $\cos\phi$ and $\cos2\phi$ terms in the second line of eq.~(\ref{th:SIDIS-L1}),
that  is, there are contributions to the structure functions $F_{UU}^{\cos\phi}$ and $F_{UU}^{\cos2\phi}$, see also \cite{Anselmino:2005nn}.
Similarly, the convolution of lepton tensor with the spin-1 part 
\begin{eqnarray}\label{th:SIDIS-L2}
\(g_T^{\mu\nu}-2\frac{q_T^\mu q_T^\nu}{q_T^2}\) L_{\mu\nu}&=&\frac{2Q^2}{1-\varepsilon}\Big[\varepsilon \cos(2\phi)
\(1-\frac{\vec p_\perp^2 \gamma^2 }{2 z^2 Q^2(1-\varsigma^2)}\)
\\\nn &&
-\cos\phi \frac{\sqrt{2\varepsilon(1+\varepsilon)\vec p_\perp^2}}{zQ}\frac{\sqrt{1-\varsigma_\perp^2}}{1-\varsigma^2}
+\frac{\vec p_\perp^2}{Q^2 z^2}\frac{\varepsilon-\frac{\gamma^2}{2}}{1-\varsigma^2}
\Big].
\end{eqnarray}
produces also contribution to the $\cos\phi$ and $\cos2\phi$ parts.

The terms $\sim \vec p_\perp^2/Q^2$  in eqs.~(\ref{th:SIDIS-L1}, \ref{th:SIDIS-L2}) can be modified by power corrections to  TMD factorization, see discussion in sec.~\ref{sec:power}. 


\subsubsection{SIDIS cross-section in TMD factorization}

Combining together the expressions for   the cross-section in eq.~(\ref{def:xSec-0}), the differential phase-space volume in eq.~(\ref{th:phase-elem-1}), the hadronic tensor in eq.~(\ref{SIDIS:Wmunu-J}), the leptonic tensor in eq.~(\ref{th:SIDIS-L1},~\ref{th:SIDIS-L2}),  and integrating over the azimuthal angles  we obtain
\begin{eqnarray}\label{SIDIS:xSec}
&&\frac{d\sigma}{dxdz dQ^2 d\vec p_\perp^2}=\frac{\pi}{\sqrt{1-\varsigma_\perp^2}}\frac{\alpha_{\text{em}}^2}{Q^4}\frac{y^2}{1-\varepsilon}\frac{z_S}{z}
\\\nn &&
\qquad\times \sum_{f}e_f^2\Big[
\(1+\frac{\vec q_T^2}{Q^2}\frac{\varepsilon-\frac{\gamma^2}{2}}{1+\gamma^2}\)W_{f_1D_1}^f(Q,\sqrt{\vec q^2_T},x_S,z_S)
+\frac{\vec q_T^2}{Q^2}\frac{\varepsilon-\frac{\gamma^2}{2}}{1+\gamma^2}W_{h_1^\perp H_1^\perp}^f(Q,\sqrt{\vec q^2_T},x_S,z_S)
\Bigg],
\end{eqnarray}
where $x_S$, $z_S$ and $\vec q^2_T$ are the functions of $\vec p_\perp^2$, $x$, and $z$ defined in eq.~(\ref{def:SIDIS-x1}),~(\ref{def:SIDIS-z1}) and eq.~(\ref{def:qT<->pT}), correspondingly. The functions $W^f_{ab}$ are defined in eq.~(\ref{def:WfD},~\ref{def:WhH}).

The final expression for the cross section in eq.~(\ref{SIDIS:xSec}) explicitly shows that part of power corrections has a kinematical origin, and therefore, it is independent of the factorization theorem and it can be taken into account in the present formalism without contradictions. As an example one can consider the factor $\sqrt{1-\varsigma_\perp^2}$ that is a part of the phase-space element, and the difference between $z_S$ and $z$ that is a consequence of the TMDFF definition. The separation between kinematical power corrections and higher orders in power expansion of the cross-section is however not neat, because a detailed study of the factorization theorem correction is still not complete.  The admixture of  these effects can be seen in the second line  of eq.~(\ref{SIDIS:xSec}), which is the present status of our understanding. In the fit  we omit the contribution $W_{h_1^\perp H_1^\perp}$ in eq.~(\ref{SIDIS:xSec}) and perform a check of the  importance of mass-corrections for the agreement with experimental data in sec.~\ref{sec:SIDIS-power-corr}. We discuss power corrections also in sec.~\ref{sec:power}.

\subsection{DY cross-section}

The Drell-Yan pair production (or DY for shortness) is defined by the process
\begin{eqnarray}
h_1(P_1)+h_2(P_2)\to l(l)+l'(l')+X,
\end{eqnarray}
where $l$, $l'$ are the lepton pair, $h_1$, $h_2$ are the colliding hadrons, and the symbols in brackets denote the momentum of each particle. In the following, we include  hadron masses and  
 we neglect  lepton masses:
\begin{eqnarray}
P_1^2=M_1^2,\qquad P_2^2=M_2^2,\qquad l^2=l'^2=m_l^2\simeq0.
\end{eqnarray}
The energies of the DY experiments are higher than the SIDIS ones, and the interference of electro-weak (EW) bosons must be included. Approximating the interactions of leptons and hadrons by a single EW-gauge boson exchange one obtains the following expression for the differential cross-section
\begin{eqnarray}\label{th:DY-xsec0}
d\sigma=\frac{2\alpha^2_{\text{em}}}{s} \frac{d^3l}{2E}\frac{d^3l'}{2E'}\sum_{GG'}  L^{GG'}_{\mu\nu}W_{GG'}^{\mu\nu}\Delta_G(q)\Delta^*_{G'}(q).
\end{eqnarray}
where $q=l+l'$, $\alpha_{\text{em}}=e^2/4\pi$ is the QED coupling constant and the index $G$ runs over gauge bosons $\gamma$, $Z$. Here, we have approximated the exact flux factor $[(s-(M_1-M_2)^2)(s-(M_1+M_2)^2)]^{-1/2}$ with $1/s$ because the corrections of order $M^2/s$ are negligibly small for any considered data set. The function $\Delta_G(q)$ is defined as
\begin{align}
\Delta_G(q)=\frac{1}{q^2+i0}\delta_{G\gamma}+\frac{1}{q^2-M_Z^2+i \Gamma_Z M_z}\delta_{GZ},
\end{align}
with $M_Z=91.188$GeV and $\Gamma_Z=2.495$GeV \cite{Olive:2016xmw}.   Finally, $L_{GG'}^{\mu\nu}$ and $W_{GG'}^{\mu\nu}$ are the leptonic and hadronic tensors that are defined as
\begin{eqnarray}\label{DY:L1}
L^{GG'}_{\mu\nu}&=&e^{-2}\langle 0|J^G_{\mu}(0)|l, l'\rangle \langle l, l'|J^{G'\dagger}_{\nu}(0)|0\rangle,
\\
W_{\mu\nu}&=&e^{-2}\int \frac{d^4x}{(2\pi)^4}e^{-i(x\cdot q)}\sum_X\langle P_1,P_2|J^{G\dagger}_{\mu}(x)|X\rangle \langle X|J^{G'}_{\nu}(0)|P_1,P_2\rangle,
\end{eqnarray}
where $e$ is the lepton charge, and $J^G_\mu$ is the current for the production of EW gauge boson $G$. 

Integrating the cross-section over a lepton momentum one finds
\begin{eqnarray}\label{th:DY-xsec1}
d\sigma=\frac{2\alpha^2_{\text{em}}}{s}d^4q\sum_{GG'}\widehat L^{GG'}_{\mu\nu}W_{GG'}^{\mu\nu}\Delta_G(q)\Delta^*_{G'}(q),
\end{eqnarray}
where $q$ is the momentum of the EW-gauge boson. The new lepton tensor is
\begin{eqnarray}\label{DY:Lhat}
\widehat L^{GG'}_{\mu\nu}=\int \frac{d^3l}{2E}\frac{d^3l'}{2E'} \delta^{(4)}(l+l'-q)L^{GG'}_{\mu\nu}.
\end{eqnarray}

\subsubsection{Kinematic variables for DY}
The relevant kinematic variable in DY read
\begin{eqnarray}
s=(P_1+P_2)^2,\qquad q^2=Q^2,\qquad y=\frac{1}{2}\ln\frac{q^+}{q^-}.
\end{eqnarray}
The transverse components are projected by a tensor $g_T^{\mu\nu}$, that is orthogonal to $P_1^\mu$ and $P_2^\mu$, identically to the SIDIS case eq.~(\ref{def:gT-SIDIS}),
\begin{eqnarray}\label{eq:gmunuDY}
g_T^{\mu\nu}=g^{\mu\nu}-\frac{2}{s}\(P_1^\mu P_2^\nu+P_2^\mu P_1^\nu\),
\end{eqnarray}
and we have dropped the negligible corrections of order of $M^2/s$. In this limit, the factorization theorem is expressed in the center-of-mass frame, the components of momenta are $P_1^+=P_2^-=\sqrt{s/2}$ and the variables $x_{1,2}$ in eq.~(\ref{def:x1x2}) are
\begin{eqnarray}
x_1=\sqrt{\frac{Q^2+\vec q_T^2}{s}}e^{+y},\qquad x_2=\sqrt{\frac{Q^2+\vec q_T^2}{s}}e^{-y}.
\end{eqnarray}
The differential phase-space element reads
\begin{eqnarray}\label{DY:dQ}
d^4q=\frac{1}{2}dQ^2dyd^2q_T=\frac{1}{4}dQ^2dy d\vec q_T^2 d\varphi,
\end{eqnarray}
where $\varphi$ is the azimuthal angle of the vector boson. 

\subsubsection{Factorization for hadronic tensor in DY}
\label{sec:DY:hadron-tensor}

The factorization for DY hadronic tensor is totally analogous to the SIDIS case. The vectors $n$ and $\bar n$ are defined by hadrons,
\begin{eqnarray}\label{def:DY-nn}
P_1^\mu&=&P_1^+ \bar n^\mu+\frac{M_1^2}{2P_1^+}n^\mu\simeq P_1^+ \bar n^\mu,\qquad P_2^\mu =P_2^- n^\mu+\frac{M_2^2}{2P_2^-}\bar n^\mu\simeq P_1^- n^\mu,
\end{eqnarray}
where on r.h.s. the small contributions $\sim M^2/s$ are neglected. The inclusion of weak-boson exchange requires the consideration of a more general current. To this purpose we define
\begin{eqnarray}\label{def:J-EW}
J^\mu_G(x)=\bar q(x)[g^G_R\gamma^\mu(1+\gamma^5)+g^G_L\gamma^\mu(1-\gamma^5)]q(x),
\end{eqnarray}
with the  EW coupling constants 
\begin{eqnarray}\label{def:gRgL}
g^\gamma_R=g^\gamma_L=\frac{e_f}{2},\qquad g_R^Z=\frac{-e_fs_W^2}{2s_Wc_W},\qquad g^Z_L=\frac{T_3-e_f s_W^2}{2s_Wc_W},
\end{eqnarray}
where $e_f$ is the electric charge of a particle (in units of $e$), $T_3$ is the third projection of weak isospin, $s_W=\sin(\theta_W)$, $c_W=\cos(\theta_W)$.

Collecting all this, the unpolarized part of the factorized hadronic tensor reads
\begin{eqnarray}\label{DY:Wmunu}
W^{\mu\nu}_{GG'}&=&\sum_{f}|C_V(-Q^2,\mu^2)|^2\int \frac{d^2b}{(2\pi)^2}e^{-i(bq_T)} \Big[
\\\nn && -2g_T^{\mu\nu}(g^G_Rg^{G'}_R+g^G_Lg^{G'}_L)\(f_{1,f\ot h_1}f_{1,\bar f\ot h_2}+f_{1,\bar f\ot h_1}f_{1,f\ot h_2}\)
\\\nn &&
-\frac{g_T^{\mu\nu}b^2-2b^\mu b^\nu}{2}M_1M_2(g^G_Rg^{G'}_R+g^G_Lg^{G'}_L)\(h^\perp_{1,f\ot h_1}h^\perp_{1,\bar f\ot h_2}+h^\perp_{1,\bar f\ot h_1}h^\perp_{1,f\ot h_2}\)
\\\nn &&
-2i\epsilon_T^{\mu\nu}(g^G_Rg^{G'}_R-g^G_Lg^{G'}_L)\(f_{1,f\ot h_1}f_{1,\bar f\ot h_2}-f_{1,\bar f\ot h_1}f_{1,f\ot h_2}\)
\\\nn &&
+i(\epsilon_T^{\mu\alpha} b_\alpha b^\nu+\epsilon_T^{\nu\alpha} b_\alpha b^\mu)\frac{M_1M_2}{2}(g^G_Rg^{G'}_L-g^G_Lg^{G'}_R)\(h^\perp_{1,f\ot h_1}h^\perp_{1,\bar f\ot h_2}-h^\perp_{1,\bar f\ot h_1}h^\perp_{1,f\ot h_2}\)\Big]
\\\nn && + O\(\frac{q_T^2}{Q^2}\),
\end{eqnarray}
where $f$ runs through all quark flavours. 
The functions $f_{1,f\ot h_1}$ are the TMDPDF and the functions $h^\perp_{1,f\ot h_1}$ are the Boer-Mulders functions, defined  in eq.~(\ref{def:f1}, \ref{def:h1}). In eq.~(\ref{DY:Wmunu}) we have omitted the arguments of the TMD distributions for brevity, however they can be included with substitutions like  e.g.
\begin{eqnarray}
f_{1,f\ot h_1}f_{1,\bar f\ot h_2}\to f_{1,f\ot h_1}(x_1,b;\mu,\zeta_1)f_{1,\bar f\ot h_2}(x_2,b;\mu,\zeta_2),
\end{eqnarray}
and proceed similarly for all products of TMD distributions. The variables $x_1$ and $x_2$ measure the collinear fractions of parton momenta,
\begin{eqnarray}\label{def:x1x2}
x_1=\frac{q^+}{P_1^+},\qquad x_2=\frac{q^-}{P_2^-}.
\end{eqnarray}
The flavor indices $f$, $\bar f$ run through all flavors of quarks and anti-quarks respectively. Here, the flavor index $\bar f$ refers to the anti-parton of $f$.  Note that, in the case of W-boson, the constants $g_{L/R}^W$ mix the flavors of quarks. 

In the factorized hadronic tensor in eq.~(\ref{DY:Wmunu}), different terms are not equally important. In fact, the fifth line of eq.~(\ref{DY:Wmunu}) vanishes identically due to the peculiar combination of $g$-constants that is null for any electro-weak channel. The forth line can contribute only to $ZZ$ and $Z\gamma$-channels, that have an anti-symmetric part of the leptonic tensor. However, the resulting expression is anti-symmetric in the rapidity parameter, and thus it vanishes when the rapidity is measured/integrated on symmetric intervals. In principle, this part can contribute to a cross-section when experiments perform very asymmetric kinematic cuts on the detected leptons (e.g. at LHCb). However, even in this case the resulting integral is suppressed as $q_T^2/Q^2 e^{-2|y|}$ and it is numerically very small, e.g in some bins it can give a $10^{-6}-10^{-8}$-size relative to the leading contribution. Thus, in the following we do not consider contributions of the last two lines in eq.~(\ref{DY:Wmunu}).

Performing the integration over angles we  obtain a result formally similar to the SIDIS case in eq.~(\ref{SIDIS:Wmunu-J}),
\begin{eqnarray}\label{DY:Wmunu-J}
W^{\mu\nu}_{GG'}&=&\frac{1}{2\pi}\sum_{f}2(g^G_Rg^{G'}_R+g^G_Lg^{G'}_L)\Big[-g_T^{\mu\nu} W^f_{f_1f_1}(Q,|q_T|,x_1,x_2)
\\\nn && \qquad\qquad +\(g_T^{\mu\nu}-2\frac{q_T^\mu q_T^\nu}{q_T^2}\)W^f_{h_1^\perp h_1^\perp}(Q,|q_T|,x_1,x_2)\Big] + O\(\frac{q_T^2}{Q^2}\),
\end{eqnarray}
where $g_T^{\mu\nu}$ is now given in eq.~(\ref{eq:gmunuDY}) and 
\begin{eqnarray}\label{def:Wff}
W^f_{f_1f_1}(Q,q_T,x_1,x_2)&=&|C_V(-Q^2,\mu^2)|^2
\\\nn && \qquad\int_0^\infty db\,bJ_0(bq_T)f_{1,f\ot h_1}(x_1,b;\mu,\zeta_1)f_{1,\bar f\ot h_2}(x_2,b;\mu,\zeta_2),
\\\label{def:Whh}
W^f_{h_1^\perp h_1^\perp}(Q,q_T,x_1,x_2)&=&\frac{M_1M_2}{4}|C_V(-Q^2,\mu^2)|^2
\\\nn &&\qquad\times
\int_0^\infty db\,b^3J_2(bq_T)h^\perp_{1,f\ot h_1}(x_1,b;\mu,\zeta_1)h^\perp_{1,\bar f\ot h_2}(x_2,b;\mu,\zeta_2).
\end{eqnarray}

\subsubsection{Lepton tensor and fiducial cuts in DY}
\label{sec:DY-fiducial}

In experiments not all final state leptons are collected in the measurements and fiducial cuts are for instance performed at LHC. We use the same implementation of cuts as in~\cite{Scimemi:2017etj,Bertone:2019nxa}. However, here we give a more general discussion to see how they affect power suppressed parts of the cross section.

The lepton tensor of unpolarized DY formally written in  eq.~(\ref{DY:L1}) is
\begin{eqnarray}
L_{\mu\nu}^{GG'}&=&8\Big[(l^\mu l'^\nu+l^\nu l'^\mu-g^{\mu\nu}(ll'))\(g_{G}^Rg_{G'}^R+g_{G}^Lg_{G'}^L\)
 + i \epsilon^{\mu\nu \alpha\beta}l_\alpha l'_\beta \(g_{G}^Rg_{G'}^R-g_{G}^Lg_{G'}^L\)\Big],
\end{eqnarray}
where $g_G^{R}$($g_G^{L}$) are the couplings of right (left) components of a lepton field to EW current as in eq.~(\ref{def:J-EW}). In the case of $W$ boson, these couplings also carry flavor indices. As  discussed in sec.~\ref{sec:DY:hadron-tensor}, the anti-symmetric part does not contribute visibly to the unpolarized cross-section even in the presence of asymmetric fiducial cuts.

\begin{figure}
\begin{center}
\includegraphics[width=0.4\textwidth]{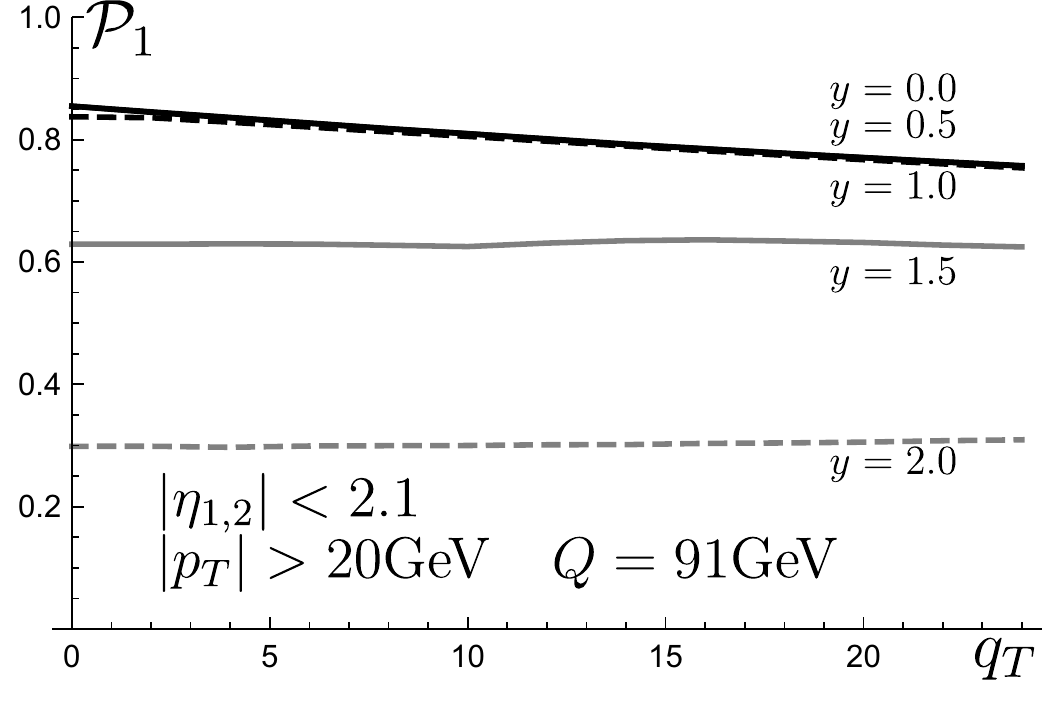}
~~
\includegraphics[width=0.4\textwidth]{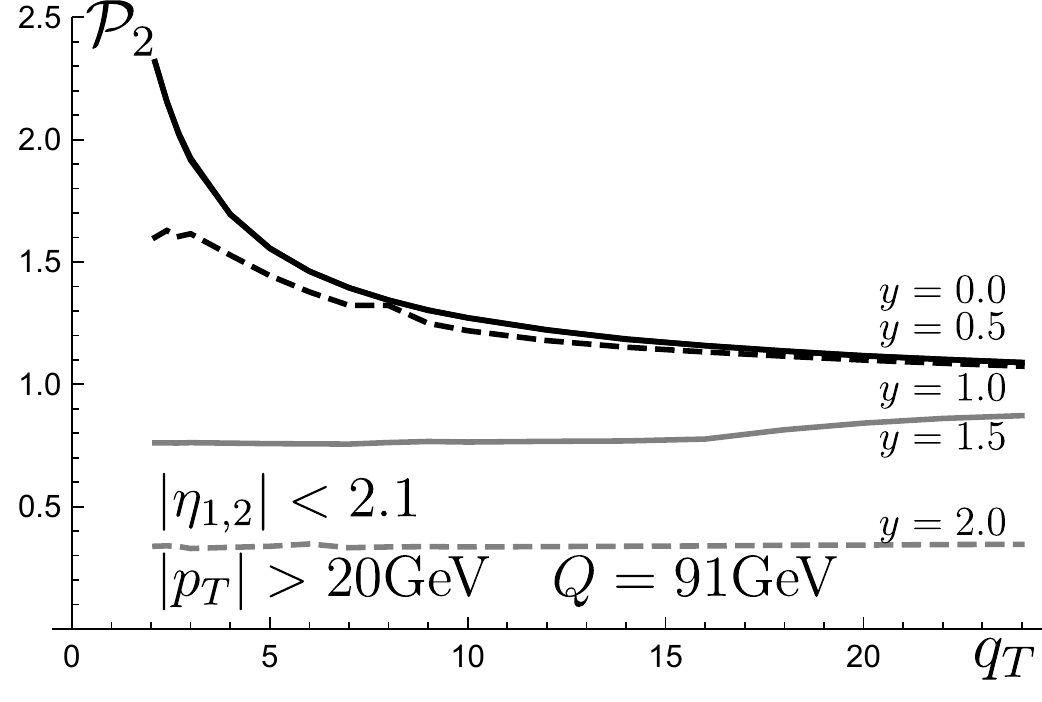}
\end{center}
\caption{\label{fig:cut-factors} Plot of cut-factors $\mathcal{P}_{1,2}$ (\ref{DY:P1},\ref{DY:P2}) versus $q_T$[GeV] in the case of fiducial cuts of ATLAS and CMS Z-boson measurement \cite{Aad:2014xaa,Aad:2015auj,Chatrchyan:2011wt,Khachatryan:2016nbe}, at $Q=91.$GeV and different values of $y$. The lines with $y=0.0$ and $y=0.5$ are very close to each other.}
\end{figure}

The DY cross-section contains the lepton tensor integrated over the lepton momenta with $l+l'=q$, in eq.~(\ref{DY:Lhat}), and this gives
\begin{align}\label{DY:lG}
(-g_T^{\mu\nu})\hat{L}^{GG'}_{\mu\nu}&=16\(g_{G}^Rg_{G'}^R+g_{G}^Lg_{G'}^L\)\int \frac{d^3l}{2E}\frac{d^3l'}{2E'}\delta^{(4)}(l+l'-q)((ll')-(ll')_T)
\\\nn &=\[2\(g_{G}^Rg_{G'}^R+g_{G}^Lg_{G'}^L\)\]\frac{4\pi}{3}Q^2\(1+\frac{\vec q_T^2}{2Q^2}\),\\
\label{DY:lGG}
(g_T^{\mu\nu}-2\frac{q_T^\mu q_T^\nu}{q_T^2})\hat{L}^{GG'}_{\mu\nu}&= -32\(g_{G}^Rg_{G'}^R+g_{G}^Lg_{G'}^L\)
\\\nn &\times\int \frac{d^3l}{2E}\frac{d^3l'}{2E'}\delta^{(4)}(l+l'-q)
\frac{2l_T^2l'_T+(ll')_Tl_T^2+(ll')_T{l'}^2_T}{q_T^2}
\\\nn &=\[2\(g_{G}^Rg_{G'}^R+g_{G}^Lg_{G'}^L\)\]\frac{4\pi}{3}Q^2\frac{\vec q_T^2}{Q^2}.
\end{align}
 The cuts on the lepton pair at LHC are usually reported as 
\begin{eqnarray}
\eta_{\text{min}}<\eta,\eta'<\eta_{\text{max}},\qquad l_T^2>p_1^2,\qquad {l'}^2_T>p_2^2,
\end{eqnarray}
where $\eta$ and $\eta'$ are pseudo-rapidity of the leptons. In the presence of these cuts the integration volume of the leptonic tensor can be done only numerically. To account this effect we introduce cut factors as
\begin{eqnarray}\label{DY:P1}
\mathcal{P}_1&=&\int \frac{d^3l}{2E}\frac{d^3l'}{2E'}\delta^{(4)}(l+l'-q)((ll')-(ll')_T)\theta(\text{cuts})\Big/\[\frac{\pi}{6}Q^2\(1+\frac{\vec q_T^2}{2Q^2}\)\]^{-1},
\\\label{DY:P2}
\mathcal{P}_2&=&\frac{12}{\pi}\int \frac{d^3l}{2E}\frac{d^3l'}{2E'}\delta^{(4)}(l+l'-q)(2l_T^2l'^2_T+(ll')_Tl_T^2+(ll')_T{l'}^2_T)\theta(\text{cuts}).
\end{eqnarray}
These factors are equal to one in the absence of cuts. The impact of these cuts  at LHC is extremely important  and  depends on the rapidity interval and the value of the vector boson transverse momentum. We  show $\mathcal{P}_{1,2}$ for ATLAS experiment in  fig.~\ref{fig:cut-factors}. One can see that the factor $\mathcal{P}_2$ is enhanced at smaller $q_T$ and in general these factors are very different from 1.

\subsubsection{DY cross-section in TMD factorization}

Collecting the expressions for  the differential phase-space element in eq.~(\ref{DY:dQ}), the hadronic tensor eq.~(\ref{DY:Wmunu-J}), the leptonic tensor (\ref{DY:lG},~\ref{DY:lGG}) with the fiducial cuts in eq.~(\ref{DY:P1},~\ref{DY:P2}), we obtain the final cross-section in the TMD factorization. For the case of neutral vector boson (i.e. Z- and $\gamma$- bosons) it reads
\begin{eqnarray}\label{DY:xSec}
&&\frac{d\sigma}{dQ^2dy d\vec q_T^2}=\frac{2\pi}{3N_c}\frac{\alpha^2_{\text{em}}}{sQ^2}\sum_{f}\Big[\(1+\frac{\vec q_T^2}{2Q^2}\)\mathcal{P}_1W_{f_1f_1}^f(Q,\sqrt{\vec q_T^2})+\frac{\vec q_T^2}{Q^2}\mathcal{P}_2W_{h_1^\perp h_1^\perp}^f(Q,\sqrt{\vec q_T^2})\Big]
\\\nn && \qquad
\times \Big[z^{\gamma\gamma}_lz_f^{\gamma\gamma}+z^{\gamma Z}_lz_f^{\gamma Z}\frac{2Q^2(Q^2-M_Z^2)}{(Q^2-M_Z^2)^2+\Gamma_Z^2M_Z^2}+
z^{ZZ}_lz_f^{ZZ}\frac{Q^4(Q^2-M_Z^2)}{(Q^2-M_Z^2)^2+\Gamma_Z^2M_Z^2}\Big],
\end{eqnarray}
where functions $W^f$ are defined in (\ref{def:Wff},~\ref{def:Whh}), $M_Z$ and $\Gamma_Z$ are mass and width of Z-boson. The factors $z$ are the  combinations of couplings $g^{R,L}$ for quarks and for leptons (\ref{def:gRgL}):
\begin{eqnarray}
z_f^{\gamma\gamma}&=&e_f^2,
\\
z_f^{\gamma Z}&=&\frac{T_3-2e_fs_W^2}{2s_W^2c_W^2},
\\
z_f^{Z Z}&=&\frac{(1-2|e_f|s_W^2)^2+4e_f^2s_W^4}{8s_W^2c_W^2}.
\end{eqnarray}
The term $W_{h_1^\perp h_1^\perp}^f$ describes the contributions of the Boer-Mulders functions and we omit this term in  the rest of the fit as motivated  in section~\ref{sec:power}.

\subsection{Power corrections and higher twist structure functions}
\label{sec:power}

The cross-section of SIDIS and DY given by eq.~(\ref{SIDIS:xSec},~\ref{DY:xSec}) 
contains a variety of power suppressed contributions, which have different origin, as listed in the following:
\begin{itemize}
\item Power corrections to TMD factorization. These corrections appear during the factorization procedure for the hadronic tensors, see eq.~(\ref{SIDIS:Wmunu},~\ref{DY:Wmunu}). One can distinguish two kinds of power corrections: corrections that are proportional to the leading structure functions $W_{ab}$, which arise through the so-called Wandzura–Wilczek terms (in the case of SIDIS, this part of cross-section has  been studied recently in \cite{Bastami:2018xqd});  corrections that involve genuine ``twist-3'' TMD distributions (some part of these corrections is discussed in~\cite{Balitsky:2017gis});
\item  Mass and $\vec q_T^2$ dependence within the momentum fraction  variables $(x_S,z_S)$ (SIDIS), $(x_1,x_2)$ (DY), see eq.~(\ref{def:x1z1}, \ref{def:x1x2}). Despite the fact that the corrections in the momentum fraction can be interpreted as  part of  power corrections to TMD factorization (contributing to the Wandzura–Wilczek terms), we consider them on their own. These corrections come from the field-modes separation and the definition of the scattering plane, and they can be seen as the ``Nachtmann-variable for TMD factorization''. The usage of these variables is also in agreement with expected large-$q_T$ structure of cross-section, which has different form, but uses similar variables, e.g. see \cite{Boer:2006eq}.
\item Fiducial cuts for DY. The cut factors for the DY lepton tensor in eq.~(\ref{DY:P1},~\ref{DY:P2}) are a source of power corrections and they can mix  different structure functions. They are accumulated in  separate factors, and have totally auxiliary nature. They must be accounted for the proper description of LHC data.
\item  Mismatch between factorization and laboratory frames in SIDIS. The azimuthal angles and transverse planes are defined differently in the factorization and laboratory frames see eq.~(\ref{def:gPerp-SIDIS}, \ref{def:gT-SIDIS}). This introduces target-mass, produced-mass, and $q_T$-corrections. A good example is the  $\vec p_\perp$-linear contribution to the structure function $F_{UU}^{\cos\phi}$ (\ref{th:SIDIS-L1}, \ref{th:SIDIS-L2}), which is a purely a frame-dependent effect. 
\item  Cross-section phase-space volume in SIDIS. In the case of a non-negligible mass for the detected particle, the phase-volume contains  power corrections. They are accumulated in a universal factor in eq.~(\ref{th:phase-elem-1}), and are  part of the definition of the observable. 
\end{itemize}
Some of the power corrections of this list can be accounted exactly (e.g. the corrections to the phase-space, the collinear momentum fractions, the relation between $\vec q^2_T$ and $\vec p_\perp^2$), while some are absolutely unknown (i.e.  the power correction to the TMD factorization). 

The problem of power corrections to TMD factorization is unsolved and should be addressed in future studies. We resume it here for the interested readers in the DY case. The hadronic tensor defined in eq.~(\ref{DY:Wmunu-J}) is expressed in terms of the tensor $g^{\mu\nu}_T$ defined in eq.~(\ref{eq:gmunuDY}) and it is transverse to a plane containing hadrons. The appearance of the tensor $g_T^{\mu\nu}$ is the consequence of the TMD factorization approach. This tensor is not transverse to the vector boson momentum, and as a result whenever one uses the leading term of factorized formula for the cross section one finds 
\begin{eqnarray}\label{th:QED-ward}
q_\mu W^{\mu\nu}_{\text{TMD fact.}}\neq 0,
\end{eqnarray}
which demonstrates the violation of QED Ward identity. The violation can be  accounted for as a power-suppressed contribution, since $q_\mu W^{\mu\nu}\sim q_T$. Accounting of the linear power correction $(\sim q_T/Q)$ would correct the QED Ward identity to this order (i.e. one would obtain $q_\mu W^{\mu\nu}\sim q^2_T\neq 0$). In order to get a hadron tensor completely transverse to $q_\mu$ one has to account for the full chain of power corrections. This problem is well known and it has been addressed several times in the literature in DY and SIDIS cases~\cite{Mulders:1995dh,Bacchetta:2004jz,Bacchetta:2006tn,Arnold:2008kf,Collins:2011zzd,Nefedov:2018vyt}. All the suggested solutions extend the TMD factorization in some model-dependent way and they provide different expressions for the cross-section. A systematic solution is still not available. It is also often assumed that the resummation of Sudakov logarithms and the matching to the perturbative expansion of the cross section can interpolate between the TMD factorization region and the perturbative region. This method however presents its own limitations because in practice not all sources of power corrections listed above are usually taken into account and a more systematic work in this sense is still missing.


In the present work we adopt a different strategy. We first observe that  power suppressed terms have not a single origin and that part of them are calculable, so that they can be included in our computations. The TMD factorization provides the cross section for DY and SIDIS in terms of  4 structure functions $W_{ab}$ defined in  eq.~(\ref{def:WfD},~\ref{def:WhH},~\ref{def:Wff},~\ref{def:Whh}) and each of them is a  Hankel convolution of two TMD distributions times a hard coefficient function. We remark that the TMD include all the non-perturbative information of the process, and it is different from the one contained in a collinear PDF. The unknown parts in eq.~(\ref{SIDIS:xSec},~\ref{DY:xSec}) come from higher twist matrix elements $W_{h_1^\perp H_1^\perp}$ and $W_{h_1^\perp h_1^\perp}$ which are expected to  contribute at larger values of $q_T$. 

The structure functions $W_{h_1^\perp H_1^\perp }$ and $W_{h_1^\perp h_1^\perp }$ are formally of higher dynamical twist with respect to the others. While higher twist contributions are in principle accompanied by $\vec q_T^2/Q^2$ factors, the complex kinematics of the experiments (especially in the SIDIS case) makes it hard to distinguish  purely non-perturbative higher-twist effects from the kinematical ones. For instance, the azimuthal angles measured in the lab frames and in the Breit frame for SIDIS are different and some non-perturbative QCD effects can be overlooked when we pass from one frame to the other. The only way to solve this problem would be a complete inclusion of higher power corrections to the cross section, which goes beyond the scope of the present work. For this reason, while we  consider the exact kinematics, as described in the previous section,  we also put
\begin{eqnarray}
W_{h_1^\perp H_1^\perp }^f(Q,q_T,x,z)=0,\qquad W_{h_1^\perp h_1^\perp }^f(Q,q_T,x,x')=0.
\end{eqnarray}
The effect of this assumption must be very small at $\vec q_T^2\ll Q^2$, and this justifies the conservative data sets  used in the present fit (see sec.~\ref{sec:data}).

The $Q$ dependence of $W^f_{ab}$ is dictated by the TMD evolution, and it is discussed in the next section~\ref{sec:evolution}. The asymptotic limit of high $q_T$ allows for a perturbative matching of TMD distributions to collinear ones and it is discussed in sec.~\ref{sec:matching}. The non-perturbative inputs on top of the large-$q_T$ asymptotic limit are discussed in sec.~\ref{sec:ansatzNP}.  Finally, we summarize all theoretical inputs in sec.~\ref{sec:summary-theory}.

In sec. \ref{sec:DY:x12} and \ref{sec:SIDIS-power-corr}, we test the influence of the power corrections to the fit quality. This test provides us an estimation of the systematic error due to the presence of unknown power correction.

\subsection{TMD evolution and optimal TMD distributions}
\label{sec:evolution}

While the differential evolution  equations for TMD  are fixed by the factorization theorem, the boundary conditions of their solution are a matter of choice. They clearly determine the convergence of the perturbative series and the success of the theoretical description of DY and SIDIS spectrum. In this paper, we work with the so-called $\zeta$-prescription described in \cite{Scimemi:2018xaf}, and including the improvement found in \cite{Vladimirov:2019bfa}. The prescription consists in defining the  TMD distribution on a null-evolution line. The null-evolution line has the defining property of keeping the evolution factor for TMD distributions is equal to one for all values of the impact parameter $b$. Because of this property, the $\zeta$-prescription is conceptually different from other popular prescriptions, where the reference scales do not belong to a null-evolution line. In this case, the resulting (reference) TMD distribution includes an admixture with the perturbative evolution factor evaluated at different values of $b$. Thus it appears that  the $\zeta$-prescription has an important advantage that the resulting TMD distribution is independent of any perturbative parameter, i.e. it is completely non-perturbative and one can freely parameterize a distribution without any reference to perturbative order. For a detailed description and analyses of TMD evolution and the $\zeta$-prescription we refer to \cite{Scimemi:2018xaf}, whereas here we present only the final expressions without derivation.

The system of TMD evolution equations is 
\begin{eqnarray}\label{def:TMD_ev_UV}
\mu^2 \frac{d}{d\mu^2} F(x,b;\mu,\zeta)&=&\frac{\gamma_F(\mu,\zeta)}{2}F(x,b;\mu,\zeta),
\\\label{def:TMD_ev_RAP}
\zeta\frac{d}{d\zeta}F(x,b;\mu,\zeta)&=& -\mathcal{D}(\mu,b)F(x,b;\mu,\zeta),
\end{eqnarray}
where $F$ is any TMD distribution ($f_1$ or $D_1$ in the present case). The TMD evolution equations are not sensitive to the flavor of a parton\footnote{The TMD evolution is sensitive to the color-representation. Since in this work we deal only with quark channels, we do not write the corresponding labels.} and thus we omit flavor indices in this section for simplicity. The eq.~(\ref{def:TMD_ev_UV}) is a standard renormalization group equation, which comes from the renormalization of the ultraviolet divergences, with the function $\gamma_F(\mu,\zeta)$ being  the anomalous dimension. The eq.~(\ref{def:TMD_ev_RAP}) results from the factorization of rapidity divergences. The function $\mathcal{D}(\mu,b)$ is called the rapidity anomalous dimension (RAD). The RAD is a generic non-perturbative function that can be computed at small values of $b$ in  perturbation theory. The perturbative expression for the RAD and $\gamma_F$ can be found in the literature (e.g. see appendix of ref.~\cite{Echevarria:2016scs}). In this work we use the resummed version of RAD \cite{Echevarria:2012pw}. The resummed expressions are also given in appendix \ref{app:RAD} (see also appendix B in ref.~ \cite{Bizon:2018foh}).

The scales $\mu$ and $\zeta$ have an independent origin, and this has important consequences. To start with, the TMD evolution takes  place in the plane $(\mu,\zeta)$. The solution of equations eq.~(\ref{def:TMD_ev_UV},~\ref{def:TMD_ev_RAP}) for the evolution from a point $(\mu_f,\zeta_f)$ to a point $(\mu_i,\zeta_i)$ is
\begin{eqnarray}\label{def:exp[evol]}
F(x,b;\mu_f,\zeta_f)=\exp\[\int_P \(\gamma_F(\mu,\zeta)\frac{d\mu}{\mu}-\mathcal{D}(\mu,b)\frac{d\zeta}{\zeta}\)\] F(x,b;\mu_i,\zeta_i)
\end{eqnarray}
where $P$ is any path in $(\mu,\zeta)$-plane that connects initial $(\mu_i,\zeta_i)$ and final points $(\mu_f,\zeta_f)$. The value of evolution is (in principle) independent on the path, thanks to integrability condition (also known as Collins-Soper (CS) equation \cite{Collins:1981va})
\begin{eqnarray}\label{def:integrability}
-\zeta\frac{d\gamma_F(\mu,\zeta)}{d\zeta}=\mu\frac{d\mathcal{D}(\mu,b)}{d\mu}=\Gamma_{\text{cusp}}(\mu),
\end{eqnarray}
where $\Gamma_{\text{cusp}}(\mu)$ is the cusp anomalous dimension. This equation dictates the logarithmic structure of anomalous dimensions. In particular, the TMD anomalous dimension is
\begin{eqnarray}
\gamma_F(\mu,\zeta)=\Gamma_{\text{cusp}}(\mu)\ln\(\frac{\mu^2}{\zeta}\)-\gamma_V(\mu).
\end{eqnarray}
The formal path-independence of eq.~(\ref{def:exp[evol]}) is violated at any fixed order of perturbation theory. The penalty term is proportional to the area surrounded by paths, and can be huge in the case of very separated scales. Nevertheless, the path dependence decreases with the increase of the perturbative order and it is numerically small at N$^3$LO \cite{Scimemi:2018xaf}.

The final scales of the evolution are binded to the hard scale of factorization such that $\mu^2_f\sim Q^2$ and $\zeta_{1f}\zeta_{2f}=Q^4$. In particular, we choose the symmetric point
\begin{eqnarray}
\mu^2_f= Q^2,\qquad \zeta_{1f}=\zeta_{2f}=Q^2.
\end{eqnarray}
The TMD initial (or defining) scale is chosen with the $\zeta$-prescription and deserves some explanation.
In the $\zeta$-prescription the scales $\mu$ and $\zeta$ belong to a null-evolution line, that we parameterize as $(\mu,\zeta_\mu(b))$. To find the null-evolution line, we recall that the system of eq.~(\ref{def:TMD_ev_UV},~\ref{def:TMD_ev_RAP}) is a two-dimensional gradient equation ($\pmb\nabla F=\mathbf{E}F$) with the field $\mathbf{E}=(\gamma_F(\mu,\zeta)/2,-\mathcal{D}(\mu,b))$. Therefore, the null-evolution line is simply an equipotential line of the field $\mathbf{E}$. It provides the equation that define $\zeta_\mu(b)$ such that
\begin{eqnarray}\label{th:special-line}
\Gamma_{\text{cusp}}(\mu)\ln\(\frac{\mu^2}{\zeta_\mu(b)}\)-\gamma_V(\mu)=2\mathcal{D}(\mu,b)\frac{d\ln \zeta_\mu(b)}{d\ln \mu^2},
\end{eqnarray}
A TMD distribution does not evolve between scales belonging to the same equipotential line by definition. 

\begin{figure}
\begin{center}
\includegraphics[width=0.95\textwidth]{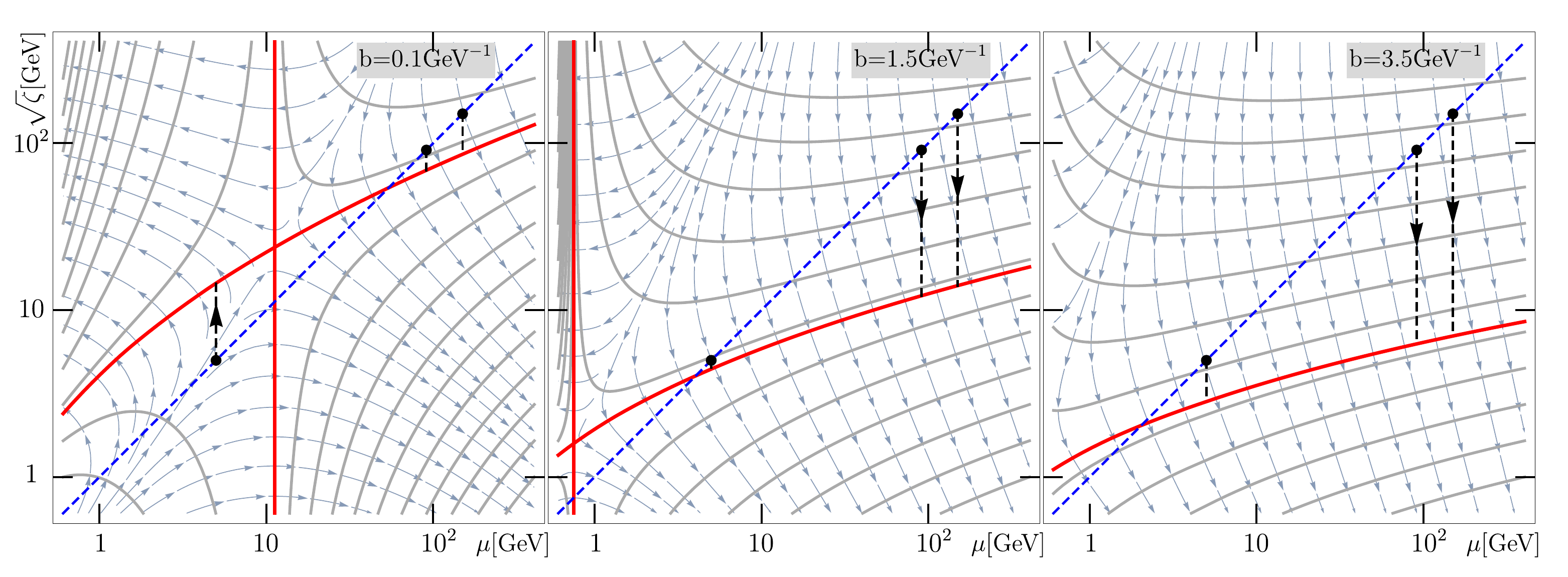}
\caption{\label{fig:zeta-line} In  the $(\zeta,\mu)$ plane we show the force-lines of the TMD evolution field $\mathbf{E}$ at different values of $b$ (in grey, with arrows). The thick continuous gray lines are  null-evolution (equipotential) lines. Red lines are the  equipotential lines that define the saddle point. The red line which crosses each panel from left to right is the special evolution curve where the TMD are defined.  The blue dashed lines in each plot correspond to the final scale choice ($\mu_f, \zeta_f$) for typical experimental measurements. The black points indicate the initial evolution scales for $Q=5$, $91$ and $150$ GeV cases. Black dashed lines with arrows are paths of evolution implemented in eq.~(\ref{def:TMD-evolved}).
  }
\end{center}
\end{figure}

Among equipotential lines there is a special line that passes through the saddle point $(\mu_0,\zeta_0)$ of the field $\mathbf{E}$. The values $(\mu_0,\zeta_0)$ are defined as
\begin{eqnarray}\label{th:saddle-point}
\mathcal{D}(\mu_0,b)=0,\qquad \gamma_F(\mu_0,\zeta_0)=0.
\end{eqnarray}
The special equipotential line is preferable for the definition of TMD scales for two important reasons. First, there is only one saddle point in the evolution field, and thus, the special null-evolution line is unique. Second, the special null-evolution line is the only null-evolution line, which has finite $\zeta$ at all values of $\mu$ (bigger than $\Lambda_\text{QCD}$). These properties follow from its definition and they are very useful. In fig.~\ref{fig:zeta-line} we show the force-lines of the evolution field $\mathbf{E}$  (in grey, with arrows), null-evolution lines, (thick grey lines, orthogonal to the force-lines), and the lines that cross at the saddle point (in red) at different values of $b$. In this figure the special line is the one that goes from left to right in each panel. 

The concept of $\zeta$ prescription has been introduced in ref.~\cite{Scimemi:2017etj} and elaborated in \cite{Scimemi:2018xaf}. Presently we use a form slightly different from the original version of refs.~\cite{Scimemi:2017etj,Scimemi:2018xaf}. Here we follow the updated realization introduced in ref.~\cite{Vladimirov:2019bfa} that has been used for the description of the pion-induced DY process. In refs.~\cite{Scimemi:2017etj,Bertone:2019nxa} the $\zeta$-lines has been taken perturbative for all ranges of $b$ (with slight deformations due to the Landau pole). Notwithstanding, such definition introduces an undesired correlation between the non-perturbative parts of the TMD distribution and RAD. In ref.~\cite{Vladimirov:2019bfa} a new simple solution has been found for the values of special null-evolution line at large $b$ that accurately incorporates non-perturbative  effects, without adding new parameters to the fit. In appendix~\ref{app:zeta-line} we present the expression for the special line as it is used in this fit.

A TMD distribution $F(x,b;\mu,\zeta_\mu)$ with $\zeta_\mu$ belonging to the special line is called optimal TMD distribution, and denoted by $F(x,b)$ (without scale arguments), to emphasize its uniqueness and independence on scale $\mu$. The exact independence of optimal TMD distribution on scale $\mu$, allows us to select the simplest path for the evolution exponent in eq.~(\ref{def:exp[evol]}), that is, the path at fixed value of $\mu=Q$ along $\zeta$ from the value $\zeta_f=Q^2$ down to any point of $\zeta_i=\zeta_Q(b)$. In fig.~\ref{fig:zeta-line} this path is visualized by black-dashed lines. The resulting expression for the evolved TMD distributions is exceptionally simple
\begin{eqnarray}\label{def:TMD-evolved}
F(x,b;Q,Q^2)=\(\frac{Q^2}{\zeta_Q(b)}\)^{-\mathcal{D}(b,Q)}F(x,b).
\end{eqnarray}
We recall that this expression is same for all (quark) TMDPDFs and TMDFF. Substituting (\ref{def:TMD-evolved}) into the definition of structure functions $W$ we obtain,
\begin{eqnarray}\label{def:Wff-final}
W_{f_1f_1}^f(Q,q_T;x_1,x_2)&=&|C_V(-Q^2,Q^2)|^2
\\\nn &&\qquad\times \int_0^\infty db\, bJ_0(bq_T)f_{1,f\ot h}(x_1,b)f_{1,\bar f\ot h}(x_2,b)\(\frac{Q^2}{\zeta_Q(b)}\)^{-2\mathcal{D}(b,Q)},
\\\label{def:WfD-final}
W_{f_1D_1}^f(Q,q_T;x_S,z_S)&=&|C_V(Q^2,Q^2)|^2
\\\nn &&\qquad\times \int_0^\infty db\, bJ_0(bq_T)f_{1,f\ot h}(x_S,b)D_{1,f\to h}(z_S,b)\(\frac{Q^2}{\zeta_Q(b)}\)^{-2\mathcal{D}(b,Q)}.
\end{eqnarray}
These are the final expressions used to extract the NP functions.

\begin{figure}[t]
\begin{center}
\includegraphics[width=0.32\textwidth]{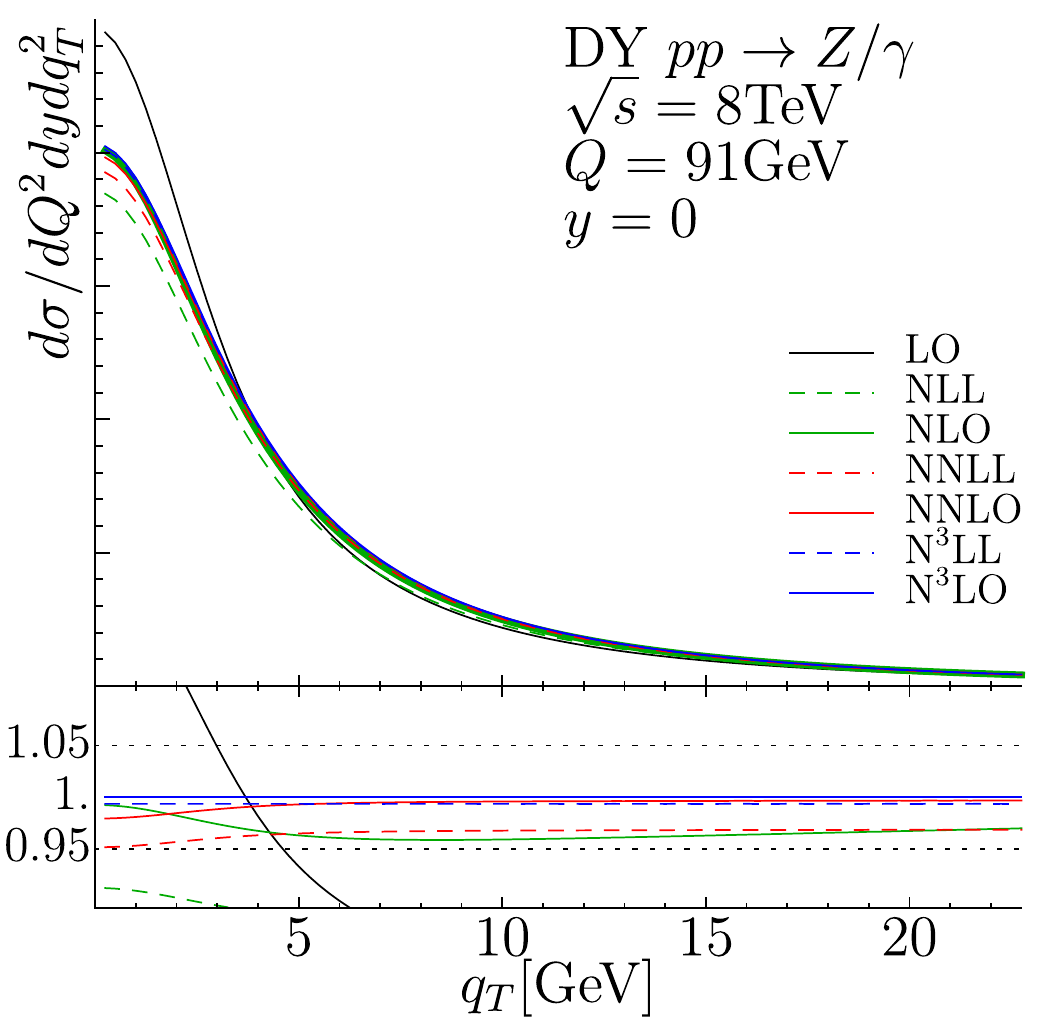}
\includegraphics[width=0.32\textwidth]{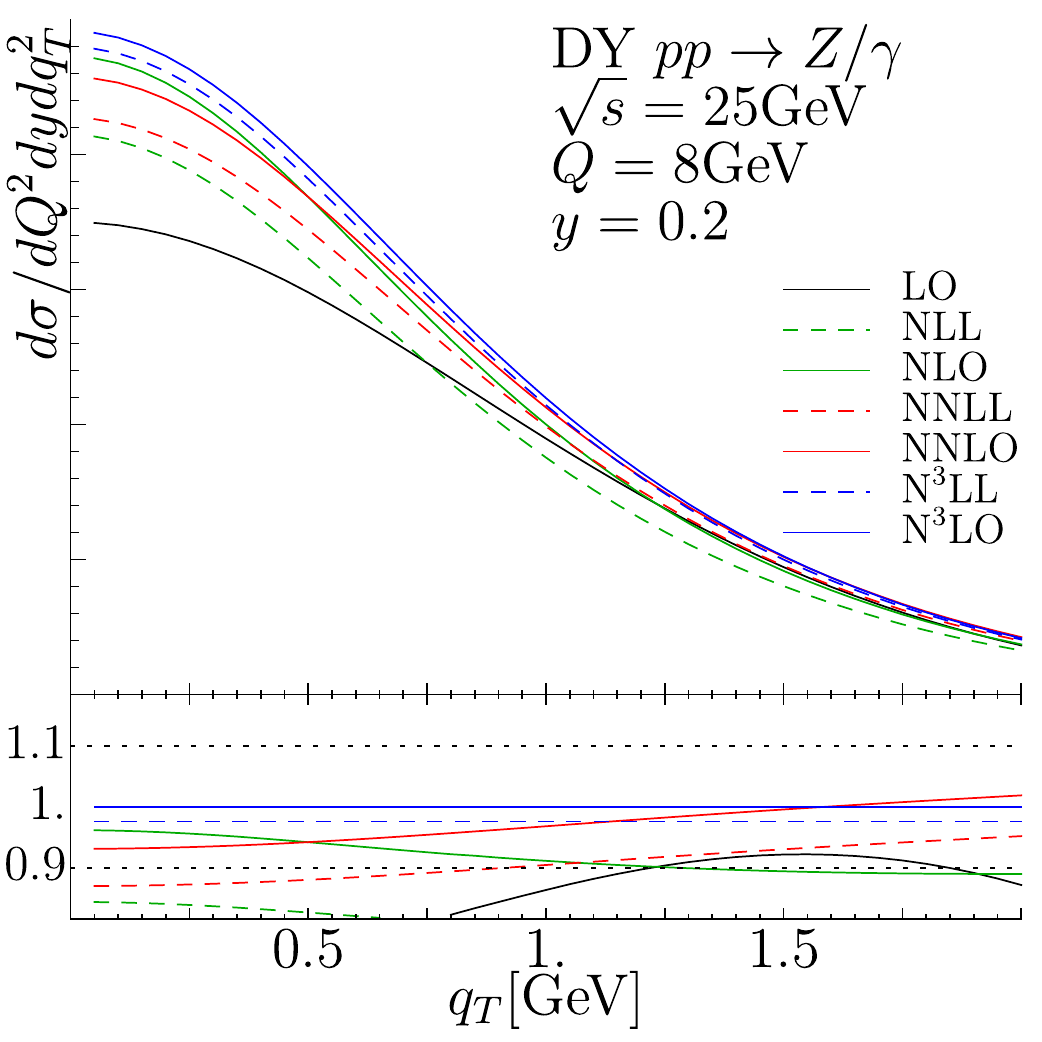}
\includegraphics[width=0.32\textwidth]{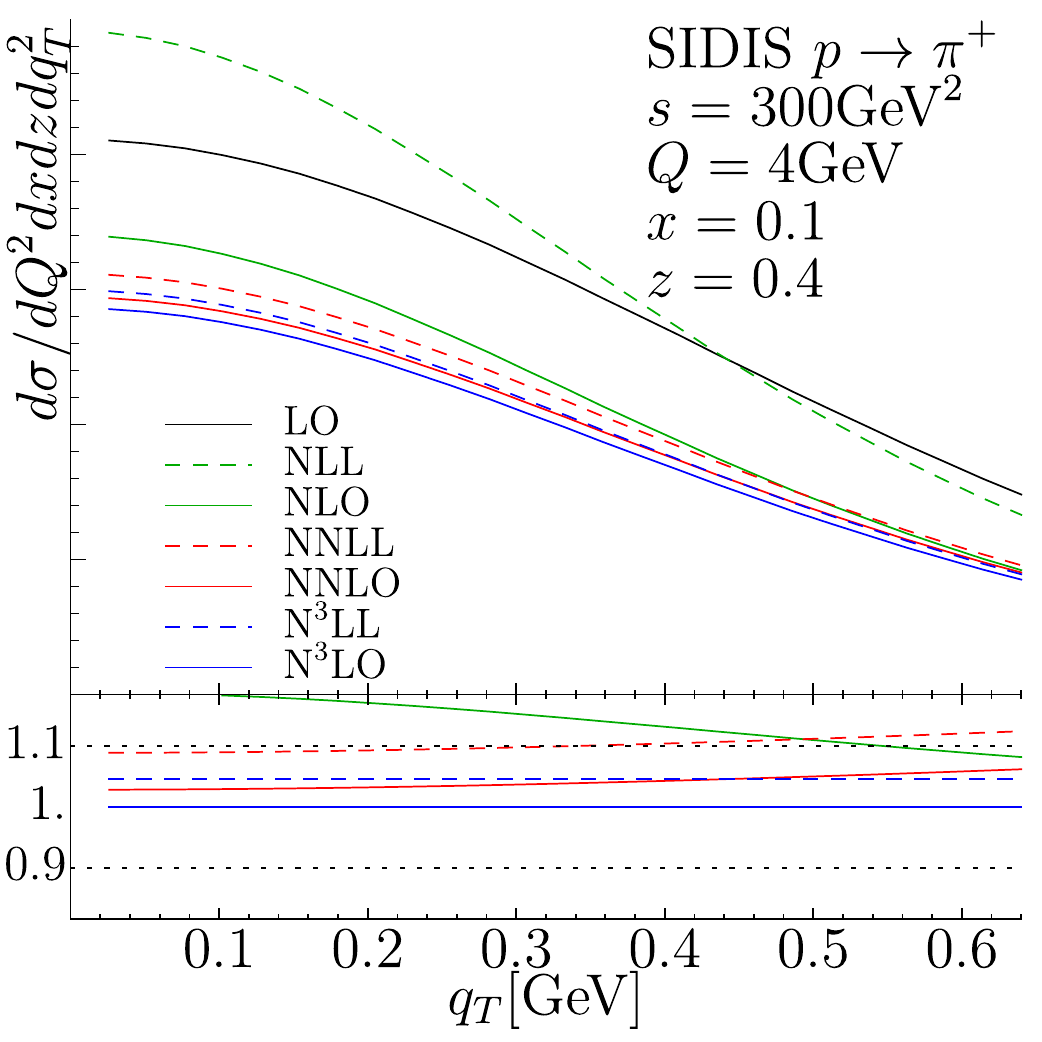}
\caption{\label{fig:convergence} The  cross-section at different orders of TMD factorization and for different boson energies. The  legend of the perturbative orders means that 
 N$^k$LO (N$^k$LL) incorporates $a_s^k$-order ($a_s^{k-1}$-order) of the coefficient function, $a_s^{k}$-order of anomalous dimensions with $a_s^{k+1}$-order of $\Gamma_{\text{cusp}}$. The TMD distributions and the NP part of the evolution are the same for all cases.}
\end{center}
\end{figure}

The simplicity of expressions (\ref{def:Wff-final},\ref{def:WfD-final}) is also accompanied by a good convergence of the cross section. In fig.~\ref{fig:convergence} we show the comparison of curves for DY and SIDIS cross-section at typical energies. In the plot the TMD distributions and the NP part of the evolution are held fixed while the perturbative orders are changed. The perturbative series  converges very well, and the difference between NNLO and N$^3$LO factorization is of order of percents. This is an additional positive aspect of the $\zeta$-prescription, which is due to fact that all perturbative series are evaluated at $\mu=Q$.

\subsubsection{Matching of TMD distribution to collinear distributions}
\label{sec:matching}

The TMD are generic non-perturbative functions that depend on the parton fraction $x$ and the impact parameter $b$. A fit of a two-variable function is a hopeless task due to the enormous parametric freedom. This freedom can be essentially reduced by the matching of a $b\to0$ boundary of a TMD distribution to the corresponding collinear distribution. In the asymptotic limit  of small-$b$ one has 
\begin{eqnarray}\label{def:F-match}
\lim_{b\to0}f_{1,f\ot h}(x,b)&=&\sum_{f'}\int_x^1 \frac{dy}{y} C_{f\ot f'}\(\frac{x}{y},\mathbf{L}_{\mu_{\text{\tiny OPE}}},a_s(\mu_{\text{\tiny OPE}})\)f_{1,f'\ot h}(y,\mu_{\text{\tiny OPE}}),
\\\label{def:D-match}
\lim_{b\to0}D_{1,f\to h}(z,b)&=&\sum_{f'}\int_z^1 \frac{dy}{y} \mathbb{C}_{f\to f'}\(\frac{z}{y},\mathbf{L}_{\mu_{\text{\tiny OPE}}},a_s(\mu_{\text{\tiny OPE}})\)\frac{d_{1,f'\to h}(y,\mu_{\text{\tiny OPE}})}{y^2},
\end{eqnarray}
where $f_1(x,\mu)$ and $d_1(x,\mu)$ are collinear PDF and FF, the label $f'$ runs over all active quarks, anti-quarks and a gluon, and
\begin{eqnarray}
\mathbf{L}_\mu=\ln\(\frac{\vec b^2 \mu^2}{4\exp^{-2\gamma_E}}\),\qquad a_s(\mu)=\frac{g^2(\mu)}{(4\pi)^2},
\end{eqnarray}
with $\gamma_E$ being the Euler constant and $g$ being QCD coupling constant. The extra factor $y^{-2}$ in eq.~(\ref{def:D-match}) is present due to the normalization difference of the TMD operator in eq.~(\ref{def:D1}) and the collinear operator, see e.g.~\cite{Collins:2011zzd,Echevarria:2015usa}. The coefficient functions $C$ and $\mathbb{C}$ can be calculated with operator product expansion methods (for a general review see ref.~\cite{Scimemi:2019gge}) and  in the case of unpolarized distributions the coefficient functions are known up to NNLO \cite{Gehrmann:2014yya,Echevarria:2015usa,Echevarria:2016scs,Luo:2019hmp}. 
The coefficient function $C$ has the general form
\begin{eqnarray}\label{th:match-coeff}
&&C_{f\ot f'}(x,\mathbf{L}_{\mu},a_s)=\delta(\bar x)\delta_{ff'}+a_s(\mu)\(-\mathbf{L}_\mu P^{(1)}_{f\ot f'}+C_{f\ot f'}^{(1,0)}\)
\\\nn &&\qquad\qquad+a_s^2(\mu)\Big[\frac{P^{(1)}_{f\ot k}\otimes P^{(1)}_{k\ot f'}-\beta_0 P_{f\ot f'}^{(1)}}{2}\mathbf{L}^2_\mu
-\mathbf{L}_\mu \(P^{(2)}_{f\ot f'}+C_{f\ot k}^{(1,0)}\otimes P^{(1)}_{k\ot f'}-\beta_0 C_{f\ot k}^{(1,0)}\)
\\\nn && \qquad\qquad\qquad+C_{f\ot f'}^{(2,0)}+\frac{d^{(2,0)}\gamma_1}{\Gamma_0}\delta(\bar x)\delta_{ff'}\Big]+O(a_s^3),
\end{eqnarray}
where $\bar x=1-x$, the symbol $\otimes$ denotes the Mellin convolution, and a summation over the intermediate flavour index $k$ is implied. In eq.~(\ref{th:match-coeff}) we have omitted argument $x$ of functions on left-hand-side for brevity. The functions $P^{(n)}(x)$ are the coefficients of the PDF evolution kernel $P(x)=\sum_n a_s^n P^{(n)}(x)$ (DGLAP kernel), which can be found f.i. in ref.~\cite{Moch:1999eb}. The  functions $C_{f\ot f'}^{(n,0)}(x)$ are given in \cite{Gehrmann:2014yya,Echevarria:2015usa,Echevarria:2016scs,Luo:2019hmp}. In particular, the NLO terms are
\begin{eqnarray}
C_{q\ot q}^{(1,0)}(x)=C_F\(2\bar x-\delta(\bar x)\frac{\pi^2}{6}\),\qquad C_{q\ot g}^{(1,0)}(x)=2x\bar x.
\end{eqnarray}
The last term in the square brackets of eq.~(\ref{th:match-coeff}) is the consequence of the boundary condition of eq.~(\ref{th:saddle-point}), and it consists of some coefficients of the anomalous dimension defined in eq.~(\ref{app:def:dnk},~\ref{app:gammaV}). 

In the case of TMDFF the matching coefficient $\mathbb{C}$ follows the same pattern as in eq.~(\ref{th:match-coeff}) with the replacement of the PDF DGLAP kernels $P^{(n)}_{f\ot f'}(x)$  by the FF DGLAP kernels $P^{(n)}_{f\ot f'}(z)$  (they can be found f.i. in ref.~\cite{Stratmann:1996hn}), and $C_{f\ot f'}^{(n,0)}(x)$ by $\mathbb{C}_{f\to f'}^{(n,0)}(z)$ \cite{Echevarria:2015usa,Echevarria:2016scs}. In TMDFF case, the NLO terms are
\begin{eqnarray}\label{C_FF_NLO}
\mathbb{C}_{q\to q}^{(1,0)}(z)=\frac{C_F}{z^2}\(2\bar z+\frac{4(1+z^2)\ln z}{1-z}-\delta(\bar z)\frac{\pi^2}{6}\),
&&
\mathbb{C}_{q\to g}^{(1,0)}(z)=\frac{2C_F}{z^2}\(z+2(1-\bar z^2)\frac{\ln z}{z}\).\quad
\end{eqnarray}
As a consequence of the $\zeta$-prescription the scale of operator product expansion $\mu_{\text{OPE}}$ is independent on external parameters. In particular, it has no connection to the scales of the TMD evolution, as it happens f.i. in the case of $b^*$-prescription~\cite{Collins:2011zzd,Aybat:2011zv}. In other words, in the $\zeta$-prescription, the scale $\mu_{\text{OPE}}$ is entirely encapsulated inside the convolutions in eq.~(\ref{def:F-match},~\ref{def:D-match}). This fact gives an enormous advantage to achieve a  complete decorrelation of RAD from TMD distributions (we will be  more quantitative about this point in later sections). The optimal TMD distributions as any scale-less observables, are formally, independent on the value of $\mu_{\text{OPE}}$ given the good convergence of perturbative series. So, the scale $\mu_{\text{OPE}}$ has to be selected such that on one hand, it minimizes the logarithm contributions at $b\to 0$, and on another hand, it does not hit the Landau pole at large-$b$. For TMDPDF, we use the following value
\begin{eqnarray}
\mu^{\text{PDF}}_{\text{OPE}}=\frac{2e^{-\gamma_E}}{b}+2\text{GeV},
\end{eqnarray}
whereas for TMDFF we use
\begin{eqnarray}\label{muOPE_FF}
\mu^{\text{FF}}_{\text{OPE}}=\frac{2e^{-\gamma_E}z}{b}+2\text{GeV}.
\end{eqnarray}
The extra factor $z$ in (\ref{muOPE_FF}) effectively compensates $\ln z$ terms in the matching coefficient, and in this way improve the convergence of the series (e.g. it completely neglects $\ln z$ terms in NLO expressions (\ref{C_FF_NLO})). The choice of the large-b offset of $\mu_{\text{OPE}}$ as 2 GeV is arbitrary, with the only motivation that it is a typical reference scale for PDFs (and lattice calculations). In the $\zeta$-prescription, this scale is intrinsic to the model of TMD distribution, and thus, any modifications in it would be absorbed by NP parameters discussed in the next section.

Let us note that in the $\zeta$-prescription, the coefficient functions of small-$b$ matching in eq.~(\ref{th:match-coeff}) do not contain a double-logarithm contribution. For that reason the perturbative convergence, as well as the radius of convergence improves. Both these facts make the $\zeta$-prescription highly advantageous.

\subsubsection{Ansatzes for NP functions}
\label{sec:ansatzNP}

In this work we deal with  \textit{three independent non-perturbative functions} in total. These are the unpolarized (optimal) TMDPDF, $f_1(x,b)$, the unpolarized (optimal) TMDFF, $D_1(x,b)$, and the RAD, $\mathcal{D}(b,\mu)$. The amount of perturbative and non-perturbative contributions to each function depends on the value of the impact parameter $b$. Namely, at small values of $b$ the perturbative approximation is good and the TMD distributions can be matched onto collinear functions as in  eq.~(\ref{def:F-match},~\ref{def:D-match}). In the case of the RAD the small-$b$ limit is given in appendix~\ref{app:RAD}. The small-$b$ perturbative expressions gains power corrections in even powers $\vec b^{2n}$~\cite{Scimemi:2016ffw}. Therefore, with the increase of $b$ the perturbative approximation becomes less and less correct, and must be replaced by some generic function. 

The phenomenological ansatzes for TMD distributions that satisfy this picture, can be written as following:
\begin{eqnarray}\label{def:phen-f1}
f_{1,f\ot h}(x,b)&=&\int_x^1 \frac{dy}{y}\sum_{f'}C_{f\ot f'}\(y,\mathbf{L}_{\mu_{\text{OPE}}},a_s(\mu_{\text{\tiny OPE}})\)f_{1,f'\ot h}\(\frac{x}{y},\mu_{\text{\tiny OPE}}\)f_{\text{NP}}(x,b),
\\\label{def:phen-D1}
D_{1,f\to h}(z,b)&=&\frac{1}{z^2}\int_z^1 \frac{dy}{y}\sum_{f'} y^2\mathbb{C}_{f\to f'}\(y,\mathbf{L}_{\mu_{\text{\tiny OPE}}},a_s(\mu_{\text{\tiny OPE}})\)d_{1,f'\to h}\(\frac{z}{y},\mu_{\tiny \text{\tiny OPE}}\)D_{\text{NP}}(z,b),\quad
\end{eqnarray}
where functions $f_{\text{NP}}$ and $D_{\text{NP}}$ are non-perturbative functions. Note, that in our ansatz we do not modify the value of $b$ within the coefficient function. Therefore, at large-$b$ the logarithm part of the coefficient function grows unrestrictedly. This growth is suppressed by the non-perturbative functions.

Generally, the functions $f_{\text{NP}}$ and $D_{\text{NP}}$ depend also on parton flavor $f$ and hadron type $h$. However, \textit{in the present work we use the approximation that $f_{\text{NP}}$ and $D_{\text{NP}}$ are flavor and hadron-type independent.} 
All hadron- and flavor dependence is driven by the collinear PDFs and FFs (see also sec.~\ref{sec:nuclear}). Given such an ansatz the only requirement for NP functions is that they are even-functions of $b$ that turn to unity for $b\to 0$ (see ref.~\cite{Scimemi:2016ffw} for an analysis of these processes using renormalons). We use the following parameterizations
\begin{eqnarray}\label{def:fNP}
f_{NP}(x,b)&=&\exp\(-\frac{\lambda_1(1-x)+\lambda_2 x+x(1-x)\lambda_5}{\sqrt{1+\lambda_3 x^{\lambda_4} \vec b^2}}\vec b^2\),
\\\label{def:DNP}
D_{NP}(x,b)&=&\exp\(-\frac{\eta_1 z+\eta_2 (1-z)}{\sqrt{1+\eta_3(\vec b/z)^2}}\frac{\vec b^2}{z^2}\)\(1+\eta_4 \frac{\vec b^2}{z^2}\),
\end{eqnarray}
and we extract $\lambda_i$ and $\eta_i$ from our fit. The functional form of $f_{NP}$ has been already used in \cite{Bertone:2019nxa}. It has five free parameters which grant a sufficient flexibility in $x$-space as needed for the description of the precise LHC data. The form of $D_{\text{NP}}$ has been suggested in \cite{Bacchetta:2017gcc} (albeit there are more parameters in \cite{Bacchetta:2017gcc}). In both cases the function has exponential or Gaussian form depending on the relative size of $\lambda_{1,2,5}/\lambda_3$, and $\eta_{1,2}/\eta_3$. There are natural restrictions on the parameter space $\lambda_{1,2,3}>0$, $\eta_{1,2,3}>0$, $\lambda_5\gtrsim-2(\lambda_1+\lambda_2)$, due to the request that TMD distribution is null for $b\to \infty$.

We use the following ansatz for the  NP RAD,
\begin{eqnarray}\label{NP:RAD}
\mathcal{D}(\mu,b)=\mathcal{D}_{\text{resum}}(\mu,b^*(b))+c_0 bb^*(b),
\end{eqnarray}
where
\begin{eqnarray}
b^*(b)=\frac{b}{\sqrt{1+\vec b^2/B^2_{\text{NP}}}}.
\end{eqnarray}
The the term  $c_0 bb^*(b)$ dictates the large-$b$ behavior of the RAD and its form is suggested in \cite{Bertone:2019nxa}. At large-$b$ the NP expression for RAD is linear in $b$, $\mathcal{D}\sim c_0B_{\text{NP}} b$. The linear behavior is suggested by model calculations of the RAD~\cite{Tafat:2001in,Vladimirov:2020umg}. Generally, the asymptotic behavior of RAD could vary from constant to linear \cite{Hautmann:2020cyp,Vladimirov:2020umg,Collins:2014jpa}. 

The function $\mathcal{D}_{\text{resum}}$ is the resummed perturbative expansion of RAD \cite{Echevarria:2012pw,Scimemi:2018xaf} reported in the appendix~\ref{app:RAD}. At LO it reads
\begin{eqnarray}\label{NP:RAD:LO}
\mathcal{D}^{\text{LO}}_{\text{resum}}=-\frac{\Gamma_0}{2\beta_0}\ln\(1-\beta_0 a_s(\mu)\mathbf{L}_\mu\).
\end{eqnarray}
The higher order expressions (up to N$^3$LO) are given in eq.~(\ref{app:RAD:resum}). The parameters $c_0$ and $B_{\text{NP}}$ are free positive parameters, in principle totally uncorrelated from the rest of non-perturbative parameters.

The resummed expression for RAD shows explicitly a singularity in $b$ (see e.g. eq.~(\ref{NP:RAD:LO})). The singularity designates the convergence radius of the perturbative expression. Consequently, the perturbative behavior must be turned off well before $b$ approaches the singularity. In the ansatz in eq.~(\ref{NP:RAD}), this is achieved  freezing the perturbative part at $b\sim B_{\text{NP}}$. The singularity is located at $\beta_0a_s(\mu)\mathbf{L}_\mu=1$ and thus, the value of $B_{\text{NP}}$ is restricted from above by: $B_{\text{NP}}\lesssim 2e^{-\gamma_E}\Lambda_{\text{QCD}}^{-1}\approx 4$GeV$^{-1}$.

The special null-evolution line can be incorporated both at perturbative  and non-perturbative level. In \cite{Scimemi:2017etj} and \cite{Bertone:2019nxa} the special null-evolution line included only its perturbative part for simplicity. This part is the most important one because it guarantees the cancellation of double-logarithms in the matching coefficient. However, at large-$b$, the non-perturbative corrections to the RAD are large and cannot be ignored:  in \cite{Scimemi:2017etj} they can be seen as a part of the non-perturbative model, at the price of  introducing an  undesired correlation between $f_{NP}$ and $\mathcal{D}$.  In order to adjust the null-evolution curve with a non-perturbative RAD one has to solve eq.~(\ref{th:special-line}) including the RAD in the full generality. Such solution can be found in principle, but its numerical implementation is problematic at very small-$b$, because it is very difficult to obtain the exact numerical cancellation of the \textit{perturbative} series of logarithms with an \textit{exact} solution. To by-pass this problem we use the perturbative solution at very small $b$, (and hence cancel all logarithm exactly) and turn it to an exact solution at larger $b$. This is realized by 
\begin{eqnarray}\label{NP:zeta}
\zeta_\mu(b)=\zeta^{\text{pert}}_\mu(b)e^{-\frac{\vec b^2}{B^2_{\text{NP}}}}+\zeta^{\text{exact}}_\mu(b)\(1-e^{-\frac{\vec b^2}{B^2_{\text{NP}}}}\),
\end{eqnarray}
that is, for $\vec b^2\ll B^2_\text{NP}$ we have the {\it perturbative} solution, and one turns to the {\it exact} for larger $b$. Since the RAD is entirely perturbative at small-$b$, the numerical difference between eq.~(\ref{NP:zeta}) and $\zeta^{\text{exact}}_\mu(b)$ is negligibly small. 

\subsection{Summary on theory input}
\label{sec:summary-theory}

The structure functions $W_{f_1D_1}$ and $W_{f_1f_1}$ are evaluated according to eq.~(\ref{def:Wff-final},~\ref{def:WfD-final}). The phenomenological ansatzes for the optimal unpolarized TMDPDF and TMDFF are defined in eq.~(\ref{def:phen-f1}, \ref{def:phen-D1}, \ref{def:fNP},~\ref{def:DNP}). At small-$b$ TMD distributions are matched to corresponding collinear distributions. The phenomenological ansatz for the RAD is given in eq.~(\ref{NP:RAD}).  In  table \ref{tab:pert} we list the perturbative orders used in  each factor of the cross section. The N$^3$LO perturbative composition used here is equivalent to the one used in \cite{Bizon:2018foh,Bizon:2019zgf} on the resummation side.  A total of 11 phenomenological parameters  are determined by the fit procedure. Two of these parameters describe the RAD, 5 are for  the unpolarized TMDPDF, and 4 are for the unpolarized TMDFF. Additionally, TMDPDFs and TMDFF depend on collinear distributions.  Thus collinear distributions can be seen as parameters of our model that we take from others fits. We have found that the quality of fit highly depends on the choice of collinear distributions (we can address this fact as the "PDF-bias" problem). The study of this issue is  in sec.~\ref{sec:PDF}, \ref{sec:SIDIS-FF}.

\begin{table}[h]
\begin{center}
\begin{tabular}[h]{|c|c||c|c|c|c|c|c V{4} c|}\hline
Evolution & Acronym in&$C_V$  & $\Gamma_{\text{cusp}}$ &  $\gamma_V$ & $\mathcal{D}_{\text{resum}}$ & $\zeta_\mu^{\text{pert}}$ & $\zeta_\mu^{\text{exact}}$&  $C$, $\mathbb{C}$\\
+matching &present work&&&&&&&
\\\hline 
NNLO+NNLO & NNLO &$\alpha_s^2$  & $\alpha_s^3$ ($\Gamma_2$) & $\alpha_s^2$ ($\gamma_2$) & $\alpha_s^2$ ($d_2$) & $\alpha_s^1$  ($v_1$) & $\alpha_s^1$  ($g_2$)& $\alpha_s^2$
\\\hline
N$^3$LO+NNLO & N$^3$LO&$\alpha_s^{3}$  & $\alpha_s^4$ ($\Gamma_3$) & $\alpha_s^3$ ($\gamma_3$) & $\alpha_s^3$ ($d_3$) & $\alpha_s^2$ ($v_2$) & $\alpha_s^2$  ($g_3$)& $\alpha_s^2$
\\\hline
\end{tabular}
\end{center}
\caption{\label{tab:pert} Summary of the perturbative orders used for each part of the factorized cross section. The evolution of $\alpha_s$ is provided by the LHAPDF library and comes together with PDF set (uniformly nnlo). In brackets we write \textit{the last included} term of corresponding perturbative expansion (\ref{app:def-beta-Gamma},~\ref{app:RAD:resum},~\ref{app:zeta-pert},~\ref{app:gammaV},~\ref{app:g}).}
\end{table}

\section{Data overview}
\label{sec:data}
In  the present work, we consider the  extraction of unpolarized TMD in DY  and SIDIS data, extending so the analysis of ref.~\cite{Bertone:2019nxa} and including the  theoretical improvements described in the previous sections.  The selection of data is crucial for a proper TMD extraction, because of the limits imposed by the factorization theorem. These constraints are here discussed for both type of reactions.

\subsection{SIDIS data}

\begin{table}[t]
\begin{center}
\small
\begin{tabular}{|c||c|c|c|c|c|}
\hline
Experiment & Reaction & ref. & Kinematics & \specialcellcenter{$N_{\rm pt}$\\after cuts}
\\\hline\hline
\multirow{8}{*}{HERMES} & $p\to \pi^+$ & \multirow{8}{*}{\cite{Airapetian:2012ki}} & 
\multirow{8}{*}{\specialcellcenter{
0.023<x<0.6 (6 bins)\\
0.2<z<0.8 (6 bins)\\
1.0<Q<$\sqrt{20}$GeV\\
~
\\
$W^2>10$GeV$^2$
\\
0.1<y<0.85}
}& 24
\\\cline{2-2}\cline{5-5}
& $p\to \pi^-$ &&& 24
\\\cline{2-2}\cline{5-5}
& $p\to K^+$ &&& 24
\\\cline{2-2}\cline{5-5}
& $p\to K^-$ &&& 24 
\\\cline{2-2}\cline{5-5}
& $D\to \pi^+$ &&& 24
\\\cline{2-2}\cline{5-5}
& $D\to \pi^-$ &&& 24
\\\cline{2-2}\cline{5-5}
& $D\to K^+$ &&& 24
\\\cline{2-2}\cline{5-5}
& $D\to K^-$ &&& 24
\\\hline
\multirow{2}{*}{COMPASS} & 
$d\to h^+$ & \multirow{2}{*}{\cite{Aghasyan:2017ctw}} & 
\multirow{2}{*}{\specialcellcenter{
0.003<x<0.4 (8 bins)\\
0.2<z<0.8 (4 bins)\\
1.0<Q$\simeq$ 9GeV (5 bins)}
}& 195
\\\cline{2-2}\cline{5-5}
& $d\to h^-$ &&& 195
\\&&&&
\\\hline
Total &&&& 582
\\\hline
\end{tabular}
\caption{\label{tab:SIDIS-data}Summary of the SIDIS data included in the fit. For each data set we report reference, reaction,  kinematic region, and  number of points that are left after the application of consistency cuts in eq.~(\ref{SIDIS-data:Q>2},~\ref{SIDIS-data:qT<0.25 Q}).}
\end{center}
\end{table}

\begin{figure}
\begin{center}
\includegraphics[width=0.5\textwidth]{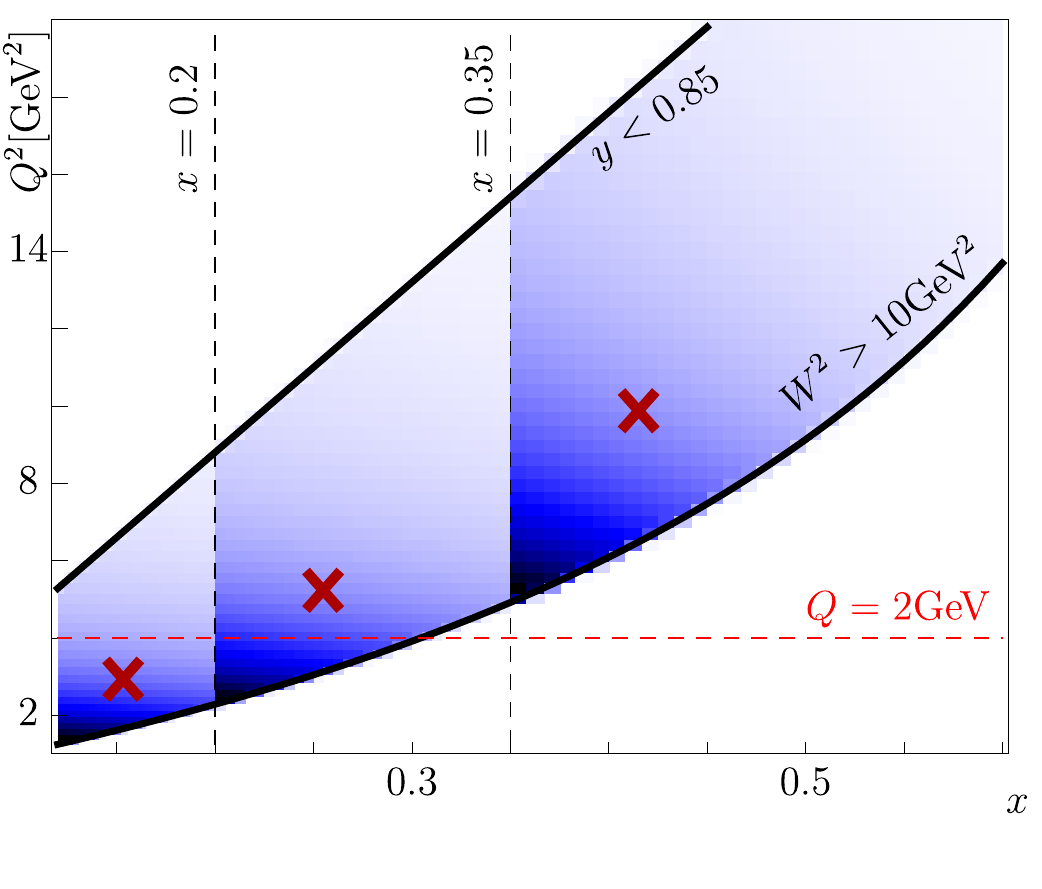}
\caption{\label{fig:bining} Illustration for bin shapes in  HERMES kinematics. Three bins are shown ($0.12<x<0.2$, $0.2<x<0.35$, $0.35<x<0.6$). Solid lines are boundaries of the fiducial region. Color density demonstrates the distribution of multiplicity value within a bin. Crosses show the averaged $(x,Q)$ over multiplicities in a bin. The bin $0.12<x<0.2$ is not included in the fit since it has $\langle Q\rangle <2$GeV.}
\end{center}
\end{figure}

In the current literature, one can find  several measurements of the unpolarized SIDIS~\cite{Derrick:1995xg,Adloff:1996dy,Asaturyan:2011mq,Airapetian:2012ki,Adolph:2013stb,Aghasyan:2017ctw} and a  total of some thousands of data points. We restrict our attention only to those data whose kinematical features are compatible with the energy scaling of  TMD factorization theorem. The first constraint comes from the di-lepton invariant mass ($Q$) and in general from the energy scale of the processes. Most of SIDIS reactions have been measured at fixed target experiments, that are typically run at low energies. Unfortunately much of these data  do not  accomplish the QCD factorization request of a high $Q$  to separate field modes.
 To secure our analysis (but still leave some data) we have used a restriction on the average $Q$ of a data point, namely
\begin{eqnarray}\label{SIDIS-data:Q>2}
\langle Q \rangle \geq 2 \text{GeV}.
\end{eqnarray}
Here, $\langle Q \rangle$ is the value of $Q$ averaged over the multiplicity value in a bin, see fig.~\ref{fig:bining}. The restriction in eq.~(\ref{SIDIS-data:Q>2}) quite reduces the pool of data. In particular, eq.~(\ref{SIDIS-data:Q>2}) completely discards the JLAB measurement published in \cite{Asaturyan:2011mq}, and cuts out the most part of HERMES data in ref.~\cite{Airapetian:2012ki}.

The second constraint comes from the TMD factorization assumptions. Namely, the TMD factorization regime is fully consistent only for  low values of $q_T/Q$ and  receives quadratic power corrections of order $(q_T/Q)^2$, see eq.~(\ref{SIDIS:Wmunu-J}) and eq.~(\ref{SIDIS:xSec}). We consider data such that
\begin{eqnarray}\label{SIDIS-data:qT<0.25 Q}
\delta\equiv\frac{\langle q_T \rangle}{\langle Q\rangle}<0.25,
\end{eqnarray}
where the value $0.25$ was deduced in \cite{Scimemi:2017etj}. From the $q_T$ interval of eq.~(\ref{SIDIS-data:qT<0.25 Q}), one can expect a $\sim4-6\%$ influence of the power corrections, which is well inside the uncertainties of the data. In sec.~\ref{sec:TMD-limit}, we have tested  cutting the condition in eq.~(\ref{SIDIS-data:qT<0.25 Q}) considering the data at different $\delta$, and found eq.~(\ref{SIDIS-data:qT<0.25 Q}) sufficient.

It should not pass unobserved that
eq.~(\ref{SIDIS-data:qT<0.25 Q}) is written in  terms of $\vec q^2_T$, that is the natural variable of  TMD factorization approach, whereas the data are presented in terms of $\vec p_\perp^2$. These variables are related by $\vec q_T^2\simeq \vec p_\perp^2/z^2$, see eq.~(\ref{def:qT<->pT}). Thus, the cut  in eq.~(\ref{SIDIS-data:qT<0.25 Q}) puts also a restriction on $z$. Altogether it makes the allowed values of $\vec p_\perp^2$ even smaller, $\vec p_\perp\lesssim 0.25 z Q$. In particular, we have to completely discard the measurements of H1 and ZEUS collaborations \cite{Derrick:1995xg,Adloff:1996dy} that are made at very small values of $z$, despite the relatively high values of $Q$.

After the application of eq.~(\ref{SIDIS-data:Q>2},~\ref{SIDIS-data:qT<0.25 Q}) we are left with the data taken by HERMES and COMPASS\footnote{We do not consider the data from \cite{Adolph:2013stb} since they have large systematic errors, and fully replaced by \cite{Aghasyan:2017ctw}.} collaborations~\cite{Airapetian:2012ki,Aghasyan:2017ctw}. For HERMES we have selected  the \texttt{zxpt-3D}-binning set due to the finer bins in $p_T$. The COMPASS data  includes the subtraction of vector-boson channel, and thus we also select the subtracted HERMES data (\texttt{.vmsub} set). In total we have 582 points that cover the region of $1.5\simeq Q\simeq 9$ GeV, $10^{-2}\simeq x<0.6$, $0.2<z<0.8$. The summary of the considered data is reported in table \ref{tab:SIDIS-data}.
\subsection{DY data}
\begin{table}[t]
\begin{center}
\small
\begin{tabular}{|c||c|c|c|c|c|c|}
\hline
Experiment & ref. 
&$\sqrt{s}$ [GeV]& $Q$ [GeV] & $y$/$x_F$ & \specialcellcenter{fiducial\\region}  & \specialcellcenter{$N_{\rm pt}$\\after cuts}
\\\hline\hline 
E288 (200) & \cite{Ito:1980ev} 
& 19.4 & \specialcellcenter{4 - 9 in\\ 1~GeV bins$^*$} & $0.1<x_F<0.7$  & -& 43
\\\hline
E288 (300) & \cite{Ito:1980ev} 
& 23.8 & \specialcellcenter{4 - 12 in \\ 1~GeV bins$^*$} & $-0.09<x_F<0.51$ & - & 53
\\\hline
E288 (400) & \cite{Ito:1980ev} 
& 27.4 & \specialcellcenter{5 - 14 in \\ 1~GeV bins$^*$} & $-0.27<x_F<0.33$ & - & 76
\\\hline\hline
E605 & \cite{Moreno:1990sf} 
& 38.8 & \specialcellcenter{7 - 18 in \\ 5 bins$^*$} & $-0.1<x_F<0.2$ & -&  53
\\\hline\hline
E772 & \cite{McGaughey:1994dx} 
& 38.8 & \specialcellcenter{5 - 15 in \\ 8 bins$^*$} & $0.1<x_F<0.3$ & - & 35
\\\hline\hline
PHENIX & \cite{Aidala:2018ajl} 
& 200 & 4.8 - 8.2 & $1.2<y<2.2$ & - & 3
\\\hline\hline
CDF (run1) & \cite{Affolder:1999jh} 
& 1800 & 66 - 116 & - & - & 33
\\\hline
CDF (run2) & \cite{Aaltonen:2012fi} 
& 1960 & 66 - 116 & - & - & 39
\\\hline\hline
D0 (run1) & \cite{Abbott:1999wk} 
& 1800 & 75 - 105 & - & - & 16
\\\hline
D0 (run2) & \cite{Abazov:2007ac} 
& 1960 & 70 - 110 & - & - & 8
\\\hline
D0 (run2)$_\mu$ & \cite{Abazov:2010kn} 
& 1960 & 65 - 115 & $|y|<1.7$ & \specialcellcenter{$p_T>15$ GeV \\ $|\eta|<1.7$} & 3
\\\hline\hline
ATLAS (7TeV) & \cite{Aad:2014xaa} 
& 7000 & 66 - 116 & \specialcellcenter{$|y|<1$\\ $1<|y|<2$ \\ $2<|y|<2.4$} & \specialcellcenter{$p_T>20$ GeV \\ $|\eta|<2.4$} & 15
\\\hline
ATLAS (8TeV) & \cite{Aad:2015auj} 
& 8000 & 66 - 116 & \specialcellcenter{$|y|<2.4$\\ in 6 bins} & \specialcellcenter{$p_T>20$ GeV \\ $|\eta|<2.4$} & 30
\\\hline
ATLAS (8TeV) & \cite{Aad:2015auj} 
& 8000 & 46 - 66 & $|y|<2.4$ & \specialcellcenter{$p_T>20$ GeV \\ $|\eta|<2.4$} & 3
\\\hline
ATLAS (8TeV) & \cite{Aad:2015auj} 
& 8000 & 116 - 150 & $|y|<2.4$ & \specialcellcenter{$p_T>20$ GeV \\ $|\eta|<2.4$} & 7
\\\hline\hline
CMS (7TeV) & \cite{Chatrchyan:2011wt} 
& 7000 & 60 - 120 & $|y|<2.1$ & \specialcellcenter{$p_T>20$ GeV \\ $|\eta|<2.1$} & 8
\\\hline
CMS (8TeV) & \cite{Khachatryan:2016nbe} 
& 8000 & 60 - 120 & $|y|<2.1$ & \specialcellcenter{$p_T>20$ GeV \\ $|\eta|<2.1$} & 8
\\\hline\hline
LHCb (7TeV) & \cite{Aaij:2015gna} 
& 7000 & 60 - 120 & $2<y<4.5$ & \specialcellcenter{$p_T>20$ GeV \\ $2<\eta<4.5$} & 8
\\\hline
LHCb (8TeV) & \cite{Aaij:2015zlq} 
& 8000 & 60 - 120 & $2<y<4.5$ & \specialcellcenter{$p_T>20$ GeV \\ $2<\eta<4.5$} & 7
\\\hline
LHCb (13TeV) & \cite{Aaij:2016mgv} 
& 13000 & 60 - 120 & $2<y<4.5$ & \specialcellcenter{$p_T>20$ GeV \\ $2<\eta<4.5$} & 9
\\\hline\hline
Total & & 
& & & & 457
\\\hline
\end{tabular}
\par
*Bins with $9\lesssim Q \lesssim 11$ are omitted due to the $\Upsilon$ resonance.
\caption{Summary table for the data included in the fit.\label{tab:data}. For each data set we report: the reference publication, the centre-of-mass energy, the coverage in $Q$ and $y$ or $x_F$, possible cuts on the fiducial region, and the number of data points that survive the cut in eq.~(\ref{DY-data:cuts}).}
\end{center}
\end{table}

The DY data are selected following the same principles as the SIDIS data,  eq.~(\ref{SIDIS-data:qT<0.25 Q}) (the rule (\ref{SIDIS-data:Q>2}) makes no sense now, because  DY processes are measured at sufficiently high-energies) with only small modifications. The changes consist in cutting some extra higher-$q_T$ data points for several specific data sets (this concerns mainly ATLAS measurements of Z-boson production). The reason for it is that the estimated size of power corrections at $q_T/Q\sim 0.25$ is of order of $5\%$, however, some highly precise data are measured with much better accuracy. So, given a data point $p\pm \sigma$, with $p$ being the central value and $\sigma$ its uncorrelated relative uncertainty, corresponding to some values of $q_T$ and $Q$, we include it in the fit only if
\begin{eqnarray}\label{DY-data:cuts}
\delta\equiv \frac{\langle q_T\rangle}{\langle Q\rangle } <0.1,\qquad \text{or}\qquad \delta<0.25\quad \text{if}\quad \delta^2<\sigma.
\end{eqnarray}
In other words, if the (uncorrelated) experimental uncertainty of a given data point is smaller than the theoretical uncertainty associated to the expected size of power corrections, we drop this point from the fit. This is the origin of the second condition in eq.~(\ref{DY-data:cuts}). 

The resulting data set contains 457 data points, and spans a wide range in energy, from $Q=4$~GeV to $Q=150$~GeV, and in $x$, from $x\sim0.5\cdot10^{-4}$ to $x\sim 1$. Table~\ref{tab:data} reports a summary of the full data set included in our fit. This selection of the data is the same as the one considered in our earlier work \cite{Bertone:2019nxa}. In the current fit, we compare the absolute values of the cross-section, whenever they are available. The only data set that require normalization factors are all CMS data, ATLAS at 7 TeV, and D0 run2 measurements. For these sets we have normalized the integral of the theory prediction to the corresponding integral over the data (see explicit expression in ref.~\cite{Scimemi:2017etj}).

\subsection{Summary of the  data set}
\label{sec:summary-data}

In total for the extraction of unpolarized TMD distribution we analyze 1039 data points that are almost equally distributes between SIDIS (582 points) and DY (457 points) processes. All these points contribute to the determination of the TMD evolution kernel $\mathcal{D}$ and unpolarized TMDPDF $f_1$. The determination of unpolarized TMDFF is based only on SIDIS data. In addition, we recall that a single DY data point is simultaneously sensitive to a larger and a smaller value of $x$. This is because the cross section is given by a pair of TMDPDFs, eq.~(\ref{def:Wff}), computed at $x_1$ and $x_2$ such that $x_1x_2 \simeq Q^2/s$. So, the statistical weight of a DY point in the determination of TMDPDF is effectively doubled.

\begin{figure}
\begin{center}
\includegraphics[width=0.6\textwidth]{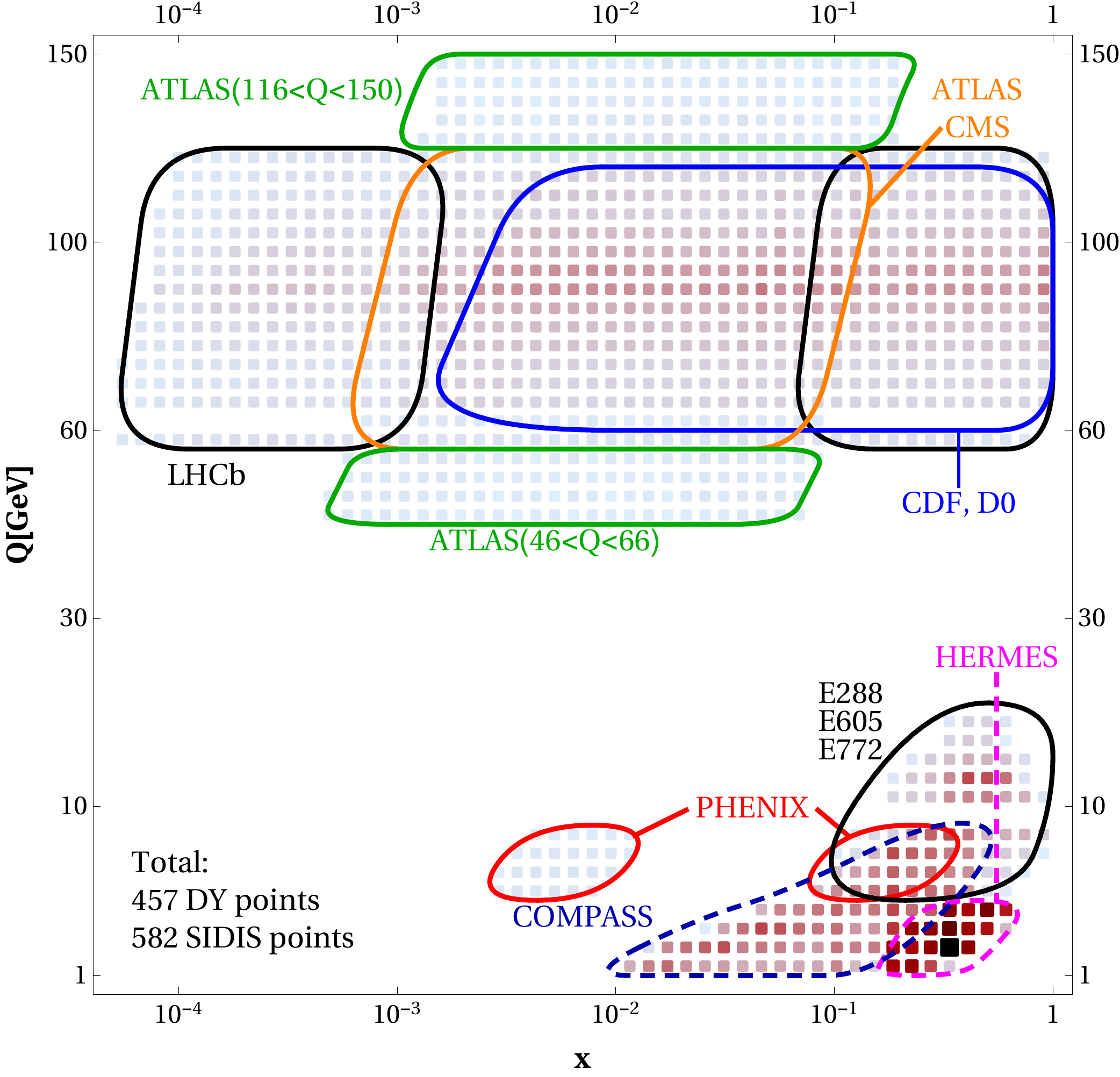}
\caption{\label{fig:dataPoints} Density of data in the plane $(Q,x)$ (a darker color corresponds to a higher density).}
\end{center}
\end{figure}

The kinematic region in $x$ and $Q$ covered by the data set and thus contributing to the determination of TMDPDF is shown in fig.~\ref{fig:dataPoints}. The boxes enclose the sub-regions covered by the single data sets.  Looking at fig.~\ref{fig:dataPoints}, it is possible to distinguish two main clusters of data: the ``low-energy experiments'', \textit{i.e.} E288, E605, E772, PHENIX, COMPASS and HERMES that place themselves at invariant-mass energies between 1 and 18~GeV, and the ``high-energy experiments'', \textit{i.e.} all those from Tevatron and LHC, that are instead distributed around the $Z$-peak region. From this plot we observe that, kinematic ranges of SIDIS and DY data do not overlap.

As a final comment  of this section let us mention that our data selection is particularly conservative because it drops points that could potentially be described by TMD factorization (see e.g. ref.~\cite{Bacchetta:2017gcc} where a  less conservative choice of cuts is used). However, our fitted data set guarantees that we operate well within the range of validity of TMD factorization. In sec.~\ref{sec:GLOBAL-fit} we show that unexpectedly our extraction can describe a larger set of  data as well.

\section{Fit procedure}
\label{sec:fitprocedure}

The experimental data  are usually provided in a form specific for each setup. In order to extract valuable information for the TMD extraction, one has to detail the methodology  that has been followed, and this is the purpose of this section. Finally, we also provide a suitable definition of the $\chi^2$ that allows for a correct exploitation of experimental uncertainties.

\subsection{Treatment of nuclear targets and charged hadrons}
\label{sec:nuclear}
The data from  E288, E605 (\textit{Cu}), E772, COMPASS, part of HERMES (isoscalar targets) come from nuclear target processes. In these cases, we perform the iso-spin rotation of the corresponding TMDPDF that simulates the nuclear-target effects. For example, we replace u-, and d-quark distributions by
\begin{eqnarray}
f_{1,u\ot A}(x,b)&=&\frac{Z}{A}f_{1,u\ot p}(x,b)+\frac{A-Z}{A}f_{1,d\ot p}(x,b),
\\
f_{1,d\ot A}(x,b)&=&\frac{Z}{A}f_{1,d\ot p}(x,b)+\frac{A-Z}{A}f_{1,u\ot p}(x,b),
\end{eqnarray}
where A(Z) is atomic number(charge) of a nuclear target. In principle, for E288, E605 data extracted from very heavy targets one should also incorporate the nuclear modification factor that depends on $x$. In the given kinematics the nuclear modification factor produces  effects of order 5-10\% in the normalization of the cross-section. The shape of cross-section is changed in much smaller amount, about $1\%$ in a point, as it is shown in f.i. \cite{Vladimirov:2019bfa,Bacchetta:2019tcu}. Simultaneously, the systematic (correlated) errors of these experiments are large 25\% and 20\%, correspondingly, as well as the  uncorrelated error (typically 2-5\%). Therefore, we are not sensitive to nuclear modification effect.

The measurements of SIDIS are made in a number of different channels. The HERMES data include $\pi^\pm$ and $K^\pm$, and COMPASS data are for charged hadrons,  $h^\pm$. Pions and kaons are described by an individual TMDFFs. However, charged hadrons are a composition of different TMDFFs. According eq.~(\ref{def:D1}) the TMDFF for charged hadrons is a direct sum of TMDFFs for individual hadrons:
\begin{eqnarray}\label{def:h+-}
D_{1,f\to h^\pm}(x,b)=\sum_{h\in h^\pm}D_{1,f\to h}(x,b)=D_{1,f\to \pi^\pm}(x,b)+D_{1,f\to K^\pm}(x,b)+...~,
\end{eqnarray}
where dots denote the higher-mass hadron states. At COMPASS energies, this sum is dominated by the pion ($65-75\%$), and the kaon ($15-20\%$) contributions. The residual term is lead by proton/antiproton contribution ($2-5\%$). The contribution of other particles is smaller (for discussion and references see \cite{Bertone:2018ecm,Bertone:2017tyb}). Thus, in our study we use the first two terms of eq.~(\ref{def:h+-}) to simulate the charged hadron fragmentation.

The SIDIS measurements in refs.~\cite{Airapetian:2012ki,Aghasyan:2017ctw} are given in form of multiplicities. The SIDIS multiplicity is defined as
\begin{eqnarray}\label{def:multiplicity}
\frac{dM^h(x,Q^2,z,\vec p_\perp^2)}{dz d \vec p_\perp^2}=\(\frac{d\sigma}{dxdz dQ^2 d\vec p_\perp^2}\)\Big/\(\frac{d\sigma_{\text{DIS}}}{dxdQ^2}\),
\end{eqnarray}
where $d\sigma_{\text{DIS}}$ is the differential cross-section  for DIS. It reads
\begin{eqnarray}
\frac{d\sigma_{\text{DIS}}}{dxdQ^2}=\frac{4\pi \alpha_{\text{em}}^2}{xQ^4}\Big[\(1-y-\frac{y^2\gamma^2}{4}\)F_2(x,Q^2)+xy^2 F_1(x,Q^2)\Big],
\end{eqnarray}
where $F_1$ and $F_2$ are DIS structure functions. The DIS cross-section cannot be computed  starting from TMD factorization, but it is described by the collinear factorization theorem. In order to evaluate the multiplicity we have pre-computed the DIS cross-section (integrated over the given bin) by the \texttt{APFEL}-library \cite{Bertone:2013vaa}, and then divided the TMD prediction according to eq.~(\ref{def:multiplicity}).

\subsection{Bin integration in SIDIS and DY}

The majority of SIDIS data is measured at relatively low-Q and in large bins. The cross-section value changes greatly within a bin, and so, binning effects are known to be strong. For a measured cross-section $d\sigma/dx dz dQ^2d \vec p_\perp^2$, a bin is specified by $\{x_{\text{min}},x_\text{max}\}$, $\{z_{\text{min}},z_\text{max}\}$, $\{Q_{\text{min}},Q_\text{max}\}$ and $\{\vec p_{\text{min}},\vec p_\text{max}\}$. The binning constraints impose certain cuts on the measured phase space. Typically, these cuts are given as intervals of the variable $y$ and of the invariant mass of photon-target system $W^2=(P+q)^2$, which belong to ranges $\{y_{\text{min}},y_\text{max}\}$ and $\{W^2_\text{min},W^2_\text{max}\}$. Both these variables are connected to $x$ and $Q^2$,
\begin{eqnarray}
W^2=M^2+Q^2 \frac{1-x}{x},\qquad y=\frac{Q^2}{x(s-M^2)}.
\end{eqnarray}
where $s$ is the Mandelshtam variable $s=(P+l)^2$. So, in the presence of fiducial cuts in SIDIS the bin boundaries are
\begin{eqnarray}
\hat x_{\text{min}}(Q)&=&\max\{x_{\text{min}},\frac{Q^2}{y_{\text{max}}(s-M^2)},\frac{Q^2}{Q^2+W^2_{\text{max}}-M^2}\},
\\
\hat x_{\text{max}}(Q)&=&\min\{x_{\text{max}},\frac{Q^2}{y_{\text{min}}(s-M^2)},\frac{Q^2}{Q^2+W^2_{\text{min}}-M^2}\},
\\
\hat Q^2_{\text{min}}&=&\max\{Q^2_{\text{min}},x_{\text{min}}y_{\text{min}}(s-M^2),\frac{x_{\text{min}}}{1-x_{\text{min}}}(W^2_{\text{min}}-M^2)\},
\\
\hat Q^2_{\text{max}}&=&\min\{Q^2_{\text{max}},x_{\text{max}}y_{\text{max}}(s-M^2),\frac{x_{\text{max}}}{1-x_{\text{max}}}(W^2_{\text{max}}-M^2)\}.
\end{eqnarray}
An example of effects of cuts in the bins is shown in fig.~\ref{fig:bining}. 
In the case of multiplicity measurements the  bin effects are taken into account with the cross-section
\begin{eqnarray}
&&\frac{dM^h(x,Q^2,z,\vec p_\perp^2)}{dz d \vec p_\perp^2}\Bigg|_{\text{bin}}=(z_\text{max}-z_{\text{min}})^{-1}(\vec p^2_\text{max}-\vec p^2_{\text{min}})^{-1}
\\\nn &&\qquad \times\int_{\vec p_{\text{min}}}^{\vec p_{\text{max}}}2\vec p_\perp d \vec p_\perp  \int_{z_{\text{min}}}^{z_{\text{max}}} dz \int_{\hat Q_\text{min}}^{\hat Q_\text{max}}2Q dQ \int_{\hat x_{\text{min}}(Q)}^{\hat x_{\text{max}}(Q)}dx \frac{d\sigma}{dxdz dQ^2 d\vec p_\perp^2}
\\\nn &&\qquad \Big/\int_{\hat Q_\text{min}}^{\hat Q_\text{max}}2Q dQ \int_{\hat x_{\text{min}}(Q)}^{\hat x_{\text{max}}(Q)}dx \frac{d\sigma_{\text{DIS}}}{dxdQ^2 },
\end{eqnarray}
where the expression in the first line is the volume of $(z,\vec p_\perp^2)$-bin.

In the case of DY the binning effects are also extremely important. The difference in the value of the cross section between center-of-bin and the averaged/integrated value can reach tenth of percents, especially, for very low-energy bins (where the change in $Q$ is rapid), and for very wide bins (such as Z-boson measurement). 
We have used the definition
\begin{eqnarray}
\frac{d\sigma}{dQ^2dy\vec q_T^2}\Big|_\text{bin}=(Q^2_{\text{max}}-Q^2_\text{min})^{-1}(\vec q^2_{\text{max}}-\vec q^2_\text{min})^{-1}(y_{\text{max}}-y_\text{min})^{-1}
\\\nn \times \int_{\vec q_\text{min}}^{\vec q_\text{max}} 2\vec q_T d\vec q_T \int_{Q_\text{min}}^{Q_\text{max}} 2Q dQ
\int_{y_\text{min}}^{y_\text{max}} dy \frac{d\sigma}{dQ^2dy\vec q_T^2}.
\end{eqnarray}

\subsection{Definition of $\chi^2$-test function and estimation of uncertainties}

To test the theory prediction against the experimental measurement we compute the $\chi^2$-test function
\begin{equation}\label{eq:chi2cov}
\chi^2=\sum_{i,j=1}^{n}\(m_i-t_i\right)V_{ij}^{-1}\(m_j-t_j\),
\end{equation}
where $m_i$ is the central value of $i$'th measurement, $t_i$ is the theory prediction for this measurement and $V_{ij}$ is the covariance matrix. An accurate definition of the covariance matrix is  essential for a correct exploitation of experimental uncertainties. In order to build the covariance matrix we distinguish, uncorrelated and correlated uncertainties. For example, a typical data point has the structure
\begin{equation}
m_i\pm \sigma_{i,\text{stat}} \pm \sigma_{i,\text{unc}} \pm \sigma_{i,\text{corr}}^{(1)}\pm\dots \pm \sigma_{i,\text{corr}}^{(k)},
\end{equation}
where $m_i$ the reported central value, $\sigma_{i,\text{stat}}$ is (uncorrelated) statistical uncertainty, $\sigma_{i,\text{unc}}$ is uncorrelated systematic uncertainty, and $\sigma_{i,\text{corr}}^{(k)}$ are correlated systematic uncertainties. Uncorrelated uncertainties give an estimate of the degree of knowledge of a particular data point irrespective of the other measurements of the data set. Instead, correlated uncertainties provide an estimate of the correlation between the statistical fluctuations of two separate data points of the same data set. With this information at hand, one can construct the covariance matrix $V_{ij}$ as follows (for more detailed discussion on this definition see refs.~\cite{Ball:2008by,Ball:2012wy}):
\begin{equation}\label{eq:covmat}
V_{ij}=\(\sigma_{i,\text{stat}}^2 +\sigma_{i,\text{unc}}^2\)\delta_{ij} + \sum_{l=1}^{k}\sigma_{i,\text{corr}}^{(l)}\sigma_{j,\text{corr}}^{(l)}.
\end{equation}
Equipped with this definition of covariance matrix the $\chi^2$-test in eq.~(\ref{eq:chi2cov}) takes into account the nature of the experimental uncertainties leading to a faithful estimate of the agreement between data and theoretical predictions. 

To estimate the error propagation  from the experimental data to the extracted values of TMD distributions we have used the replica method. This method is described in details in ref.~\cite{Ball:2008by}. It consists in the generation of $N$ replicas of  pseudo-data, and the minimization of the $\chi^2$ on each replica. The resulting set of $N$ vectors of NP parameters is distributed in accordance to the distribution law of the data. And thus, it represents a Monte Carlo sample that is used to evaluate mean values, standard deviation and correlations of the NP parameters. For the estimation of error propagation we consider $N=100$ replicas. The procedure of $\chi^2$-minimization for each replica is the most computationally heavy part of the fit.

The proper treatment of correlated uncertainties is essential in global analysis. The presence of sizable correlated uncertanties could result into a misleading visual disagreement between theory prediction and the (central values of) data points. Namely, the theory prediction for a data set could be globally shifted by significant amount, that is nonetheless in agreement with correlated experimental uncertainty. To quantify the effects of correlated shifts we use the nuisance parameter method presented in \cite{Ball:2008by,Ball:2012wy}. Within the nuisance parameter method one is able to determine the shift $d_i$ of a theory prediction $t_i$ for the $i$'th data point, such that $\bar t_i=t_i+d_i$ contributes only to the uncorrelated part of the $\chi^2$-value. The value $d_i$ is interpreted as a shift caused by the correlated uncertainties. It is computed as 
\begin{eqnarray}
d_i=\sum_{l,m=1}^k \sigma_{i,\text{corr}}^{(l)}\,A^{-1}_{lm}\,\rho_m,
\end{eqnarray}
where
\begin{eqnarray}
A_{lm}=\delta_{lm}+\sum_{i=1}^n\frac{\sigma_{i,\text{corr}}^{(l)}\sigma_{i,\text{corr}}^{(m)}}{\sigma_{i,\text{stat}}^2 +\sigma_{i,\text{unc}}^2},\qquad
\rho_l=\sum_{i=1}^n \frac{m_i-t_i}{\sigma_{i,\text{stat}}^2 +\sigma_{i,\text{unc}}^2}\sigma_{i,\text{corr}}^{(l)}~.
\end{eqnarray}
It also instructive to check the average systematic shift, which we define as
\begin{eqnarray}\label{def:d/sigma}
\langle d/\sigma \rangle = \frac{1}{n}\sum_{i=1}^n \frac{d_i}{m_i}.
\end{eqnarray}
It shows a general deficit/excess of the theory with respect to the data for a given data set.

Let us note that the multiplicities in SIDIS  are  experimentally convenient because the systematic uncertainties related to the measurement efficiency and the beam luminosity cancel in the ratio. However, theoretically, the  multiplicities are not so well defined, since the denominator and the numerator of multiplicity ratio (\ref{def:multiplicity}) need a completely different theoretical treatment. In order to account this effect, we have computed the uncertainty of theory prediction for DIS cross-section for each bin and added it as a fully correlated error for each data set. We should admit that the theory uncertainty for DIS cross-section is negligibly small (typically, $0.1-2.0\%$) in comparison to systematic uncertainties of experiment. 
As a result the values of $\chi^2$ change very little on the level of $\pm10^{-2}$ per point.

\subsection{Artemide}

The computation of the cross-section is made with the code \texttt{artemide} that is developed by us. \texttt{Artemide} is organized as a package of Fortran 95 modules, each devoted to evaluation of a single theory construct, such as the TMD evolution factor, a TMD distribution, or their combinations such as structure functions $W$ and cross-sections. The \texttt{artemide} also evaluates all necessary procedures needed for the comparison with the experimental data, such as bin-integration routines and cut factors. For simplicity of data analysis \texttt{artemide} is equipped by a python interface, called \texttt{harpy}. The \texttt{artemide} package together with the \texttt{harpy} is available in the repository~\cite{web}.

The module organization of \texttt{artemide} allows for flexible use. In particular, it gives to a user a full access to non-perturbative ansatzes and models. Although \texttt{artemide} is based on the $\zeta$-prescription, it also includes other strategies for TMD evolution, such as CSS evolution \cite{Aybat:2011zv}, $\gamma$-improved evolution \cite{Scimemi:2018xaf} and their derivatives. The user has full control on the perturbative orders, and can set each individual part to a particular (known) order. Currently, \texttt{artemide} can evaluate unpolarized TMD distributions, and linearly polarized gluon distributions together with the related cross-sections, such as DY, SIDIS, Higgs-production (for application see \cite{Gutierrez-Reyes:2019rug}), etc. In future, we plan to include more processes and distributions.

The evaluation of a single cross-section point that is to be compared with the experimental one, implies the evaluation of a number of integrals: two Mellin convolutions for small-$b$ matching eq.~(\ref{def:phen-f1}, ~\ref{def:phen-D1}), the Hankel-type integral for the structure function $W$ eq.~(\ref{def:WfD-final}, \ref{def:Wff-final}), and 3(in DY case)/4(in SIDIS case) bin-integrations. Note, that in the $\zeta$-prescription one does not need to evaluate integrations for TMD evolution, which is its additional positive point. Altogether, it makes the evaluation of TMD cross-section rather expensive in terms of computing time. \texttt{Artemide} uses adaptive integration routines to ensure the  required computation accuracy. To speed-up the evaluation, \texttt{artemide} precomputes the tables of Mellin convolutions for TMD distributions that are the most time-consuming integrations. The code presently takes about 4.5 (3.2) minutes to evaluate a single $\chi^2$ value for the full data set of DY and  SIDIS given in sec.~\ref{sec:data} on an average 8-core (12-core) processor (2.5GHz) depending on the NP-values. Therefore, the minimization $\chi^2$ and especially the computation of error-propagation are especially long. Due to that we are restricted in certain important directions of studies (e.g. error-propagation of PDF sets, and flavour dependence).

\section{Fit of DY}
\label{sec:DY-fit}

The data-set and the functional input for the DY fit is inherited from our earlier study~\cite{Bertone:2019nxa}. The only modification is the update of the functional form of the special null-evolution line in eq.~(\ref{NP:zeta}), which in the present case matches the exact solution at large-$b$. This update leads relatively minor formal changes, while some values of the model parameter are changed as a result of the fit. The value of $\chi^2$ (per 457 points) is reduced from $1.174\; {\text{\cite{Bertone:2019nxa}}}\to1.168\; ({\text{this work}})$. The main impact takes place at low-energies. In particular, the typical deficit in the cross-section for low-energy experiments is reduced by 5-6\% (compare table 3 in~\cite{Bertone:2019nxa} with table~\ref{tab:final}), which however does not significantly affects the $\chi^2$ values due to the large correlated uncertainties of fixed-target DY measurements. 

In this section, we present the fit of DY data-set only. Since the general picture is similar to ref.~\cite{Bertone:2019nxa}, we concentrate on the sources of systematic uncertainties of our approach. We discuss the dependence on the collinear PDF, that serves as a boundary for TMDPDF, and the effects of $q_T$ corrections in the definitions of $x_{1,2}$.

\subsection{Dependence on PDF}
\label{sec:PDF}

\begin{table}[t]
\begin{center}
\begin{tabular}{|l|l|c|c|}
Short name & Full name & Ref. & LHAPDF id.
\\\hline\hline
NNPDF31 & NNPDF31\_nnlo\_as\_0118 & \cite{Ball:2017nwa} &303600 
\\\hline
HERA20 & HERAPDF20\_NNLO\_VAR & \cite{Abramowicz:2015mha} &61230 
\\\hline
MMHT14 & MMHT2014nnlo68cl & \cite{Harland-Lang:2014zoa} &25300
\\\hline
CT14 & CT14nnlo & \cite{Dulat:2015mca} & 13000
\\\hline
PDF4LHC & PDF4LHC15\_nnlo\_100 &  \cite{Butterworth:2015oua} & 91700
\\\hline
\end{tabular}
\end{center}
\caption{\label{tab:PDFs} List of collinear PDF used as the boundary for unpolarized TMDPDF.}
\end{table}

The collinear PDF is an important part of our model for TMDPDF, e.g. eq.~(\ref{def:phen-f1}). The issue of PDF-bias of our result can be stated in the following terms. The small-$b$ matching essentially reduces the number of NP parameters for TMDPDF and guarantees the asymptotic agreement of the TMDPDF with the collinear observables. The small-$b$ part of the Hankel integral gives a sizable contribution to the cross-section, especially for $q_T\sim 10$-$20$ GeV. Therefore, the quality of our fit and the values of the extracted NP parameter are robustly correlated with the collinear PDF set. This observation has been made earlier, e.g. see discussion in \cite{Signori:2013mda,Bacchetta:2017gcc,Bertone:2019nxa}, but it has not been systematically studied. Ideally, the PDF set and TMDPDF are to be coherently extracted in a global fit of collinear and TMD observables. Meanwhile, we treat the collinear inputs as independent parameters that we cannot control and we test various sets available in the literature.

\begin{figure}[t]
\begin{center}
\includegraphics[width=0.5\textwidth]{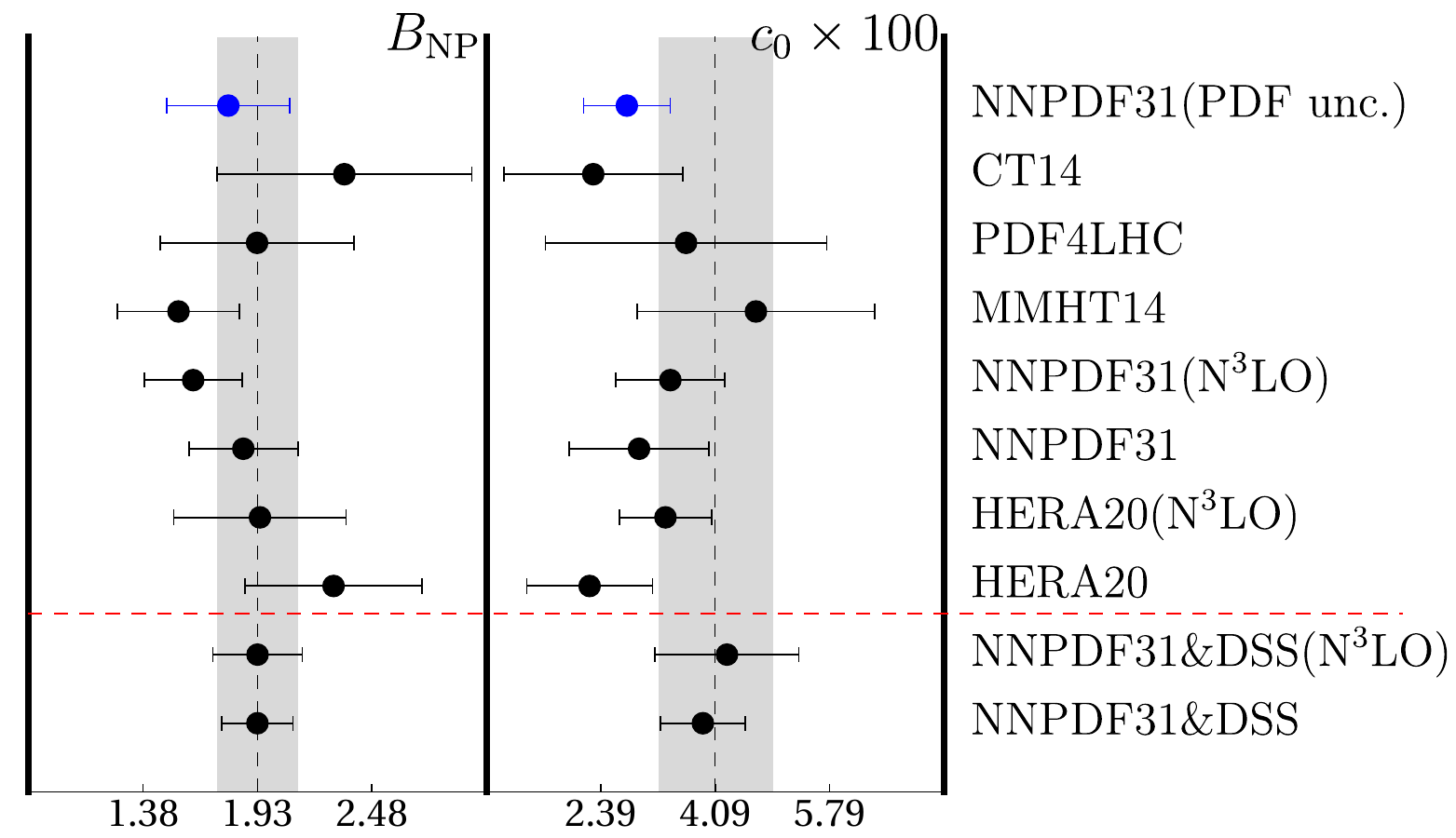}
\end{center}
\caption{\label{fig:DNP-param} Comparison of NP parameter for TMD evolution extracted in different fits. The values above the red-dashed line are extracted in fit of DY data, see table \ref{tab:DY-PDF-chi}. The values below the red-dashed line are extracted in global fit of DY and SIDIS data, see table \ref{tab:NP-param-final}. The vertical dashed lines and gray boxes correspond to average mean and standard deviation of the results of the global fit. The blue points and their error-bars correspond to the estimation of the uncertainties from the collinear PDF (see sec.\ref{sec:PDFuncert}). The input collinear distributions are marked in the right column.}
\begin{center}
\includegraphics[width=1.0\textwidth]{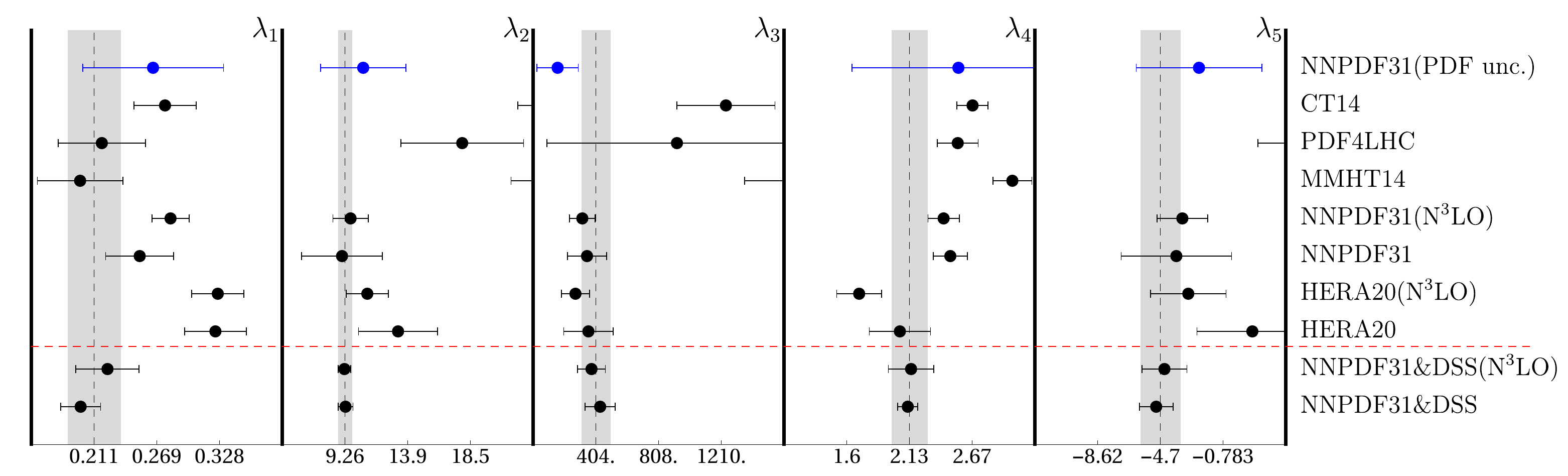}
\end{center}
\caption{\label{fig:f1-param} Comparison of NP parameter for unpolarized TMDPDF extracted in different fits. The values above the red-dashed line are extracted in fit of DY data, see table \ref{tab:DY-PDF-chi}. The values below the red-dashed line are extracted in global fit of DY and SIDIS data, see table \ref{tab:NP-param-final}. The vertical dashed lines and gray boxes correspond to average mean and standard deviation of the results of the global fit. The blue points and their error-bars correspond to the estimation of the uncertainties from the collinear PDF (see sec.~\ref{sec:PDFuncert}). The input collinear distributions are marked in the right column. For CT14 and MMHT14 some NP parameters are beyond the plot region.}
\end{figure}

There is an enormous amount of available PDF sets. We have tested some of the most popular sets that are recently extracted at NNLO accuracy, see table \ref{tab:PDFs}. All sets have LHAPDF interface \cite{Buckley:2014ana}.  For each PDF set we have performed the full fit procedure with the estimation of the error-propagation. In the fit, the central value of PDFs are used, see sec.~\ref{sec:PDFuncert} for a discussion uncertainties induced by PDFs. The values of the $\chi^2/(N_{pt}=457)$ and the NP parameters are reported in  table \ref{tab:DY-PDF-chi} for each PDF set in table \ref{tab:PDFs}. The visual comparison of the parameter values is shown in fig.~\ref{fig:DNP-param} and \ref{fig:f1-param}. The  parameters of RAD ($B_{\text{NP}}$ and $c_0$) are rather stable with respect to input PDF, and in agreement with each other (note, that $B_{\text{NP}}$ and $c_0$ are anti-correlated, see sec.~\ref{sec:NP-RAD-final}). 

Contrary to the RAD, the parameters $\lambda_i$ show a significant dependence on the collinear PDF (see fig.~\ref{fig:f1-param}). This fact is expected, since a different collinear PDF dictates a different shape in $x$, while the $Q$-dependence is not changed. The parameters $\lambda_1$ and $\lambda_2$ do not change significantly with different PDFs, while a bigger change is  provided by the parameters $\lambda_{3,4,5}$. This is because the parameters $\lambda_{1,2}$ dictate the main shape of $f_{NP}$ at middle values of $b$, whereas other parameters are responsible for the large-$b$ tale ($\lambda_{3,4}$) or fine-tuning of $x$-shape $(\lambda_5)$.

In  table \ref{tab:DY-PDF-chi}, fits are ordered according to the $\chi^2/N_{pt}$ value obtained in the DY fit. The distribution of the values of $\chi^2$ between experiments changes for different PDFs. For example, NNPDF31 demonstrates some tension between ATLAS and LHCb subsets (see table 3 in ref.~\cite{Bertone:2019nxa}, and also table \ref{tab:final}). In the case of HERA20 this tension reduces. The value of $\chi^2/N_{pt}$ for ATLAS measurements is practically the same in both cases, we find $2.02_{\text{NNPDF}}$ vs. $1.99_{\text{HERA}}$ for $N_{pt}=55$ (note, that the bin-by-bin distribution of $\chi^2$ changes between the sets). On contrary, the value of $\chi^2/N_{pt}$ for LHCb measurement undoubtedly differ in the two sets of PDF, as we find $2.93_{\text{NNPDF}}$ vs. $1.24_{\text{HERA}}$ for $N_{pt}=24$. The main part of the improvement happens due to the general normalization, that is lower by 3-5\% in NNPDF case, and almost exact in HERA case.

The TMD distributions with NNPDF31 and HERA20 show a $\chi^2$ value better than all the other, e.g. table \ref{tab:DY-PDF-chi}. These PDFs have also less tension between high- and low-energy data. For this reason, in the next sections we will consider only PDFs from these extractions. Nonetheless, we preferably select NNPDF31 set in the global SIDIS and DY analysis. The reason is that NNPDF31 distribution is extracted from the global pool of data, whereas HERA20 uses exclusively data from HERA. At the same time, we must admit that HERA20 distribution provides a spectacularly low values of $\chi^2$ in our global fit.

\begin{table}[b]
\begin{center}
\small
\begin{tabular}{|l V{4} c V{4} l|ll|}
\hline
PDF set & $\chi^2/N_{pt}$ & Parameters for $\mathcal{D}$& Parameters for $f_1$ &
\\\hline\hline
HERA20 & 0.97 & \specialcellleft{$B_\text{NP}=2.29\pm 0.43$ \\ $c_0=(2.22\pm0.93)\cdot 10^{-2}$} 
& \specialcellleft{$\lambda_1=0.324\pm0.029$ \\ $\lambda_2=13.2\pm2.9$}
& \specialcellleft{$\lambda_3=(3.56\pm 1.59)\cdot 10^2$ \\ $\lambda_4=2.05\pm0.26$ \\ $\lambda_5=-10.4\pm 3.5$}
\\\hline
NNPDF31 & 1.14 & \specialcellleft{$B_\text{NP}=1.86\pm 0.30$ \\ $c_0=(2.96\pm1.04)\cdot 10^{-2}$} 
& \specialcellleft{$\lambda_1=0.253\pm0.032$ \\ $\lambda_2=9.0\pm3.0$}
& \specialcellleft{$\lambda_3=(3.47\pm 1.16)\cdot 10^2$ \\ $\lambda_4=2.48\pm0.15$ \\ $\lambda_5=-5.7\pm 3.4$}
\\\hline
MMHT14 & 1.34 & \specialcellleft{$B_\text{NP}=1.55\pm 0.29$ \\ $c_0=(4.70\pm1.77)\cdot 10^{-2}$} 
& \specialcellleft{$\lambda_1=0.198\pm0.040$ \\ $\lambda_2=26.4\pm4.9$}
& \specialcellleft{$\lambda_3=(26.8\pm 13.2)\cdot 10^3$ \\ $\lambda_4=3.01\pm0.17$ \\ $\lambda_5=-23.4\pm 5.4$}
\\\hline
PDF4LHC & 1.53 & \specialcellleft{$B_\text{NP}=1.93\pm 0.47$ \\ $c_0=(3.66\pm2.09)\cdot 10^{-2}$} 
& \specialcellleft{$\lambda_1=0.218\pm0.041$ \\ $\lambda_2=17.9\pm4.5$}
& \specialcellleft{$\lambda_3=(9.26\pm 8.38)\cdot 10^2$ \\ $\lambda_4=2.54\pm0.17$ \\ $\lambda_5=-15.5\pm 4.7$}
\\\hline
CT14 & 1.59 & \specialcellleft{$B_\text{NP}=2.35\pm 0.61$ \\ $c_0=(2.27\pm1.33)\cdot 10^{-2}$} 
& \specialcellleft{$\lambda_1=0.277\pm0.029$ \\ $\lambda_2=24.9\pm2.9$}
& \specialcellleft{$\lambda_3=(12.4\pm 3.2)\cdot 10^3$ \\ $\lambda_4=2.67\pm0.13$ \\ $\lambda_5=-23.8\pm 2.9$}
\\\hline
HERA20(N$^3$LO) & 1.06 & \specialcellleft{$B_\text{NP}=1.94\pm 0.41$ \\ $c_0=(3.35\pm0.68)\cdot 10^{-2}$} 
& \specialcellleft{$\lambda_1=0.326\pm0.024$ \\ $\lambda_2=10.1\pm1.6$}
& \specialcellleft{$\lambda_3=(2.73\pm 0.91)\cdot 10^2$ \\ $\lambda_4=1.70\pm0.19$ \\ $\lambda_5=-6.5\pm 2.4$}
\\\hline
NNPDF31(N$^3$LO) & 1.13 & \specialcellleft{$B_\text{NP}=1.62\pm 0.24$ \\ $c_0=(3.42\pm1.04)\cdot 10^{-2}$} 
& \specialcellleft{$\lambda_1=0.282\pm0.017$ \\ $\lambda_2=9.7\pm1.3$}
& \specialcellleft{$\lambda_3=(3.17\pm 0.83)\cdot 10^2$ \\ $\lambda_4=2.42\pm0.13$ \\ $\lambda_5=-6.1\pm 1.6$}
\\\hline
\end{tabular}
\end{center}
\caption{\label{tab:DY-PDF-chi} Values of $\chi^2$ and NP parameters obtained in the fit of DY set of the data with different PDF inputs. Each set of PDF provide the corresponding  value of $\alpha_s(M_Z)$.}
\end{table}

\subsection{Impact of exact values for $x_{1,2}$ and power corrections}
\label{sec:DY:x12}

As  discussed in sec.~\ref{sec:summary-theory}, the factorization formula  eq.~(\ref{DY:xSec}) for DY contains three types of power corrections. The corrections related to TMD factorization cannot be tested, without extra modeling. The corrections due to fiducial cuts must be included without restrictions. Thus it is possible to  test only  power corrections
 due to the presence of $q_T/Q$ terms in the exact definition  of $x_{1,2}$, eq.~(\ref{def:x1x2}). The amount of this correction is obtained  comparing the fits of the DY data with
\begin{eqnarray}\nn
(\text{exact})x_{1,2}=\sqrt{\frac{Q^2+\vec q_T^2}{s}}e^{\pm y}\qquad \text{vs.}\qquad (\text{approx.}) x_{1,2}=\frac{Q}{\sqrt{s}}e^{\pm y}.
\end{eqnarray}
The approximate values for $x_{1,2}$ lead to higher values of $\chi^2$. In particular, with the approximate $x_{1,2}$  for the NNPDF31 set we have obtained $\chi^2/N_{pt}=1.35$ and $1.27$ at NNLO and N$^3$LO respectively. In the case of HERA20 set, we obtain $\chi^2/N_{pt}=1.03$ and $1.13$. Comparing these values to the ones reported in table~\ref{tab:DY-PDF-chi} (1.14 and 1.13; 0.95 and 0.06, respectively), we conclude that the quality of fit is worse.

The deterioration of the fit quality takes place in both high- and low- energy parts of the data. In the ATLAS experiment (that is the most precise set at our disposal, with $N_{pt}=55$), we observe the changes in $\chi^2/N_{pt}$: $1.82\to 2.83$ for NNPDF31 and $1.90\to 2.27$ for HERA20. For the fixed target experiments we have $\chi^2/N_{pt}$: $0.91\to 1.31$ for NNPDF31 and $0.71\to 0.97$ for HERA30 (here $N_{pt}=260$). We have also observed that the value of $\chi^2$ worsens mainly due to the change in the shape of cross-section, whereas the normalization part slightly reduces the $\chi^2$. The values NP parameters varies within the error-bands and the change in the central values is not significant.

Therefore, we conclude that  exact values of $x_{1,2}$ (\ref{def:x1x2}) considerably improve the quality of the fit. 
This conclusion is in agreement with the theory expectations presented in sec.~\ref{sec:summary-theory}.

\subsection{Uncertainties due to collinear PDFs}
\label{sec:PDFuncert}

The model in eq.~(\ref{def:phen-f1}) is not sensitive to changes of the NP parameters at small-$b$. For this reason, the error-band on the TMD distribution vanishes for $b\lesssim 0.5$GeV$^{-1}$. The only way to modify the TMD distribution in this region is to vary the values of collinear PDF. In sec.~\ref{sec:PDF} we have demonstrated that the quality of the fit, as well as the values of extracted NP parameters, essentially depend on the collinear PDF and in our extraction we have used the central values of PDF sets, ignoring the uncertainties of PDF determination. These uncertainties are however large  and could cover the gap among  different TMD fits if taken into account. Unfortunately, the incorporation of the PDF uncertainties  into the analysis is extremely demanding in terms of  computer time, especially for the full data set. In order to provide a quantitative estimate of the PDF-bias, in this section we  consider only the NNPDF31 data set with NNLO TMD evolution for the fit of DY data. We postpone to future work a similar analysis for the other PDF sets.

Thus, we have performed a fit for each one of the 100 replicas of the NNPDF31 collinear distributions. The minimization of the $\chi^2$  is done with a simplified procedure in order to speed up the computation, because for many replicas the search of $\chi^2$-minimum took much longer time in comparison to the central value minimization. It appears that the data is very demanding on the collinear PDF input. So, for some (distant from the central) replicas the fit does not converge (yielding $\chi^2/N_{pt}>5$) or produces extreme values of NP parameters (e.g. $B_{\text{NP}}<0.7$GeV). The values of NP parameters that run into the boundary of the allowed phase space region were discarded (almost $30\%$ of total replicas). The resulting distribution of NP parameters gives an estimate of the sensitivity for PDF distribution. The NP parameters and their uncertainties that we have obtained are the following
\begin{eqnarray}
&&B_{\text{NP}}=1.7\pm 0.30,\qquad c_0=0.297\pm 0.006,
\\ \nn
&&\lambda_1=0.266\pm0.066,\qquad \lambda_2=10.6\pm 3.1,\qquad \lambda_3=158.\pm 133.,
\\
&&\lambda_4=2.55\pm 0.91,\qquad \lambda_5=-7.12\pm 3.92.
\end{eqnarray}
These values are compatible with the typical values for NP parameters presented in table \ref{tab:DY-PDF-chi}, see also fig.~\ref{fig:DNP-param} and fig.~\ref{fig:f1-param}.

\begin{figure}
\begin{center}
\includegraphics[width=0.8\textwidth]{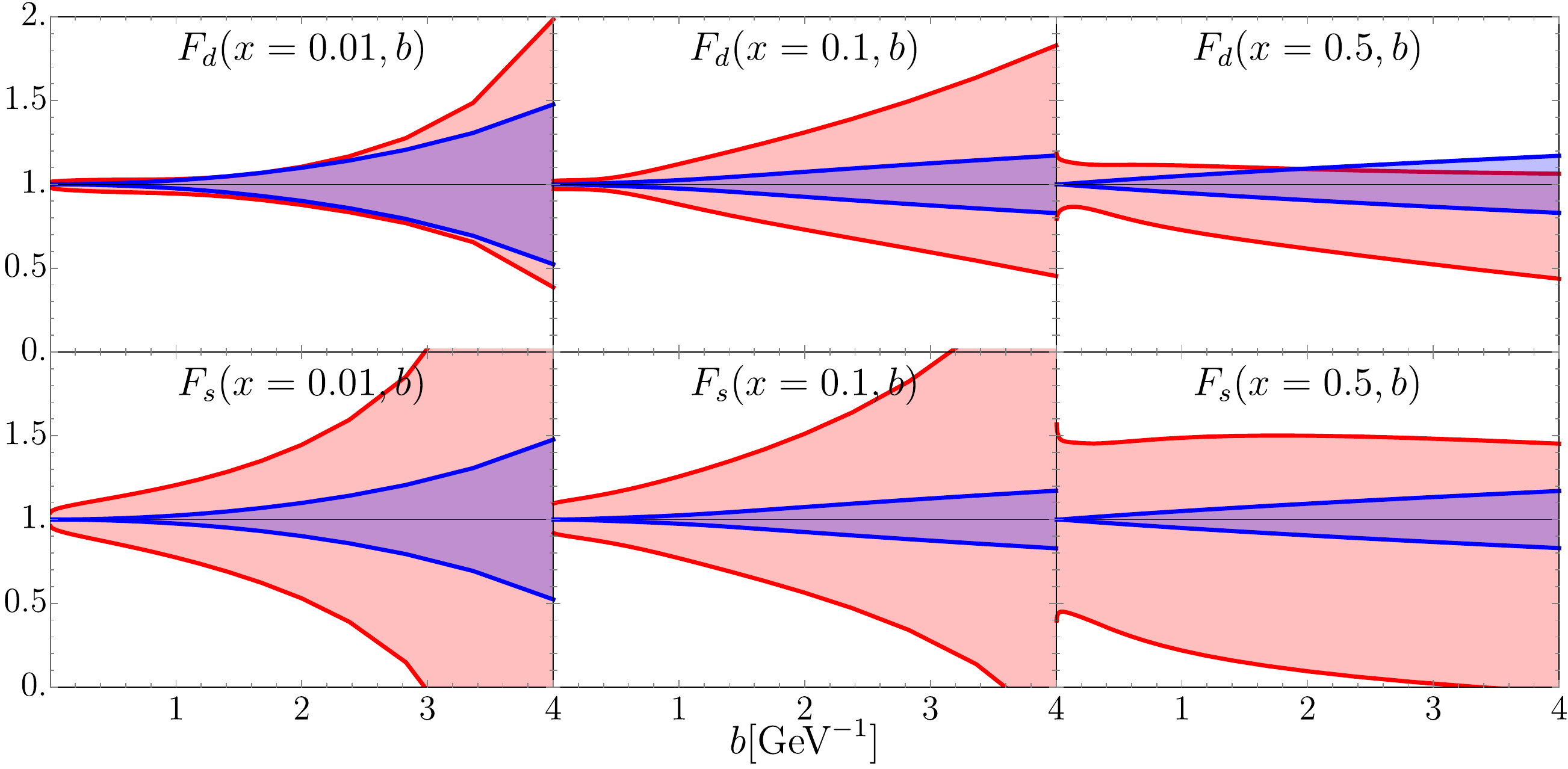}
\caption{\label{fig:PDFunc-TMD} The size of the uncertainty bands for unpolarized TMDPDFs due to collinear PDF uncertainty (red band) and due to experimental uncertainties (blue band), for d and s-quarks at different values of $x$. Both bands are weighted to the TMPDF obtained with NNPDF31 collinear distribution.}
\end{center}
\end{figure}

\begin{figure}
\begin{center}
\includegraphics[width=0.4\textwidth]{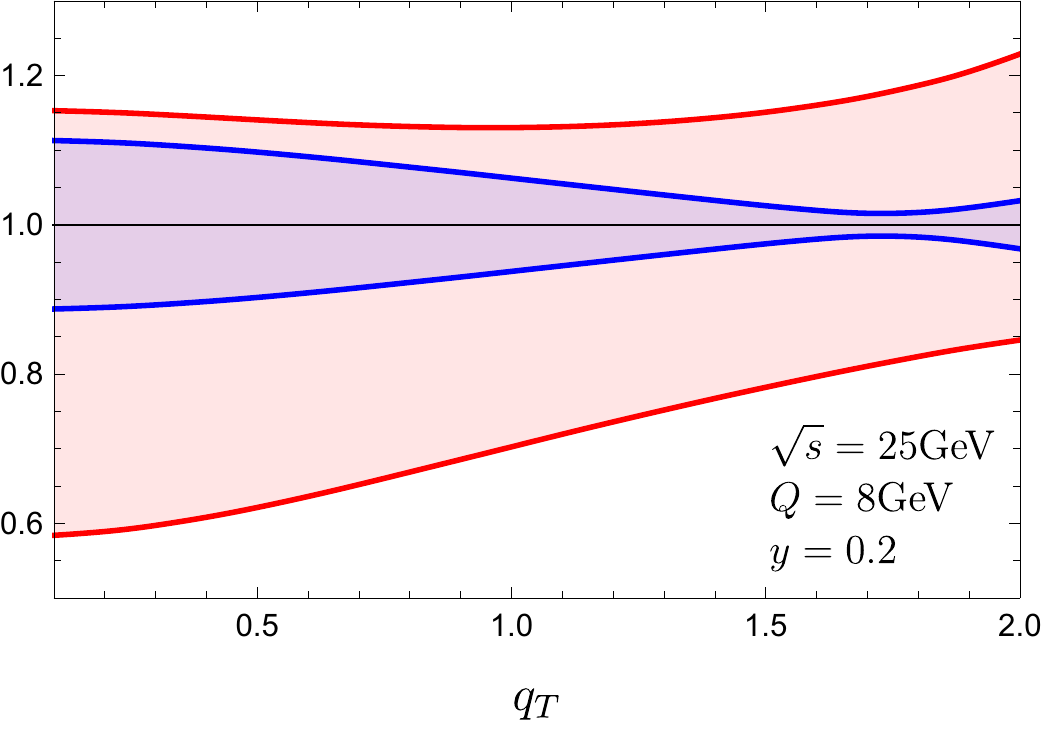}
\includegraphics[width=0.4\textwidth]{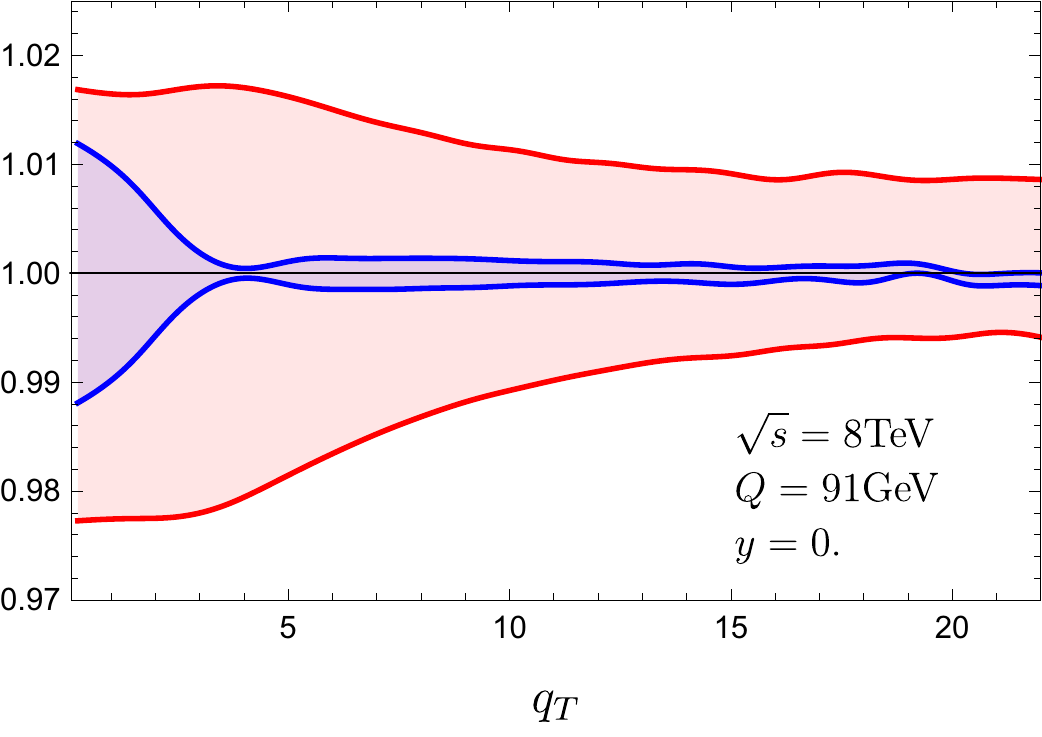}
\caption{\label{fig:PDFunc-sigma} The size of the uncertainty bands for predicted cross-section due to collinear PDF uncertainty (red band) and due to experimental uncertainties (blue band), at low and high-energies. Both bands are weighted to the prediction obtained with NNPDF31 collinear distribution.}
\end{center}
\end{figure}

In fig.~\ref{fig:PDFunc-TMD} we show the comparison of error-bands on the TMDPDF, obtained from the error-propagation from the experiment to NP parameters (blue band), and from the PDF uncertainty (red band), as described above. The main difference in these bands is that the PDF-uncertainty band is sizable already at $b=0$, and for larger $b$ these bands expand similarly. The PDF-uncertainty band is different for different flavors, and larger for non-valence partons. The resulting estimation for the (predicted) cross-section is shown in fig.~\ref{fig:PDFunc-sigma}. For the high energy case, the uncertainty is of order of $1\%$, while at low energies it reaches 20-40\%. 

The bands that we show here certainly do not accurately represent the uncertainties of TMDPDF, since many of PDF replicas do not fit the data. It implies that the TMD distributions can be used as a tool for the restriction of collinear PDFs together with the standard collinear observables. At the current stage, we can only conclude that the uncertainties of TMDPDF at small-$b$ (that are out of control in the current model) are sizable. For an accurate estimation of these errors one has to apply  more sophisticated techniques, such as reweighing of PDF values \cite{Ball:2010gb} by TMD extraction, or even joint fits of TMD distributions and collinear distributions, which are beyond the scope of the present work.

\section{Fit of SIDIS}
\label{sec:SIDIS-fit}

In this section, we use the unpolarized TMDPDF and TMD evolution, extracted in the fit of DY data, to fit the SIDIS data. The main aim is to test the universality of the TMD evolution, and TMDPDF. Namely, the SIDIS data should be easily fitted  adjusting only the parameters of TMDFF. Indeed, we have found that  the TMDFF in eq.~(\ref{def:phen-D1}) (with a 4-parameter ansatz) together with the TMDPDF and $\mathcal{D}$ (extracted  from DY data) provide a very good description of the available SIDIS data. \textit{This is one of the main results of the present work that demonstrates the complete universality of TMD factorization functions.} Another test of the TMD universality has been provided in \cite{Vladimirov:2019bfa}, that is in the fit of pion-induced DY, and it has been used in studies of the TMD distributions with jets \cite{Gutierrez-Reyes:2019msa,Gutierrez-Reyes:2019vbx,Gutierrez-Reyes:2018qez}. To our best knowledge, the test presented here  is made for the first time, because in the previous studies DY and SIDIS cases were considered or independently or  simultaneously \cite{Bacchetta:2017gcc}. Also we discuss the dependence on the collinear unpolarized FF, and  the impact of power corrections.

\subsection{Dependence on FF}
\label{sec:SIDIS-FF}

\begin{table}[b]
\begin{center}
\begin{tabular}{|l V{4} c V{4} ll|}
\hline
PDF \& FF sets & $\chi^2/N_{pt}$ & Parameters for $d_1$ &
\\\hline\hline
HERA20 \& DSS & 0.76 
& \specialcellleft{$\eta_1=0.290\pm0.014$ \\ $\eta_2=0.469\pm0.016$}
& \specialcellleft{$\eta_3=0.459\pm0.027$ \\ $\eta_4=0.496\pm0.027$}
\\\hline
HERA20 \& JAM19 & 0.93 
& \specialcellleft{$\eta_1=0.164\pm 0.012$ \\ $\eta_2=0.286\pm 0.016$}
& \specialcellleft{$\eta_3=0.223\pm 0.027$ \\ $\eta_4=0.341\pm 0.018$}
\\\hline
NNPDF31 \& DSS & 1.00 
& \specialcellleft{$\eta_1=0.257\pm0.009$ \\ $\eta_2=0.480\pm0.010$}
& \specialcellleft{$\eta_3=0.455\pm0.017$ \\ $\eta_4=0.540\pm0.020$}
\\\hline
NNPDF31 \& JAM19 & 1.65 
& \specialcellleft{$\eta_1=0.141\pm 0.012$ \\ $\eta_2=0.293\pm 0.017$}
& \specialcellleft{$\eta_3=0.224\pm 0.028$ \\ $\eta_4=0.373\pm 0.018$}
\\\hline
HERA20 \& DSS (N$^3$LO) & 0.88 
& \specialcellleft{$\eta_1=0.282\pm0.010$ \\ $\eta_2=0.466\pm0.012$}
& \specialcellleft{$\eta_3=0.468\pm0.021$ \\ $\eta_4=0.504\pm0.025$}
\\\hline
NNPDF31 \& DSS (N$^3$LO) & 1.31 
& \specialcellleft{$\eta_1=0.245\pm0.011$ \\ $\eta_2=0.475\pm0.011$}
& \specialcellleft{$\eta_3=0.463\pm0.020$ \\ $\eta_4=0.556\pm0.019$}
\\\hline
\end{tabular}
\end{center}
\caption{\label{tab:SIDIS-PDF-FF-chi} Values of $\chi^2$ and NP parameters obtained in the fit of SIDIS data with different FF inputs. The TMD evolution parameters and TMDPDF parameters are fixed from the fit of DY data (see table \ref{tab:DY-PDF-chi}), and labeled by the PDF set. The visual presentation of this table is given in fig.~\ref{fig:d1-param}.}
\end{table}

\begin{figure}[t]
\begin{center}
\includegraphics[width=0.833\textwidth]{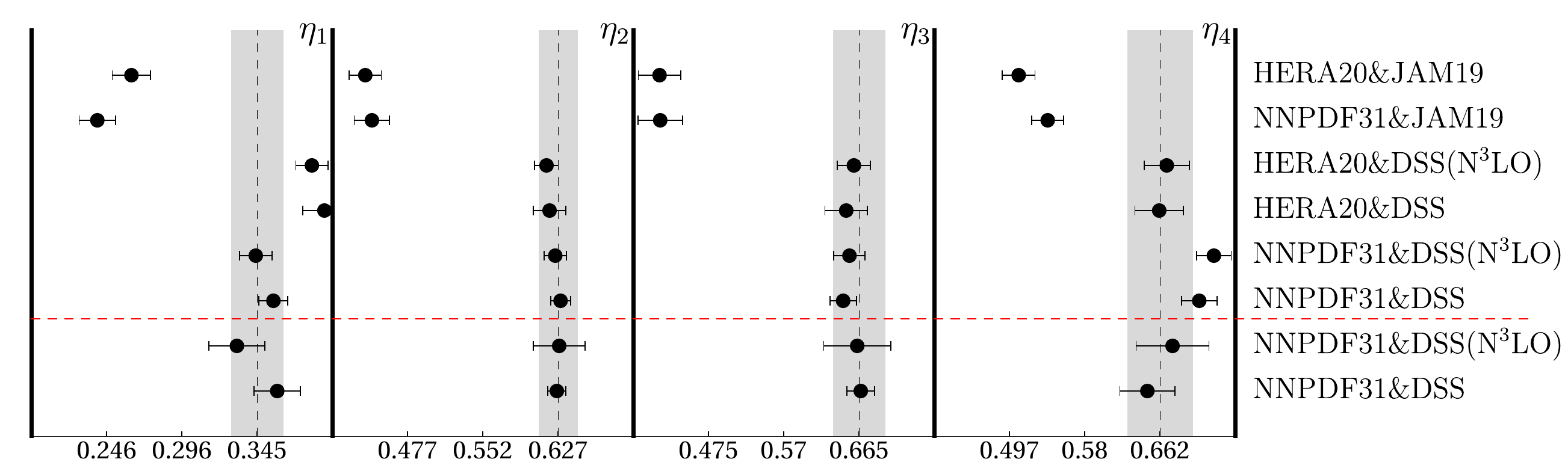}
\end{center}
\caption{\label{fig:d1-param} Comparison of NP parameter for unpolarized TMDFF extracted in different fits. The values above the red-dashed line are extracted in fit of SIDIS data with fixed TMD evolution and TMDPDF, see table \ref{tab:SIDIS-PDF-FF-chi}. The values below the red-dashed line are extracted in global fit of DY and SIDIS data, see table \ref{tab:NP-param-final}. The vertical dashed lines and gray boxes correspond to average mean and standard deviation of the results of the global fit. The input collinear distributions are marked in the right column.}
\end{figure}

In contrast to collinear PDFs, there are not too many extraction of collinear FFs. We have considered three sets of collinear FFs. Namely, DSS set\footnote{We are thankful to R.~Sassot for providing us the actual grids for DSS FFs.}(that is a composition of pion FFs from \cite{deFlorian:2014xna} (DSS14) and kaon FFs from \cite{deFlorian:2017lwf} (DSS17)), the JAM19 set \cite{Sato:2019yez} and the NNFF10 set \cite{Bertone:2017tyb}. All these extractions are made with NLO collinear evolution ($a_s^2$).

The comparison of fits with different FFs, and some of TMDPDF (together with TMD evolution) extracted in the previous section are presented in  table \ref{tab:SIDIS-PDF-FF-chi}. 
The NNFF set is not presented in the table due to the low quality of the predictions, as it is described below. 
As it is seen from  table \ref{tab:SIDIS-PDF-FF-chi}, the TMD factorization perfectly describes the low-$q_T$ SIDIS data with TMDPDF and TMD evolution fixed by DY measurements. It is one of the main result of the present analysis. 

The values of $\chi^2/N_{pt}$ are rather small (e.g. 0.76 for combination of HERA20 \& DSS), which may indicate an over-fit problem. However, this is not the case  for the following reason. The main source of low-$\chi^2$ is the COMPASS data set. The COMPASS data have very large uncorrelated systematic uncertainty for a great amount of points. Here, the systematic uncertainty is (much) larger than the statistical uncertainty, and therefore, the COMPASS data points form smooth lines with huge uncorrelated uncertainty band. As a result, the contribution of each point to the $\chi^2$-value is small.

The values of $\chi^2$ depend on the input TMDPDF and TMD evolution (compare NNLO and N$^3$LO cases) in a reasonable amount. This is mainly due to the different values of $c_0$ constant in these cases. We recall that the SIDIS measurements are made at much lower energy in comparison to DY, and thus they are more sensitive to  $\mathcal{D}$ at large-$b$. Later  in sec.~\ref{sec:GLOBAL-fit} we show that in the joint fit of SIDIS and DY data, the uncertainty of the evolution factor at  large-$b$  is reduced. 
 
 The difference between DSS and JAM collinear FFs sets is of minor importance. It is due to the fact low-energy data are less sensitive to the small-$b$ part of the TMD distributions (and thus to collinear distributions). Given in addition that the data are not very precise, the uncertainty in FF sets are compensated by the  NP function $D_{NP}$. The effect of compensation is clear from the very different values of $\eta_i$ constants for DSS and JAM19 set. Note, that in all cases we obtain a positive and sizable $b^2$-term in $D_{NP}$ (parameter $\eta_4$). It could indicate a hidden issue in the values of collinear FFs. However, we conclude that contrary to DY case, the SIDIS TMD data are not very restrictive on the values of collinear FFs.

The NNFF distributions are not able to fit the data with a $\chi^2/N_{pt}$ better than $\sim 6.8$. The reason of such an enormous discrepancy is obvious. The NNFF1.0 extraction is made from the $ee$-annihilation data only \cite{Bertone:2017tyb}, and thus is sensitive only to particular combinations of quark-flavors. The flavour separation is thus made a posteriori 
assuming  exact iso-spin symmetry. 
As a result, the FF for sea quarks have very small (and even negative) values. In the processes where the production of a hadron is dominated by the sea-quark channel, the cross-section obtained with NNFF10 collinear FF is much smaller then the experimental one. A crystal clear  example is the process $p\to K^-$, where both valence quarks of $K^-$, $\bar u s$, are sea-quarks for the proton, and thus the dominant channel is the production of $K^-$ from $u$ and $d$ quarks. However, FF for $u$ and $d$-quarks in $K^-$ are negative in NNFF extraction, and the resulting cross-section appears to be negative as well. The situation improves, if we select only the processes with dominant valence channel, e.g. $d\to\pi^\pm$, in this case we obtain $\chi^2/N_{pt}\sim 2.2$. The COMPASS measurement can be also considered separately with the NNFF1.1 set of FF for charged hadrons \cite{Bertone:2018ecm}, in this case we obtain $\chi^2/N_{pt}\sim 1.6$. In any case, we have found that NNFF sets of FF are not suitable for the description of SIDIS data.

The uncertainties on NP parameters presented in table \ref{tab:SIDIS-PDF-FF-chi} are \textit{unrealistically small}. Given the fact that the data is not very accurate, it indicates a significant underestimation of the uncertainty for TMDFF. We guess that the underestimation of uncertainties is caused mainly by the function bias of $D_{NP}$. To resolve the situation one could use  a more flexible ansatz, e.g. by inclusion of more NP-parameters. Unfortunately, this strategy is not very efficient. Already with the current set of parameters we have very low $\chi^2$, and the increase of the number of parameters could lead to an over-fit problem. Also, the computation time with a bigger number of parameters increases.

\subsection{Impact of power corrections}
\label{sec:SIDIS-power-corr}

Considering the expression for the SIDIS cross-section eq.~(\ref{SIDIS:xSec}) we distinguish four types of power corrections: \textit{(m/Q)} the corrections due to non-zero produced mass, \textit{(M/Q)} the corrections due to non-zero target mass, \textit{($q_T$/Q)} the $q_T/Q$-terms in the expression for cross-section and \textit{($x_Sz_S$)} the $q_T/Q$-terms in the expressions for $x_S$ and $z_S$, eq.~(\ref{def:x1z1}). In order to test the impact of these corrections, we have performed the (central value) fits including corrections in different combinations. The resulting values of $\chi^2/N_{pt}$ are reported in table \ref{tab:SIDIS-power}.

\begin{table}
\begin{center}
\begin{tabular}{|c||c|c|c|c|c|c|}\hline
include \textit{(m/Q)} 						& yes	& \textbf{no}	& yes			& yes			& \textbf{no}	& \textbf{no}	\\
include \textit{(M/Q)} 						& yes	& yes			& \textbf{no}	& yes			& \textbf{no}	& \textbf{no}	\\
include \textit{($q_T$/Q)} in kinematics 	& yes	& yes			& yes			& \textbf{no}	& \textbf{no}	& \textbf{no}	\\
include \textit{($q_T$/Q)} in $x_S$, $z_S$	& yes	& yes			& yes			& yes			& yes			& \textbf{no}	\\\hline
$\chi^2/N_{pt}$								& 1.00	& 1.00 			& 1.09  		& 1.06	 		& 1.16			& 1.31
\\\hline
\end{tabular}
\caption{\label{tab:SIDIS-power} Comparison of results of the fit with different combination of power suppressed terms. The fit is made only for the central values, with fixed TMD evolution and TMDPDF as in NNPDF3.1, with DSS collinear FF.}
\end{center}
\end{table}

Let us summarize the observations:
\begin{itemize}
\item \textit{Produced mass corrections}. The produced mass-corrections are not necessary extremely small, as it is typically assumed. These corrections appear in the ratio with other kinematic variables through the variable $\varsigma^2$, eq.~(\ref{def:mass-var}). In most part of data bins the value of $\varsigma^2$ is negligible, $\varsigma^2\sim 10^{-3}$, but for some  low-energy and low-z bins it can reach $\varsigma^2\sim 10^{-2}$. For example, the HERMES bin with $0.2<z<0.4$, $0.2<x<0.35$ with produced kaon has $\varsigma^2\sim 0.04$. As it is clear from  table \ref{tab:SIDIS-power}, current data are not sensitive to these corrections. The difference in $\chi^2/N_{pt}$ is of the order $10^{-3}$.
\item \textit{Target mass corrections}. The target mass corrections appear through the variable $\gamma^2$ in eq.~(\ref{def:mass-var}) and at low Q it has a rather significant size, e.g. for some bins in HERMES data $\gamma^2\sim 0.13$, for some bins in COMPASS data $\gamma^2\sim 0.06$. 
Therefore, one can expect up $10\%$ impact of $\gamma^2$ for certain bins. Note, that the dependence on $\gamma^2$ is non-linear and is different in different edges of the bin. Checking the values in table \ref{tab:SIDIS-power}, we observe that the target mass correction produces a small but visible effect on the fit quality especially for HERMES data  where the change in $\chi^2/N_{pt}$ is $1.09\to 1.24$.
\item \textit{$q_T/Q$ correction in kinematics}. This correction cannot be large due to the cuts on the data sets that we have performed.
For $q_T\sim 0.25 Q$ which is the highest  value of $q_T$  that we have considered, we can have $(q_T/Q)^2\sim 0.06$. In addition, the first correction of this type to the cross section is linear in $(q_T/Q)^2$ and it can be easily compensated by a change of the non-perturbative parameters in $D_{\text{NP}}$. Indeed, we observe that the impact on the $\chi^2$ is small.
\item \textit{$q_T/Q$ correction in $x_S$ and $z_S$}. For $q_T\sim 0.25 Q$ (which is the maximum considered $q_T$ ), the difference between exact $x_S$ and $x$ is  $\sim 0.06$, and much smaller between $z_S$ and $z$. Nonetheless, this correction changes the shape of the cross-section in a way that is difficult to compensate by NP parameters. Thus, the inclusion of this correction visibly improves the agreement. Let us note that the same conclusion has been made for the DY case, in sec.~\ref{sec:DY:x12}.
\end{itemize}
We conclude that the impact of each individual correction is rather small, but the inclusion of any of them improves the agreement between theory and  data. Most relevant effects are the  target mass correction and the ones due to  $x_S$ and $z_S$. Accounting of all effects simultaneously leads to a qualitative improvement in $\chi^2$-values. 

We also admit that the inclusion of power corrections considerably affect the values of parameters $\eta$ (especially $\eta_{1,4}$). The values of parameters $\eta_{1,4}$ varies in the range $(-10,+25)\%$. The values of parameters $\eta_{2,3}$ varies in the range $(-4,+8)\%$. It shows that our estimation of uncertainties on parameters presented in table \ref{tab:SIDIS-PDF-FF-chi} are extremely underestimated. Possibly, the main source of underestimation is the bias of our model, which is not surprising since we have only 4 parameters for all partons flavors and particle kinds. \textit{The tests of power corrections suggest that the real error-band on the extracted TMDFF is an order of magnitude larger.}

\subsection{Limits of TMD factorization for SIDIS}
\label{sec:TMD-limit}

\begin{figure}[t]
\begin{center}
\includegraphics[width=0.5\textwidth]{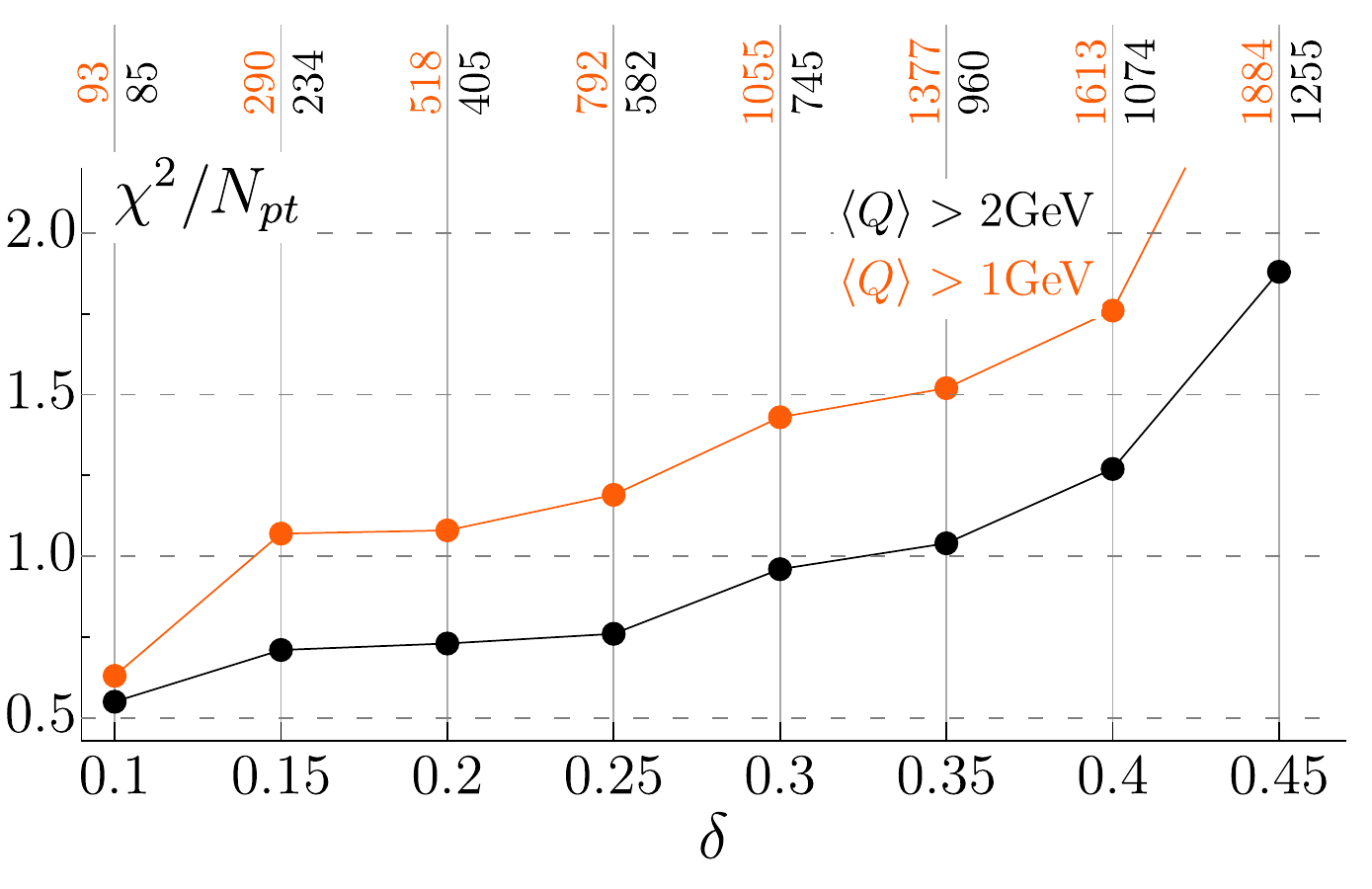}
\caption{\label{fig:chi2}The value of $\chi^2/N_{pt}$ depending on the cuts of the data for SIDIS. The theory prediction is calculated with NNPDF31 \& DSS set at NNLO. The numbers on vertical lines shows the number of points in the cut data set.}
\end{center}
\end{figure}

In ref.~\cite{Scimemi:2017etj} we tested the limits for TMD factorization using the DY data, showing that the natural limit of the leading power TMD factorization is $\delta\simeq 0.2-0.25$, where $\delta =q_T^{max}/Q$ and $q_T^{max}$ is the maximum value of the transverse momentum in the data sets included in the fit.
 We have tested the same boundary using the SIDIS data and the result of the global fit (presented in the next section) evaluating the  $\chi^2$ (without minimization) for different selections of SIDIS data. We have considered two possible cuts on data selection $\langle Q\rangle>1$ and $\langle Q\rangle>2$ , eq.~(\ref{SIDIS-data:Q>2}),  and the result is shown in fig.~\ref{fig:chi2}. 
 
The values of $\chi^2/N_{pt}$ grow when $\delta>0.25$. The same effect has been observed in ref.~\cite{Scimemi:2017etj} for DY. Therefore, we conclude that our earlier estimation of the  validity interval  of TMD factorization as $\delta\lesssim 0.2-0.25$ holds also in the SIDIS case. It is interesting to observe that the channel with the fastest growth of $\chi^2/N_{pt}$ is $d\to K^-$ (and the next is $p\to\pi^+$), which could indicate a possible tension in the description of this reaction. 
 
The inclusion of data at $\langle Q\rangle<2$GeV almost doubles the values of $\chi^2/N_{pt}$ (e.g. $\chi^2/N_{pt}=1.19$ for $\delta=0.25$).  Taking into account the large uncertainties of the COMPASS measurement, it shows that the factorization is broken down at such  low values of $Q$. This is an expected result, since in this region the power corrections dominate the cross-section. In sec.~\ref{sec:agreement}, we show data and our predictions including the low-Q bins and up to $\delta=0.4$.

\section{Global fit of DY and SIDIS data}
\label{sec:GLOBAL-fit}

The fit of  SIDIS data shows perfectly the universality of the TMD distributions and a good agreement between theory and experiment. Performing a global fit of DY and SIDIS data we essentially reduce the uncertainty for $\mathcal{D}$. The resulting sets of TMDPDF, TMDFF and RAD extracted from the global fit represent the SV19 TMD distributions (at NNLO and N$^3$LO). As an input for this set we have used NNPDF31 and DSS collinear distributions, because these sets are in  good agreement with the global set of collinear observables, and show the best values of $\chi^2$ (see discussions in the previous sections). 

\subsection{Agreement between theory and data}
\label{sec:agreement}

Table \ref{tab:final} shows the distribution of the $\chi^2$-values per individual experiments. In total we have considered 1039 points, 457 for DY and 582 for SIDIS. They form three large subsets: DY at high energy, DY at low energy, and SIDIS (at low energy). The worst $\chi^2$ values are concentrated in the high energy DY subset, because of the very high precision of Z-boson production data measured at LHC. Simultaneously, these data robustly restrict the values of TMD distributions (TMDPDF and RAD) at $b\lesssim 1$GeV$^{-1}$. The lowest $\chi^2$ is for SIDIS data  and especially for COMPASS data ($\chi^2/N_{pt}=0.65$, with $N_{pt}=390$ that is more than the third part of the total data), due to large uncorrelated systematic uncertainty (see discussion in sec.~\ref{sec:SIDIS-FF}).
\begin{table}
\small
\begin{center}
\renewcommand{\arraystretch}{0.94}
\begin{tabular}{|l | c V{4} c | c V{4} c |c V{4}}
\cline{3-6}
\multicolumn{2}{c V{4}}{ } & \multicolumn{2}{ c V{4}}{ NNLO}  & \multicolumn{2}{ c V{4}}{ N$^3$LO} \\
\hline 
Data set & $N_{pt}$ & $\chi^2/N_{pt}$ & $\langle d/\sigma \rangle$ & $\chi^2/N_{pt}$ & $\langle d/\sigma \rangle$ 
\\\hline
CDF run1          &   33 &      0.64 &   7.1\% 		& 0.65	& 6.6\%	\\
CDF run2          &   39 &      1.32 &   1.4\% 		& 1.46	& 0.6\%	\\
D0 run1           &   16 &      0.75 &   -1.3\%		& 0.81	& -1.9\% \\
D0 run2           &    8 &      1.38 &   -     		& 1.65	& -		\\
D0 run2 ($\mu$)   &    3 &      0.62 &   -	   		& 0.69	& -		\\
\hline
Tevatron total    &  99		& 0.98 &				& 1.08	& \\
\hline
ATLAS 7TeV 0.0<|y|<1.0      &    5 &      1.67 &  -	&		 0.77 &	-\\
ATLAS 7TeV 1.0<|y|<2.0      &    5 &      6.00 &  -	&		 4.10 & -\\
ATLAS 7TeV 2.0<|y|<2.4      &    5 &      1.51 &  -	&		 1.31 & -\\
ATLAS 8TeV 0.0<|y|<0.4      &    5 &      2.37 &   2.0\% &	 3.40 & 1.2\%	\\
ATLAS 8TeV 0.4<|y|<0.8      &    5 &      2.90 &   2.0\% &	 3.25 & 1.2\%   \\
ATLAS 8TeV 0.8<|y|<1.2      &    5 &      1.12 &   2.2\% & 	 1.44 &	1.3\%	\\
ATLAS 8TeV 1.2<|y|<1.6      &    5 &      1.91 &   2.8\% &	 1.39 & 1.9\%	\\
ATLAS 8TeV 1.6<|y|<2.0      &    5 &      1.23 &   3.5\% &	 0.48 &	2.6\%	\\
ATLAS 8TeV 2.0<|y|<2.4      &    5 &      2.48 &   4.2\% &	 1.91 & 3.3\%	\\
ATLAS 8TeV 46<Q<66GeV      &    3 &      0.38 &   -0.2\% &	 0.49 &	-1.1\%	\\
ATLAS 8TeV 116<Q<150GeV    &    7 &      0.76 &   0.2\%	 &	 0.95 & -0.4\% 	\\
\hline
ATLAS total & 55 & 2.04 & &									1.79 &\\
\hline
CMS 7TeV          &    8 &      1.25 &   -	&	1.25 &   -	\\
CMS 8TeV          &    8 &      0.77 &   -	&	0.76 &   -	\\
\hline
CMS total & 16 & 1.01 & &							1.00 & \\
\hline
LHCb 7TeV         &    8 &      2.32 &   4.6\%	&	2.04 &   4.0\%	\\
LHCb 8TeV         &    7 &      4.12 &   4.5\%	&	3.52 &   3.8\%	\\
LHCb 13TeV        &    9 &      0.81 &   5.1\%	& 	0.72 &   4.4\%	\\
\hline
LHCb total & 24 & 2.28 & &							1.98	&	\\
\hline\hline
\textbf{High energy DY total} & 194 & 1.44 & &	1.32 &\\
\hline\hline
PHE200        &    3 &      0.28 &   -0.1\%	&	0.30	& -0.7\%\\
E228-200      &   43 &      1.01 &  35.3\%	&	1.12	& 34.6\%\\
E228-300      &   53 &      0.91 &  28.8\%	&	1.01	& 27.8\%\\
E228-400      &   76 &      0.87 &  20.1\%	&	0.95	& 18.9\%\\
E772          &   35 &      1.86 &   8.9\%	&	1.93	& 7.9\%\\
E605          &   53 &      0.57 &  20.7\%	&	0.60	& 19.5\%\\
\hline\hline
\textbf{Low energy DY total} & 263 & 0.97 & & 1.04	&\\
\hline\hline
HERMES ($p\to \pi^+$)   &   24 &      2.20 &   1.7\%	& 3.06	& 2.2\%	\\
HERMES ($p\to \pi^-$)   &   24 &      1.12 &   0.6\%	& 1.45	& 0.9\%	\\
HERMES ($p\to K^+$)   	&   24 &      0.71 &  -0.1\%	& 0.66  & 0.0\%	\\
HERMES ($p\to K^-$)   	&   24 &      0.69 &   0.0\%	& 0.66	& 0.0\%	\\
HERMES ($d\to \pi^+$)   &   24 &      0.57 &   0.3\%	& 0.78	& 0.8\% \\
HERMES ($d\to \pi^-$)   &   24 &      0.74 &   0.5\%	& 0.97	& 0.7\% \\
HERMES ($d\to K^+$)   	&   24 &      0.52 &  -0.1\%	& 0.53  & 0.0\% \\
HERMES ($d\to K^-$)    	&   24 &      1.27 &   0.0\%	& 1.17 	& 0.1\% \\
\hline
HERMES total & 192 & 0.98 & &							1.16&\\
\hline
COMPASS ($d\to h^+$)    &  195 &      0.62 &   3.3\%	& 0.77	&	5.1\%	\\
COMPASS ($d\to h^-$)    &  195 &      0.68 &  -2.3\%	& 0.92	&	-0.5\%  \\
\hline
COMPASS total & 390 & 0.65 & &							0.85 &\\
\hline\hline
\textbf{SIDIS total} & 582 & 0.76 & &					0.95 &\\
\hline\hline
\textbf{Total} & \textbf{1039} & \textbf{0.94} & & \textbf{1.05} & \\
\hline
\end{tabular}
\caption{\label{tab:final} Distribution of the values of $\chi^2$ over data set in the global fit of SIDIS and DY. The column $\langle d/\sigma \rangle$  report the average  normalization deficit of the cross-section (multiplicity) as defined in (\ref{def:d/sigma}).}
\end{center}
\end{table}

Altogether we obtain the  global value of $\chi^2/N_{pt}=0.95$ and $1.06$ for NNLO and N$^3$LO respectively. These values can be compared to $1.55$ (for $N_{pt,\rm{total}}=8059$) and $1.02$ (for $N_{pt;\rm{SIDIS}}=477$ that is close to our data selection) obtained in the global fit of DY and SIDIS in ref.~\cite{Bacchetta:2017gcc}. The increase of  $\chi^2/N_{pt}$ between NNLO and N$^3$LO  cases does not indicate a reduction of the fit quality. This change in $\chi^2$ happens mainly because of COMPASS data, for which the $\chi^2$-value increase $0.65\to 0.85$. On the contrary, the $\chi^2$-value for ATLAS data reduces $2.12\to1.82$ (mostly due to the improvement in the total normalization). Therefore, we conclude that both NNLO and N$^3$LO fits are in agreement, although N$^3$LO shows a better agreement with high-energy data.

In  table~\ref{tab:final} we also present the values of the difference in the normalization between  theory and  data  due to the correlated shift (see definition in (\ref{def:d/sigma})). The measurements  in the table \ref{tab:final} without this value (e.g. CMS) are normalized to the total cross-section. Note, that the shift value is common to the full data subset (e.g. for all 195 point of COMPASS $d\to h^+$).

Finally, we  have some more considerations on each data set:

\begin{itemize}
\item \textit{The high energy DY} data have a common deficit of 2-5\% in the normalization, which has been already observed in \cite{Bertone:2019nxa}. It can be caused by different sources, being the main ones the collinear PDF (e.g. in the case of HERA20 PDF the deficit is much smaller, 0-3\%). Another source is the presence of corrections  due to fiducial cuts that are linear in $q_T$,  as discussed in sec.~\ref{sec:DY-fiducial}. This deficit is responsible for a larger value of $\chi^2$ for this sub-set. The nuisance parameter decomposition for high energy DY is $1.51=1.28+0.23$, where the last number is the penalty contribution to $\chi^2$ due correlated uncertainties.
\item \textit{The low energy DY} data are significantly underestimated by the TMD factorization formula. However, this underestimation is within the expected correlated systematic uncertainties of the data. This is a known issue of fixed target experiments. The underestimation has been also observed for the pion-induced DY \cite{Vladimirov:2019bfa} (E615 and E537 experiments), and for the same low-energy DY experiments (E228 and E605) in ref.~\cite{Bacchetta:2019tcu}. Note, that in ref.~\cite{Bacchetta:2019tcu} the high-$q_T$ part of the measurements has been considered (in  collinear factorization), and the observed discrepancy is an order of magnitude larger. Also, the present fit has somewhat lesser deficit in the normalization (by $5-6\%$) in comparison to previous one \cite{Bertone:2019nxa}. We connect it to the corrected shape of $\zeta$-line at large-$b$.
\item \textit{SIDIS data} do not show any problem with the total normalization. This statement is in some contradiction to the literature. In \cite{Bacchetta:2017gcc} the authors report a significant contribution of normalization to $\chi^2$ from the HERMES data (the COMPASS data was normalized exactly). In ref.~\cite{Gonzalez-Hernandez:2018ipj} an enormous discrepancy between theory and  data in the collinear factorization limit has been observed too.

\end{itemize}

\begin{figure}
\begin{center}
\includegraphics[width=1.00\textwidth]{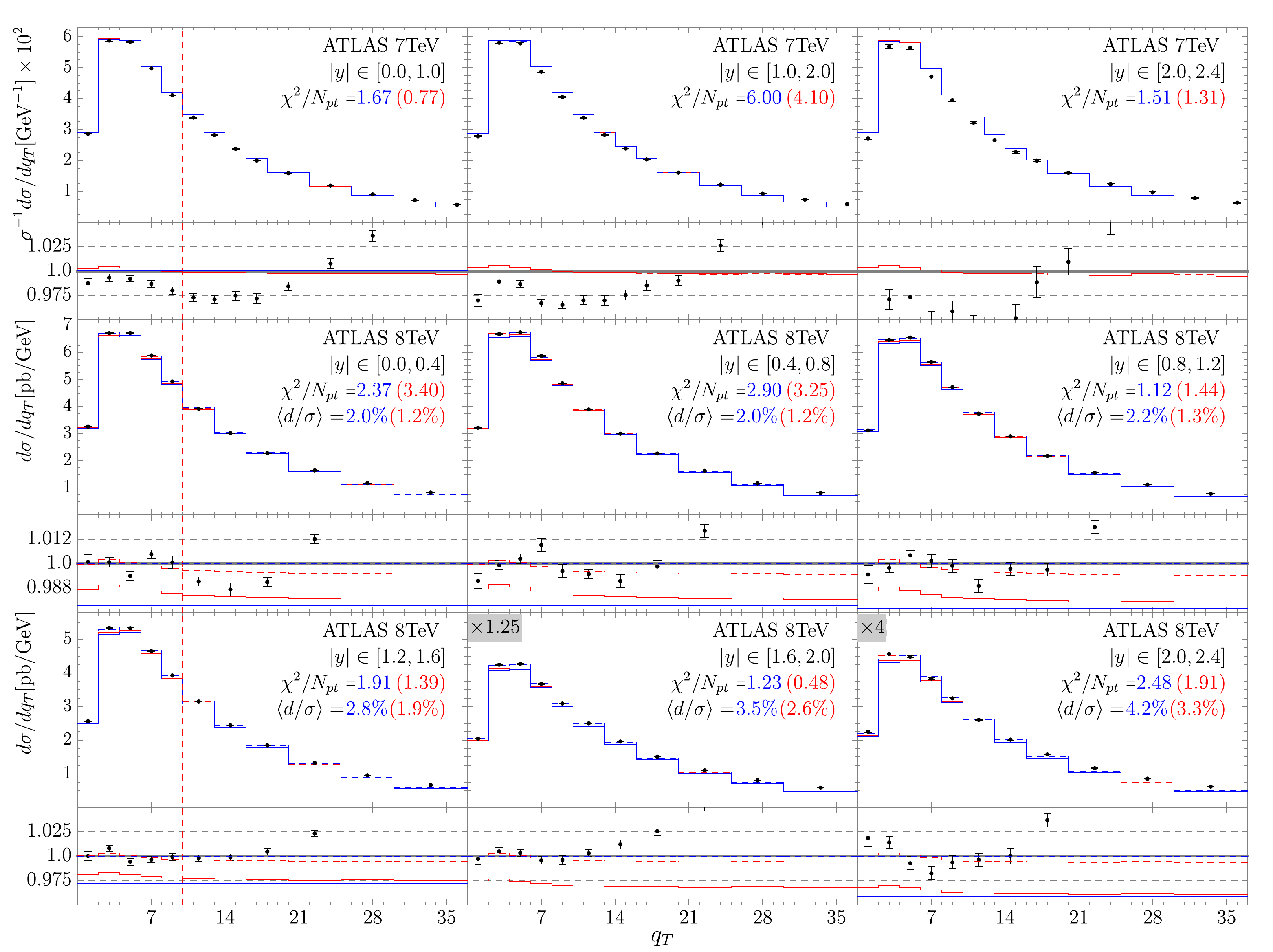}
\caption{\label{fig:ATLAS} Differential cross-section for the $Z/\gamma^*$ boson production measured by ATLAS at different values of $y$ and $s$. The solid lines are absolute prediction. The dashed lines are the theory prediction shifted by $\langle d/\sigma\rangle$ that is indicated on each case together with the values of $\chi^2/_{np}$ for given data set. Blue (red) color corresponds to the theory prediction at NNLO (N$^3$LO). The ratio boxes shows same plot weighted by the shifted theory prediction at NNLO. Vertical dashed lines show the part of the data included in the fit (to the left of the line).}
\end{center}
\end{figure}

\begin{figure}
\begin{center}
\includegraphics[width=0.68\textwidth]{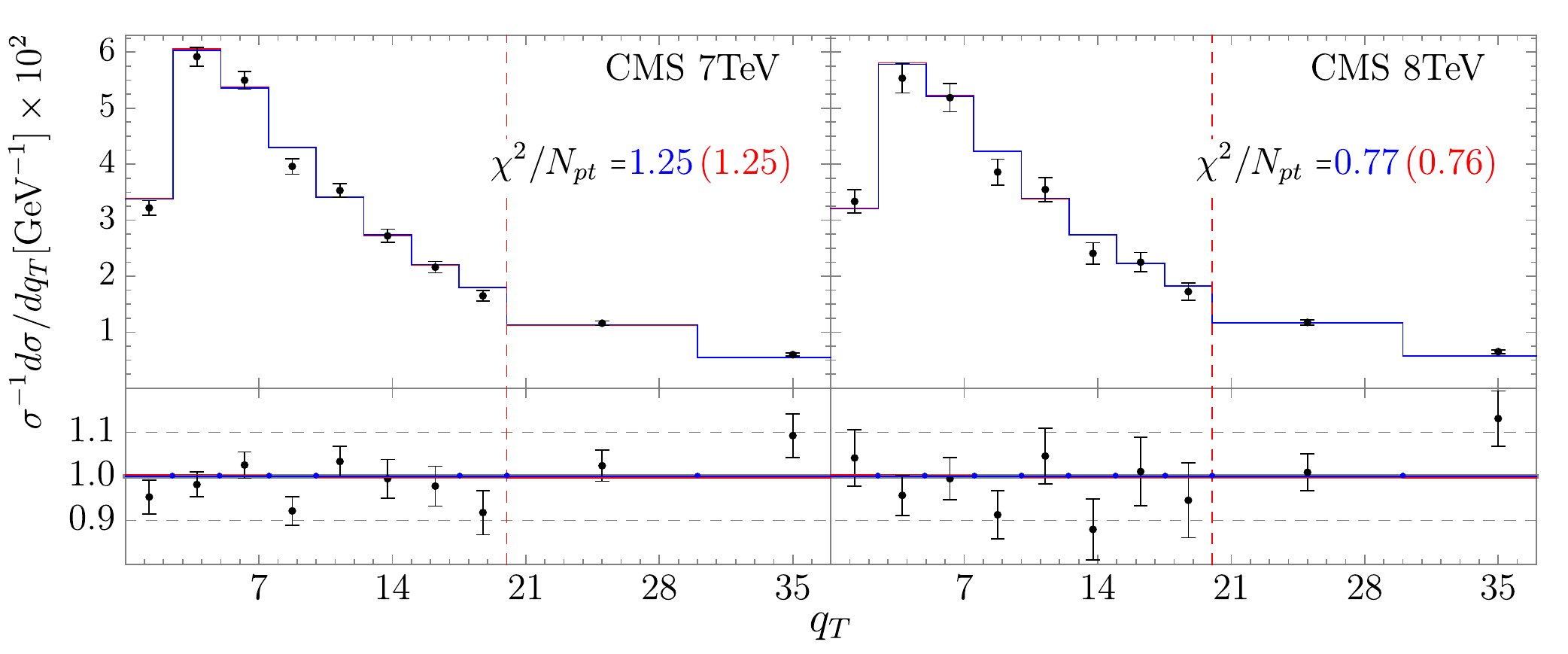}

\includegraphics[width=1.\textwidth]{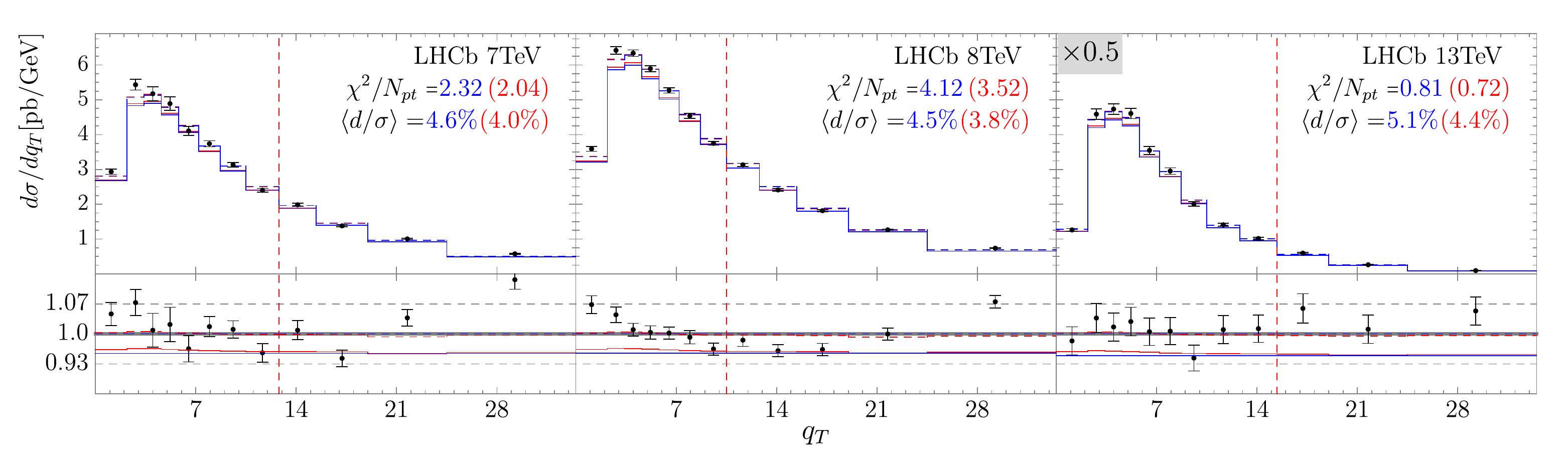}

\includegraphics[width=0.32\textwidth]{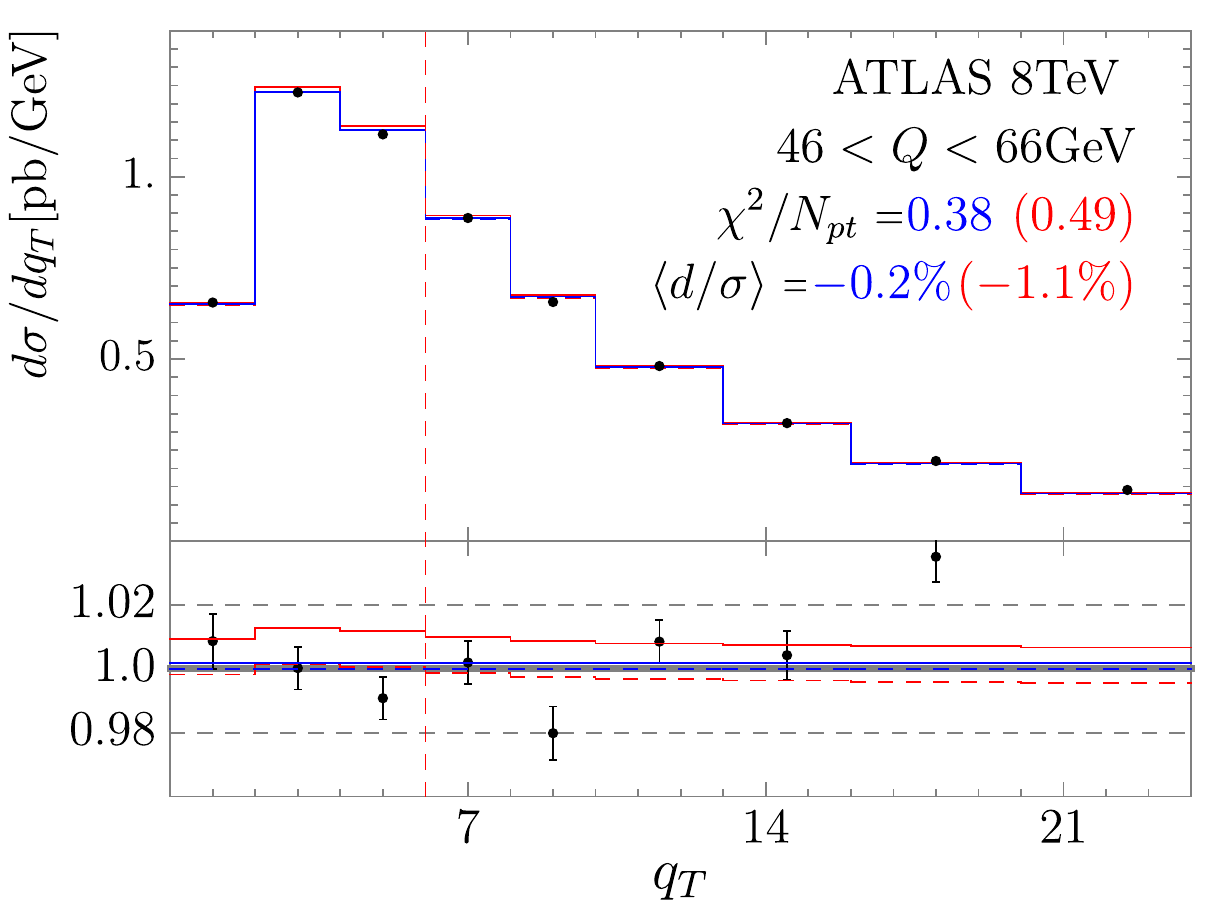}
\includegraphics[width=0.32\textwidth]{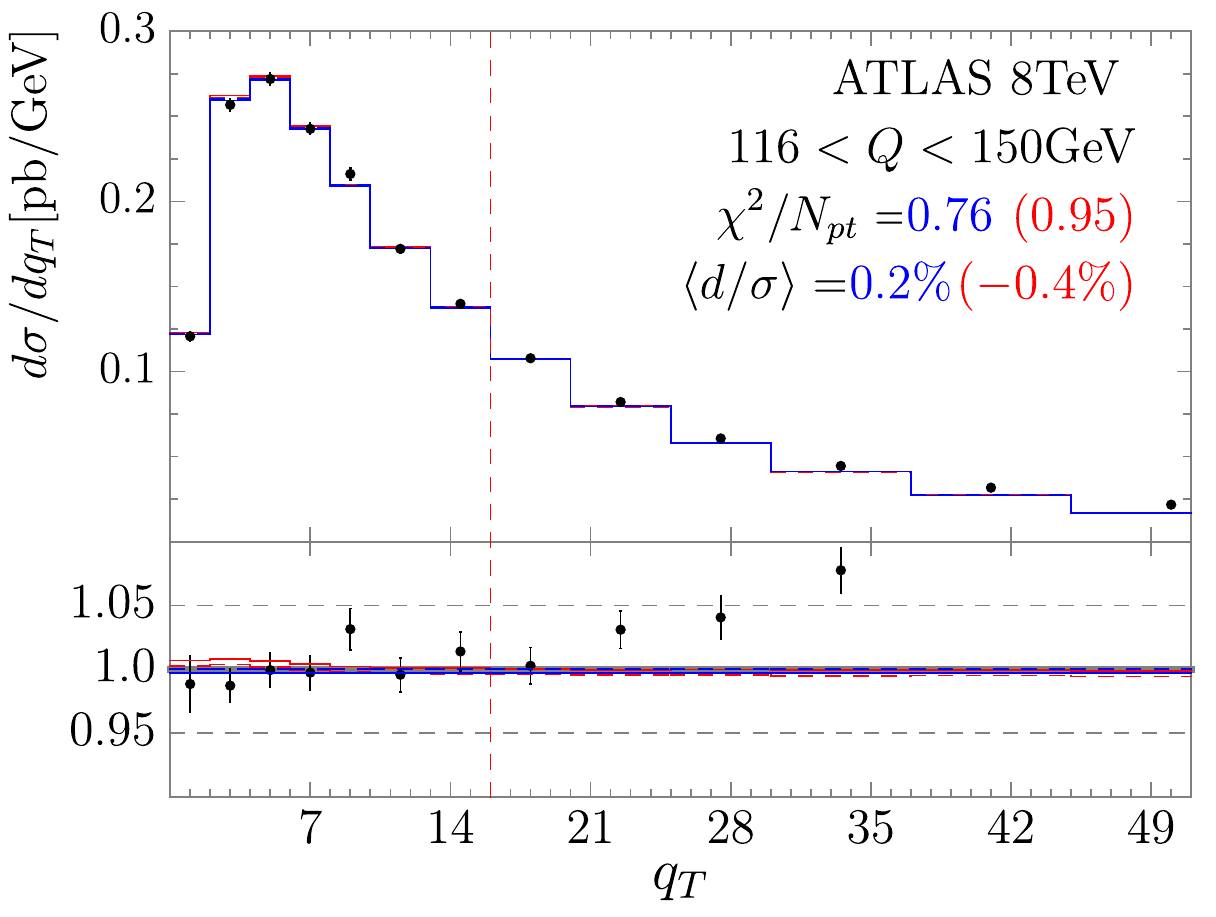}
\includegraphics[width=0.32\textwidth]{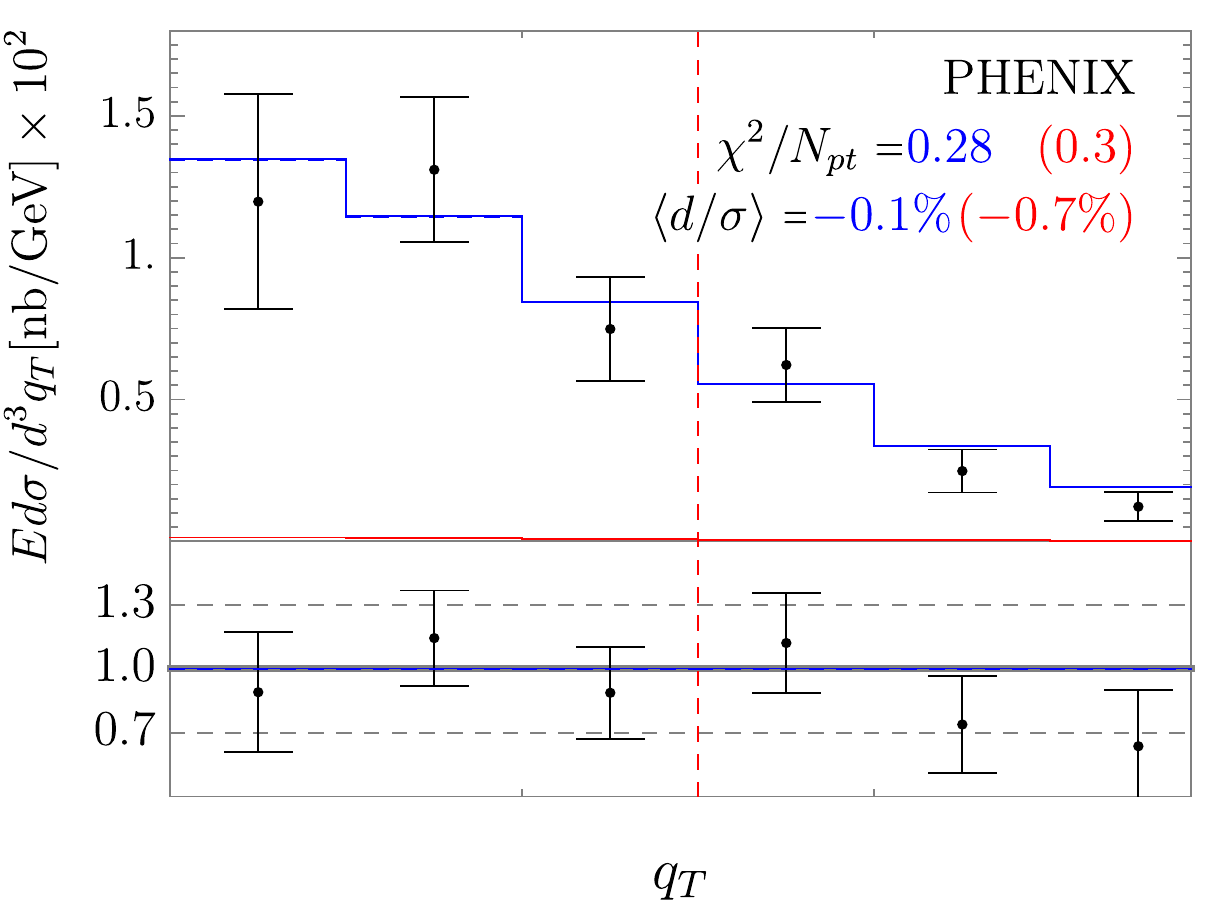}

\caption{\label{fig:CMS+LHCb} Differential cross-section for the $Z/\gamma^*$ boson production measured by ATLAS, CMS, LHCb and PHENIX at different values of $s$ and Q. The figure elements are the same as in fig.~\ref{fig:ATLAS}}
\end{center}
\end{figure}

\begin{figure}
\begin{center}
\includegraphics[width=1.00\textwidth]{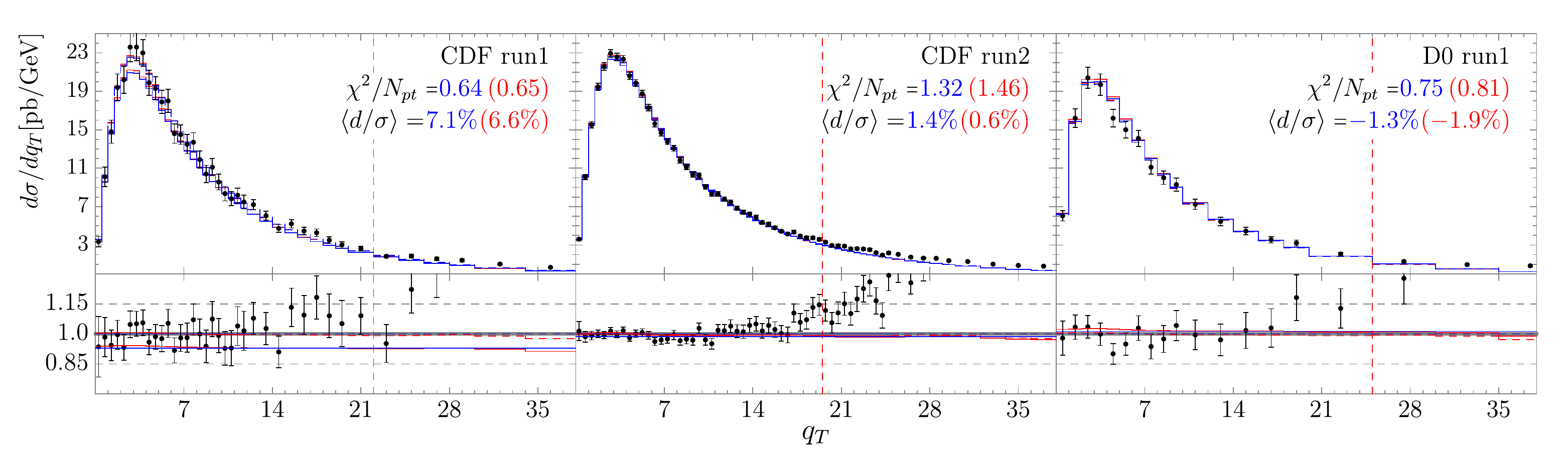}

\includegraphics[width=0.69\textwidth]{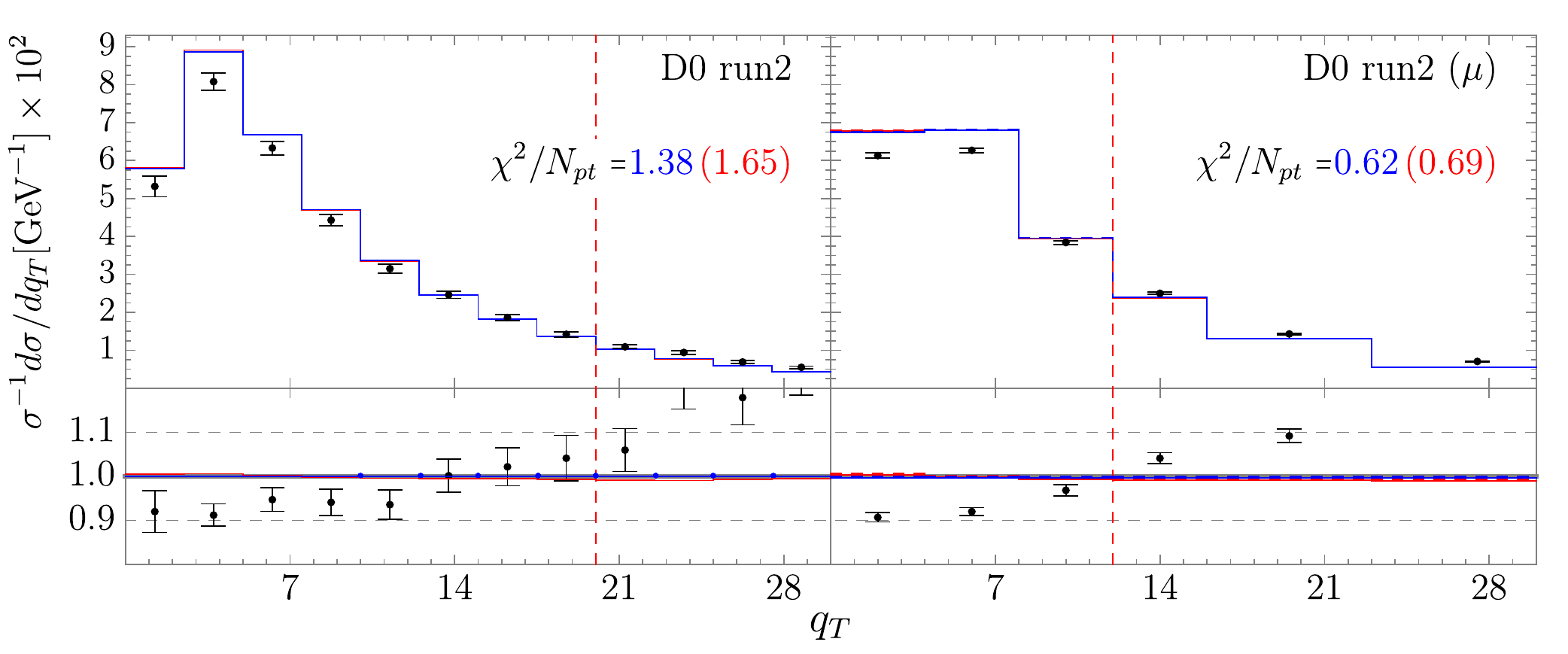}
\caption{\label{fig:tevatron} Differential cross-section for the $Z/\gamma^*$ boson production measured by CDF and D0 at different values of $s$. The figure elements are the same as in fig.~\ref{fig:ATLAS}}
\end{center}
\end{figure}

\begin{figure}
\begin{center}
\includegraphics[width=1.00\textwidth]{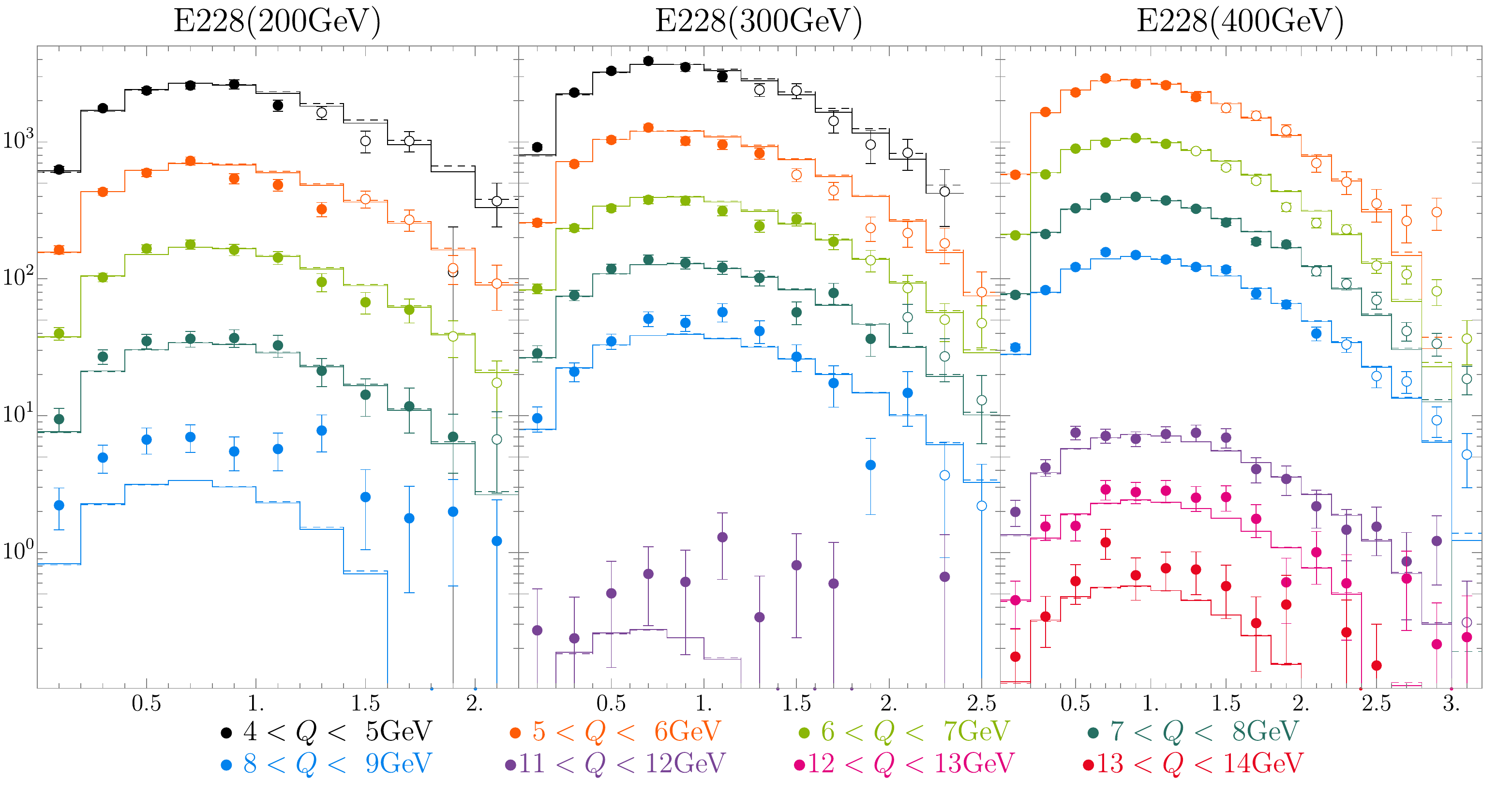}
\caption{\label{fig:E228} Differential cross-section of DY process ($d\sigma/dq_T$[fb/GeV] vs. $q_T$[GeV]) measured by E288 at different values of $s$ and $Q$. The solid (dashed) lines are the theory prediction at NNLO (N$^3$LO) shifted by the average systematic shift (see table \ref{tab:final}). Filled (empty) point were (not) included in the fit of NP parameters.}
\end{center}
\end{figure}

\begin{figure}
\begin{center}
\includegraphics[width=0.33\textwidth]{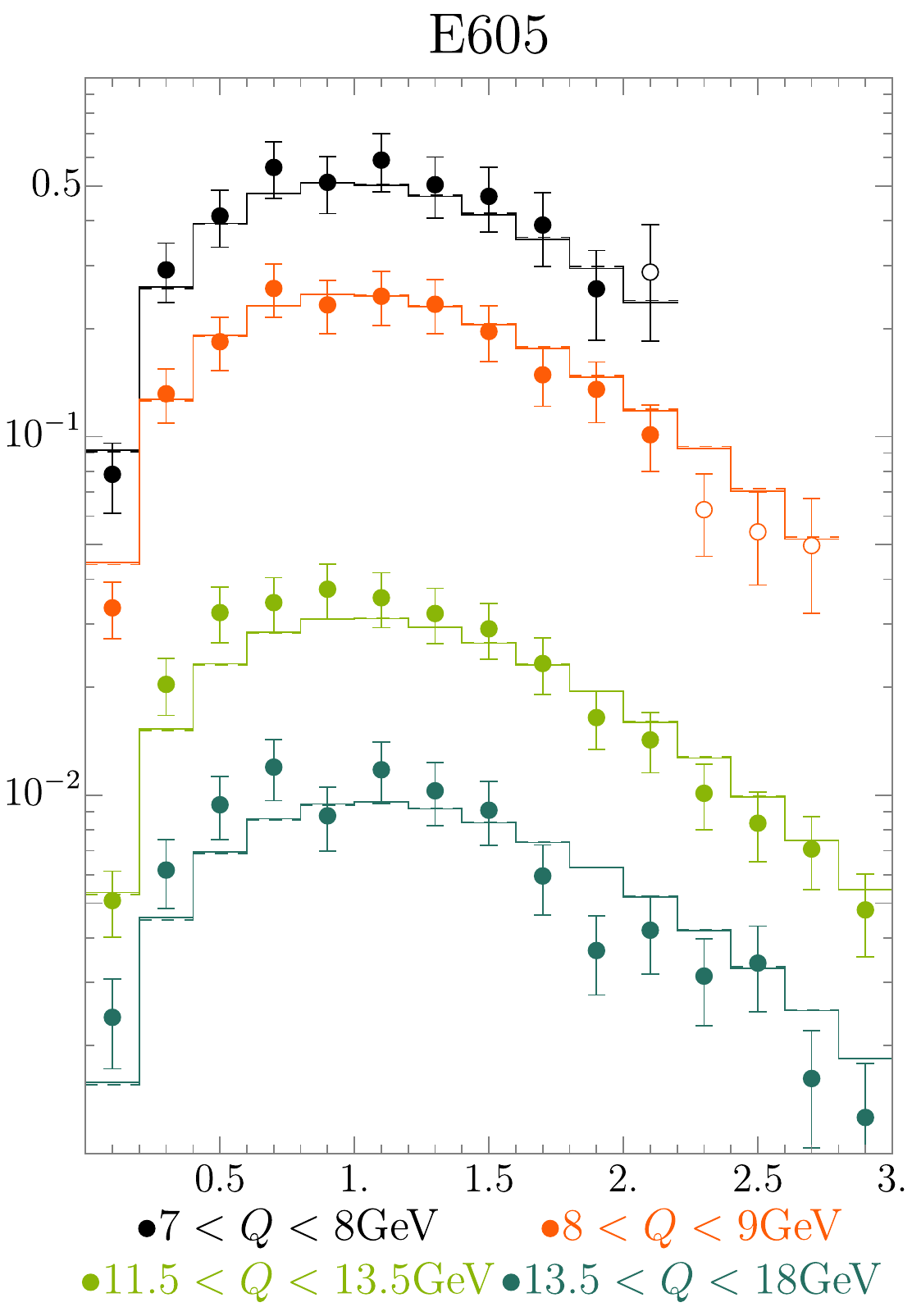}
\includegraphics[width=0.32\textwidth]{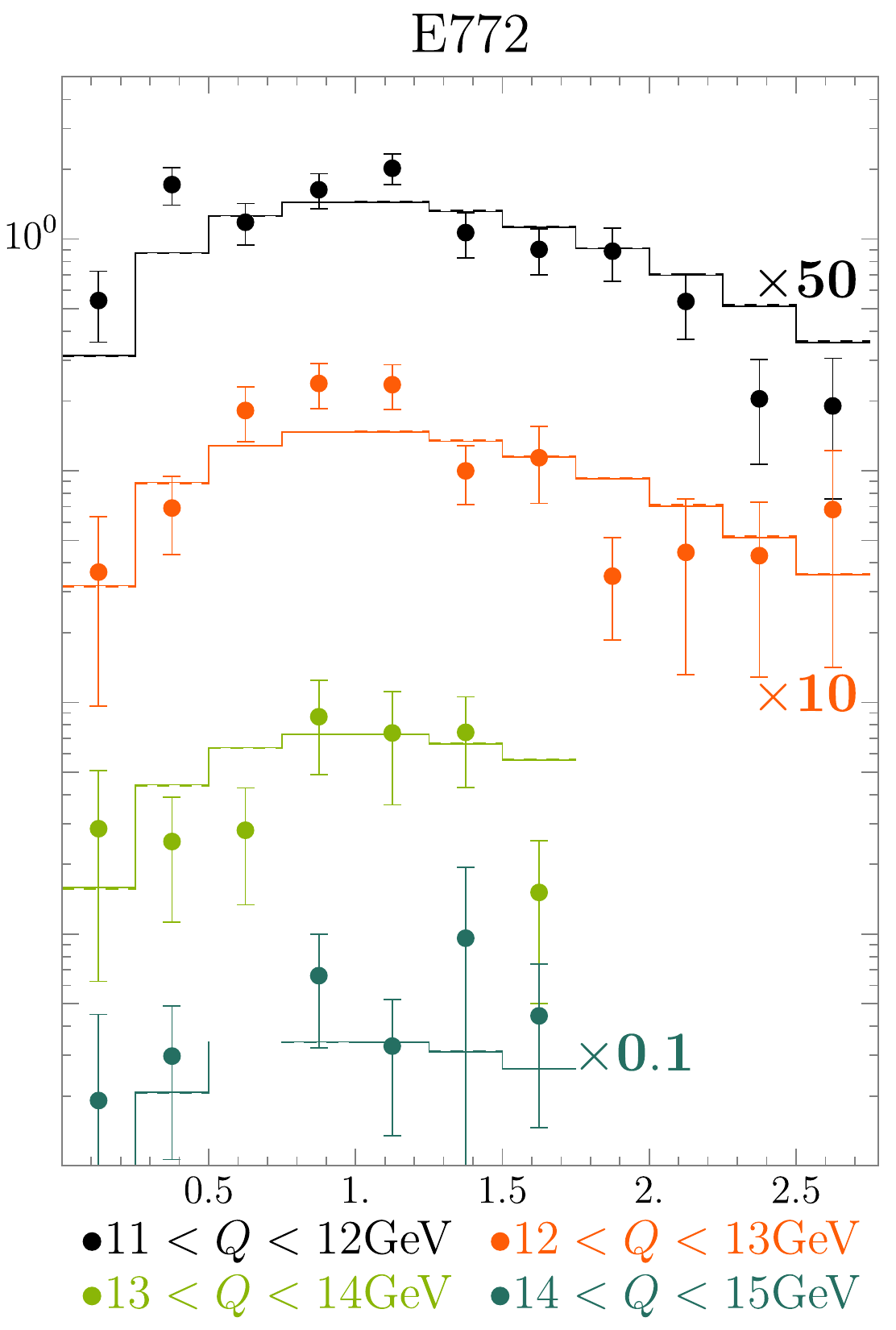}
\caption{\label{fig:E772+E605} Differential cross-section of DY process ($d\sigma/dq_T$[fb/GeV] vs. $q_T$[GeV]) measured by E605 and E772 at different values of $s$ and $Q$. The solid (dashed) lines are the theory prediction at NNLO (N$^3$LO) shifted by the average systematic shift (see table \ref{tab:final}). Filled (empty) point were (not) included in the fit of NP parameters. For clarity the data of E772 is multiplied by the factors indicated in the plot.}
\end{center}
\end{figure}

\begin{figure}
\begin{center}
\includegraphics[width=1.00\textwidth]{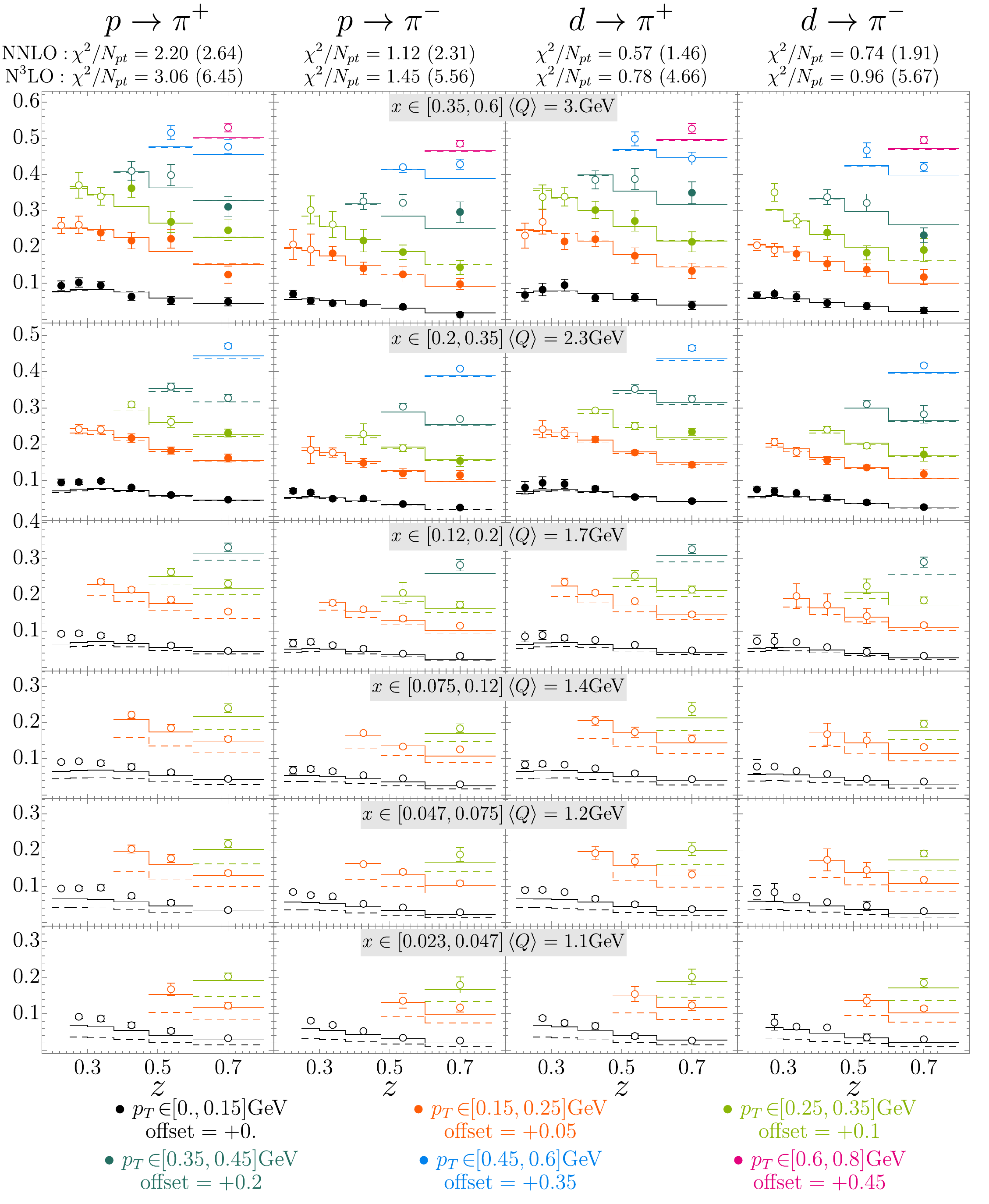}
\caption{\label{fig:HERMES-pi} Unpolarized SIDIS multiplicities (\ref{def:multiplicity}) (multiplied by $z^2$) for production of pions off proton/deuteron measured by HERMES in different bins of $x$, $z$ and $p_T$. Solid (dashed) lines show the theory prediction at NNLO (N$^3$LO). Filled (empty) point were (not) included in the fit of NP parameters. On the top of the table the value of $\chi^2/N_{pt}$ for each channel is presented, the value in brackets being the $\chi^2/N_{pt}$ for shown set of the data (empty and filled points together). For clarity each $p_T$ bin is shifted by an offset indicated in the legend.}
\end{center}
\end{figure}

\begin{figure}
\begin{center}
\includegraphics[width=1.00\textwidth]{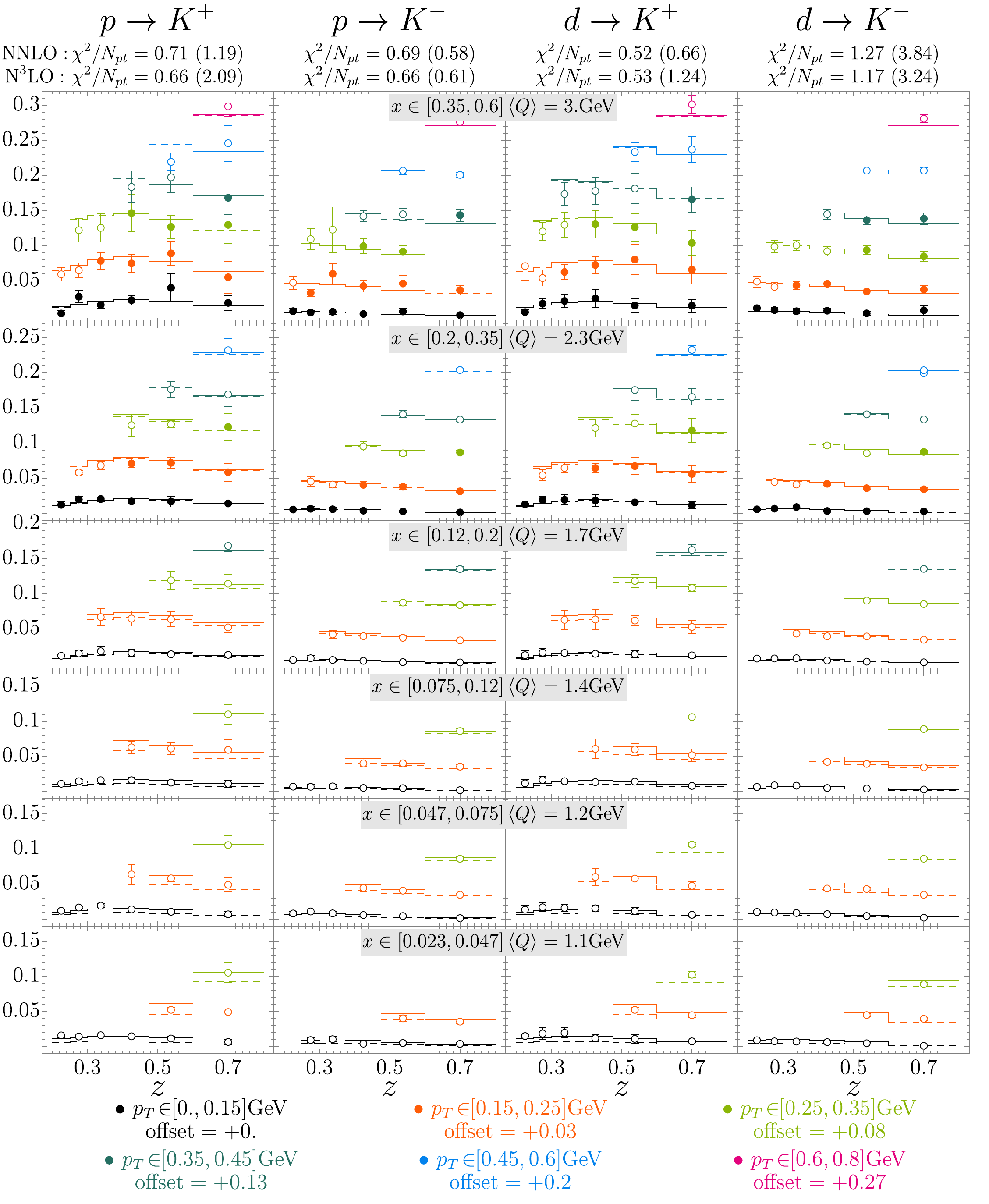}
\caption{\label{fig:HERMES-K} Unpolarized SIDIS multiplicities (\ref{def:multiplicity}) (multiplied by $z^2$) for production of kaons off proton/deuteron measured by HERMES in different bins of $x$, $z$ and $p_T$. Solid (dashed) lines show the theory prediction at NNLO (N$^3$LO). Filled (empty) point were (not) included in the fit of NP parameters. On the top of the table the value of $\chi^2/N_{pt}$ for each channel is presented, the value in brackets being the $\chi^2/N_{pt}$ for shown set of the data (empty and filled points together).  For clarity each $p_T$ bin is shifted by an offset indicated in the legend.}
\end{center}
\end{figure}

\begin{figure}
\begin{center}
\includegraphics[width=1.00\textwidth]{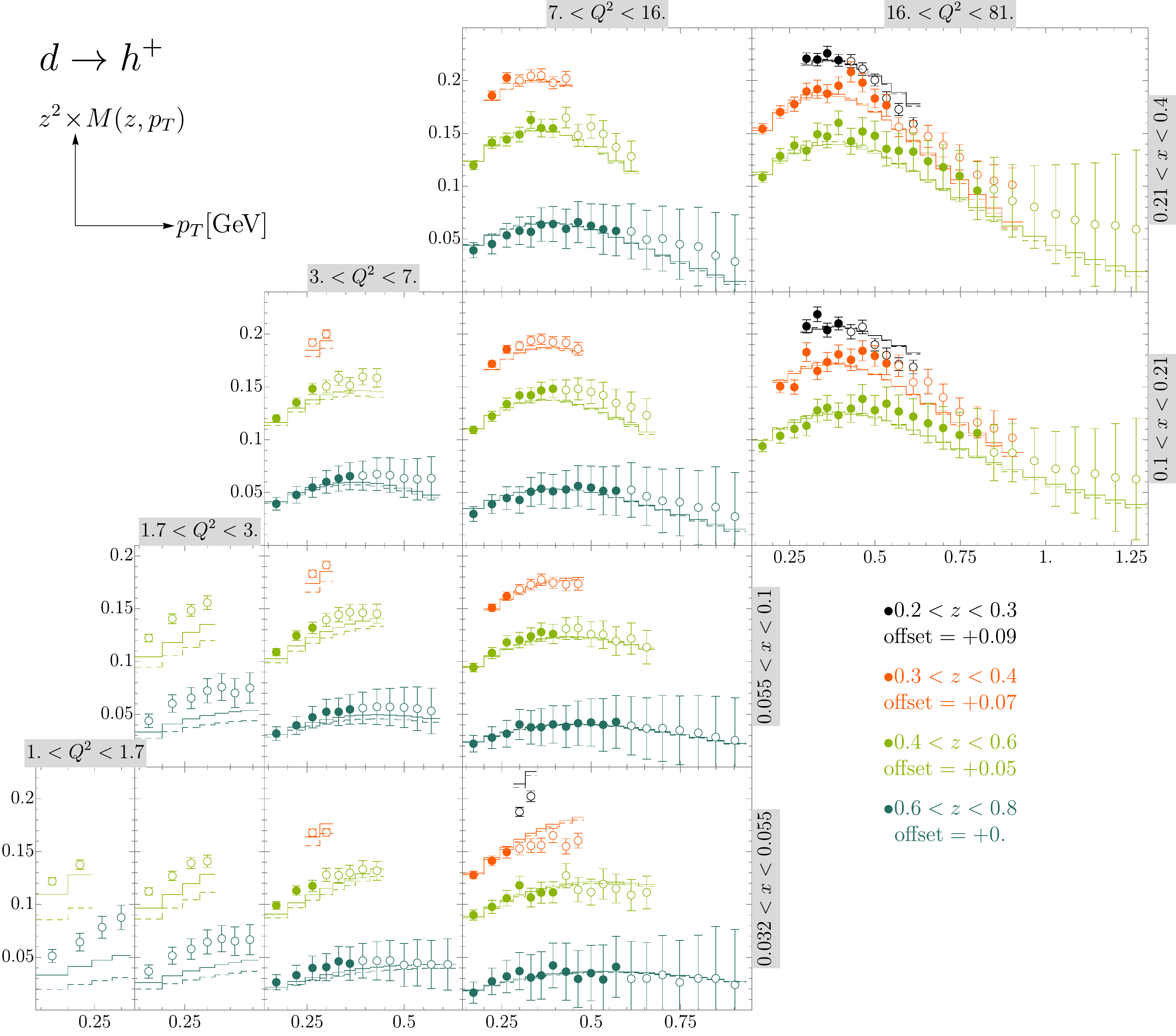}
\caption{\label{fig:COMPASS1} Unpolarized SIDIS multiplicities (\ref{def:multiplicity}) (multiplied by $z^2$) for production of positively charged hadrons off deuteron measured by COMPASS in different bins of $x$, $z$, $Q$ and $p_T$. Solid (dashed) lines show the theory prediction at NNLO (N$^3$LO). Filled (empty) point were (not) included in the fit of NP parameters. For clarity each $p_T$ bin is shifted by an offset indicated in the legend. The continuation of the picture is in fig.~\ref{fig:COMPASS3}.}
\end{center}
\end{figure}

\begin{figure}
\begin{center}
\includegraphics[width=1.00\textwidth]{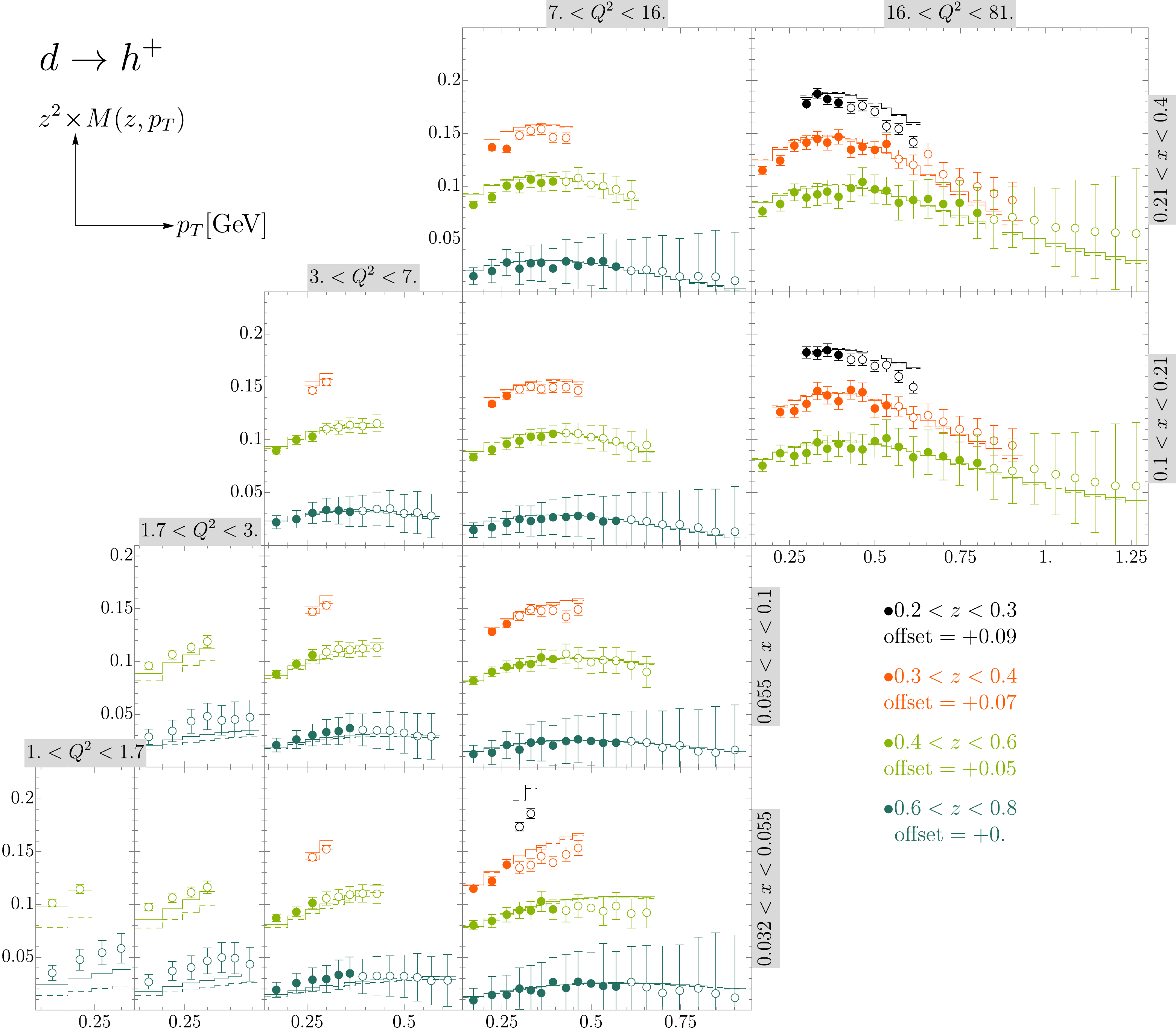}
\caption{\label{fig:COMPASS2} Unpolarized SIDIS multiplicities (\ref{def:multiplicity}) (multiplied by $z^2$) for production of negatively charged hadrons off deuteron measured by COMPASS in different bins of $x$, $z$, $Q$ and $p_T$. Solid (dashed) lines show the theory prediction at NNLO (N$^3$LO). Filled (empty) points were (not) included in the fit of NP parameters. For clarity each $p_T$ bin is shifted by an offset indicated in the legend. The continuation of the picture is in fig.~\ref{fig:COMPASS3}.}
\end{center}
\end{figure}

\begin{figure}
\begin{center}
\includegraphics[width=0.45\textwidth]{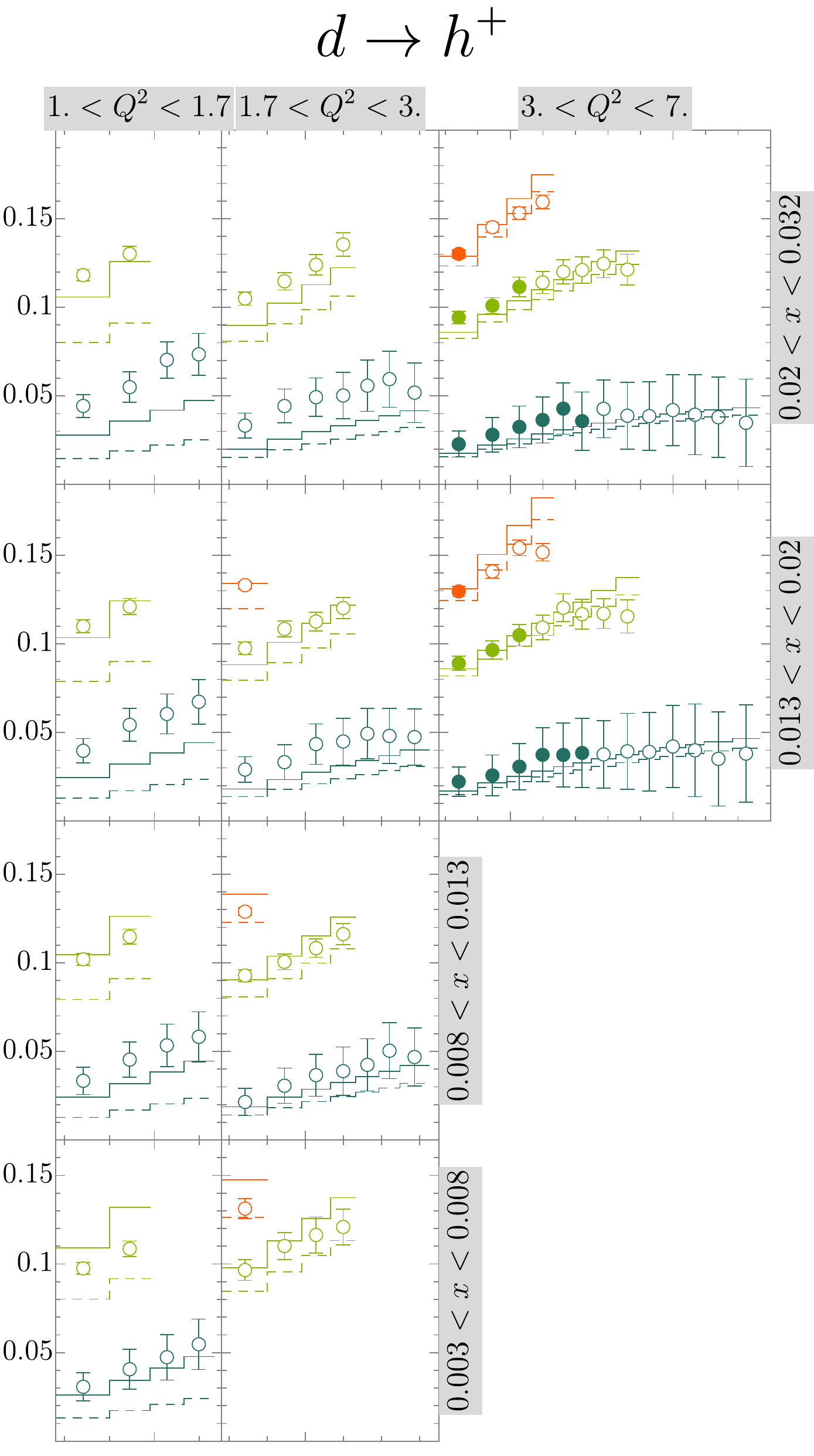}
\includegraphics[width=0.45\textwidth]{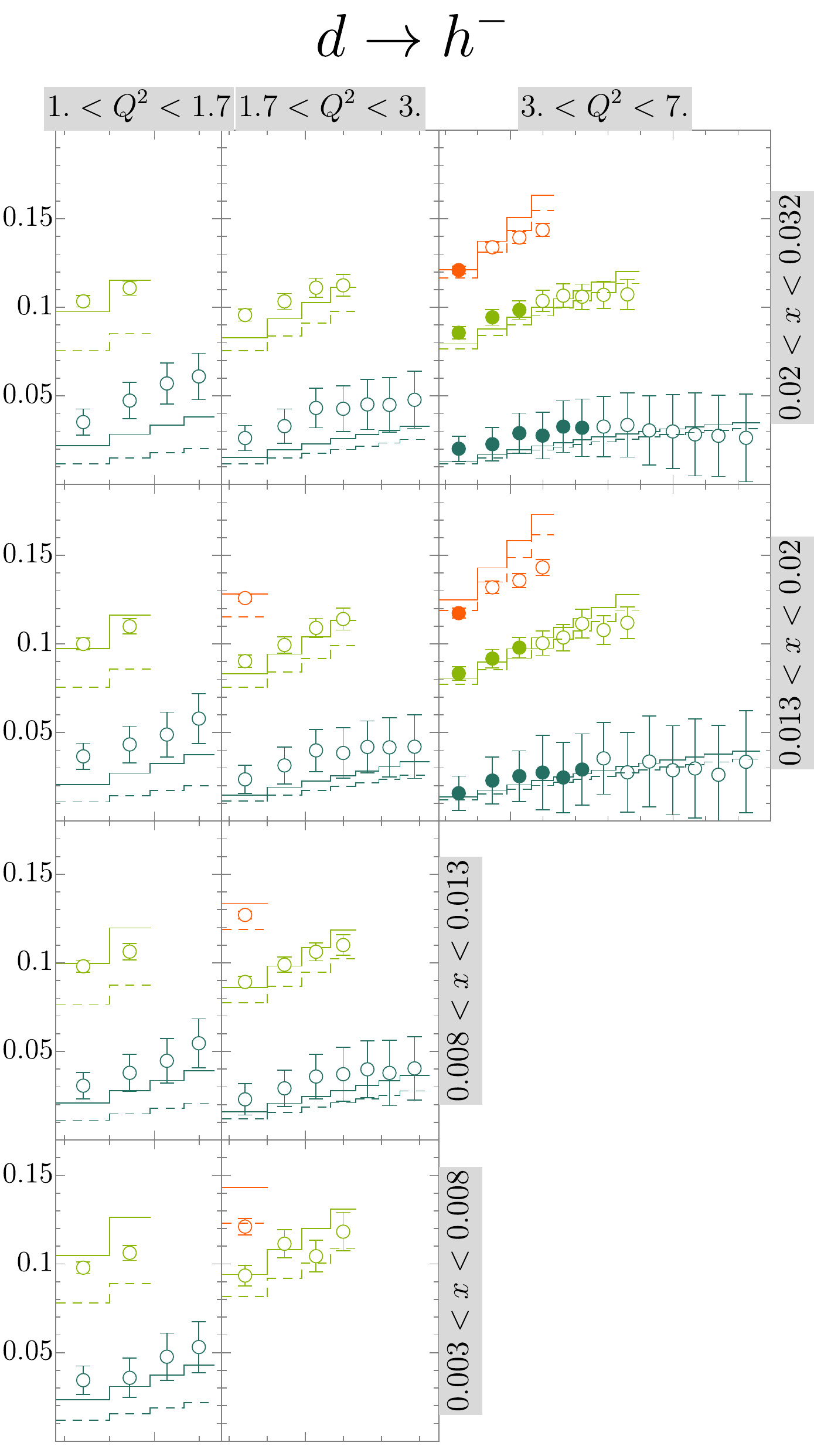}
\caption{\label{fig:COMPASS3} Continuation of the plots \ref{fig:COMPASS1} and \ref{fig:COMPASS2}.}
\end{center}
\end{figure}

In figures \ref{fig:ATLAS}-\ref{fig:COMPASS3} we present  all the data points used in the fit together with the theory prediction lines. In these figures we also show the data points that were not included in the fit due to the cutting conditions eq.~(\ref{SIDIS-data:Q>2}, \ref{SIDIS-data:qT<0.25 Q}, \ref{DY-data:cuts}) in order to demonstrate the behavior of TMD factorization beyond its limits.

\subsection{Values of NP parameters}

The extracted values of NP parameters are  in the table~\ref{tab:NP-param-final}. The central values of parameters do not shift much with respect to individual fits of DY and SIDIS data. The main effect of the global fit is the reduction of uncertainties for RAD and TMDPDF by $\sim 40-50\%$. In figures \ref{fig:DNP-param}, \ref{fig:f1-param} and \ref{fig:d1-param} we show the values of NP parameters obtained in all fits of this work. Generally, the parameters obtained in different fits are in agreement, except $\lambda_{3,4,5}$ that mainly serve for the fine-tune of TMDPDF to LHC data.

\begin{figure}[t]
\begin{center}
\includegraphics[width=0.4\textwidth]{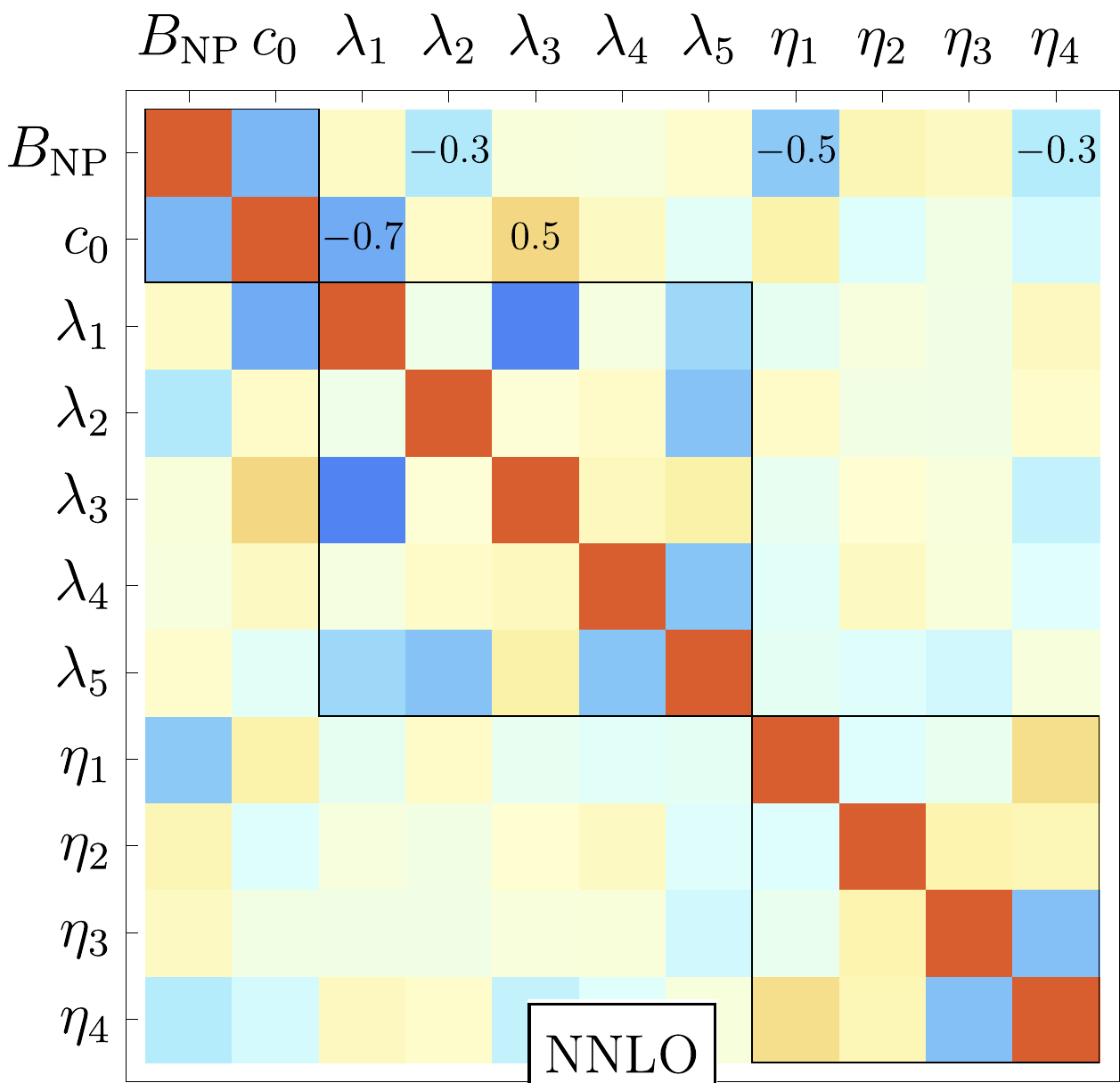}
\includegraphics[width=0.4\textwidth]{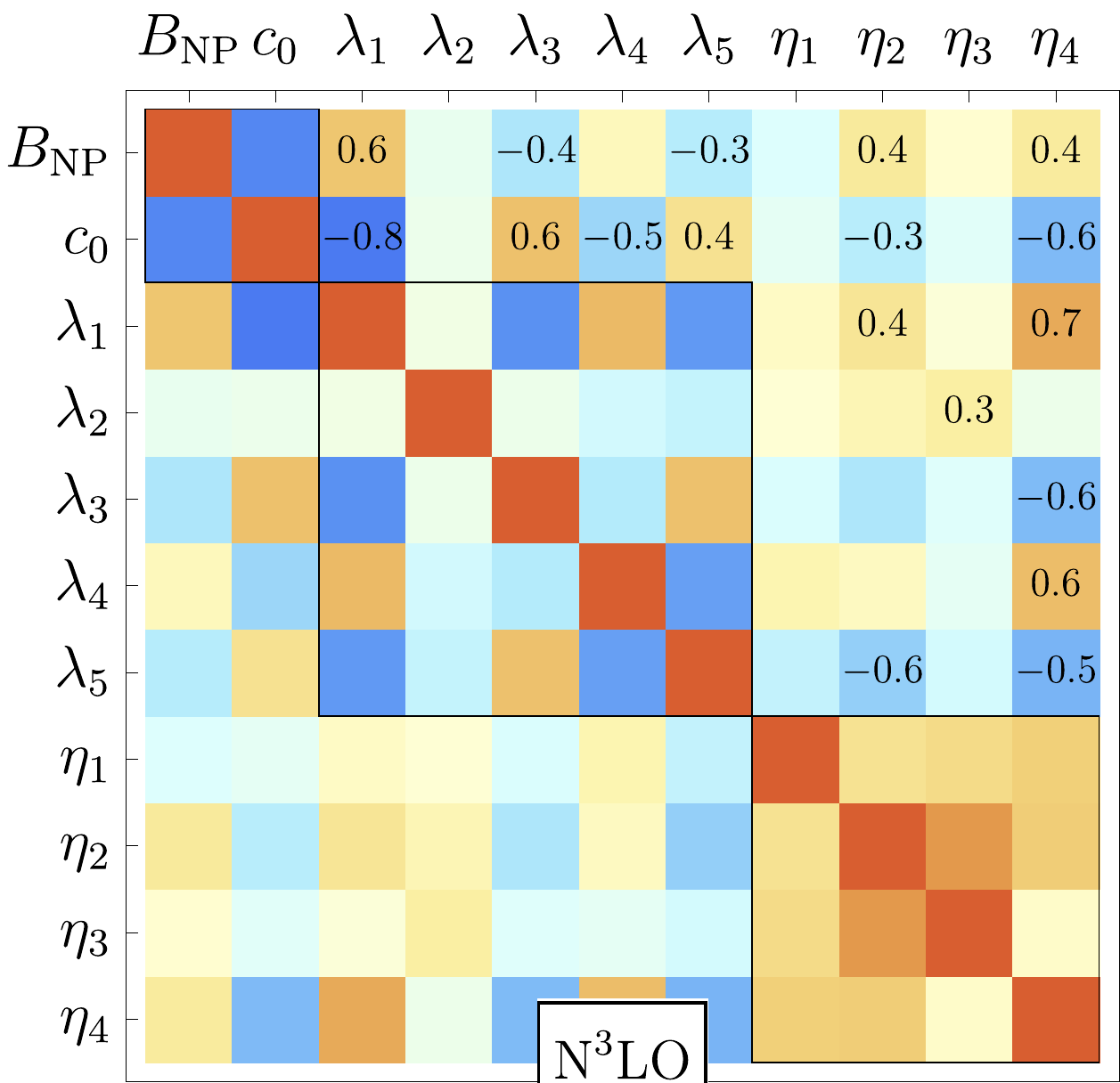}
\includegraphics[width=0.0567\textwidth]{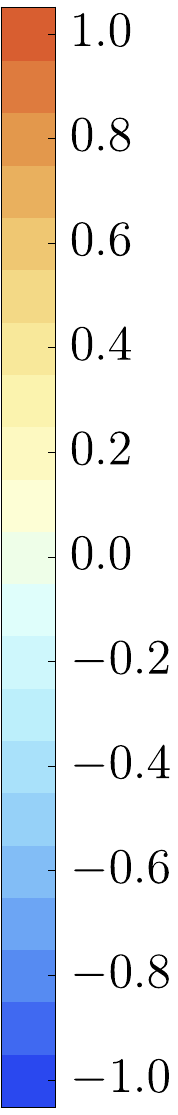}
\caption{\label{fig:correlation}The correlation matrices for NP parameters obtained in the global fit of DY and SIDIS. Numbers indicate the values of matrix elements with correlation higher then $0.3$.}
\end{center}
\end{figure}

All NP parameters are correlated. The correlation matrices for NP parameters are shown in fig.~\ref{fig:correlation}. The explicit numeric expression for correlation matrices is given in appendix \ref{app:correlation}. Ideally, one would expect the complete independence of NP parameters contributing to RAD, TMDPDF and TMDFF. In this case the correlation matrices would have a block-diagonal form. In reality, we observe  correlations among the blocks related to independent functions. In the case of NNLO these correlations are not large, and the block-diagonal structure is evident. The biggest (anti-)correlation is between $c_0$ and $\lambda_1$, with the correlation matrix element $-0.67$, with the rest being much smaller. The source of this correlation is evident -- it is due to the precise Z-boson production measurements by LHC. In the N$^3$LO case the correlation are much stronger. The biggest (anti-)correlation is between $c_0$ and $\lambda_1$, with correlation matrix element $-0.82$, with some other elements reaching $\pm 0.5$ and  it indicates a possible tension in our description of the data at N$^3$LO.

\begin{table}[b]
\begin{center}
\begin{tabular}{|c |l|lll|}
\hline
$\chi^2/N_{pt}$ & \multicolumn{4}{c|}{ NP-parameters}
\\\hline\hline
\multirow{5}{*}{0.95 (NNLO)} & RAD & $B_\text{NP}=1.93\pm0.17$ & $c_0=(3.91 \pm 0.63)\times 10^{-2}$ &
\\ \cline{2-5}
  & \multirow{2}{*}{TMDPDF} & $\lambda_1=0.198\pm 0.019$ & $\lambda_2=9.30\pm0.55$ & $\lambda_3=431.\pm96.$ \\
  &							& $\lambda_4=2.12\pm0.09$ & $\lambda_5=-4.44\pm1.05$ & \\
  \cline{2-5}
  & \multirow{2}{*}{TMDFF} 	& $\eta_1=0.260\pm0.015$ & $\eta_2=0.476\pm0.009$ & \\
  &							& $\eta_3=0.478\pm0.018$ & $\eta_4=0.483\pm0.030$ &
\\\hline
\multirow{5}{*}{1.06 (N$^3$LO)} & RAD & $B_\text{NP}=1.93\pm0.22$ & $c_0=(4.27 \pm 1.05)\times 10^{-2}$ &
\\ \cline{2-5}
  & \multirow{2}{*}{TMDPDF} & $\lambda_1=0.224\pm 0.029$ & $\lambda_2=9.24\pm0.46$ & $\lambda_3=375.\pm89.$ \\
  &							& $\lambda_4=2.15\pm0.19$ & $\lambda_5=-4.97\pm1.37$ & \\
  \cline{2-5}
  & \multirow{2}{*}{TMDFF} 	& $\eta_1=0.233\pm0.018$ & $\eta_2=0.479\pm0.025$ & \\
  &							& $\eta_3=0.472\pm0.041$ & $\eta_4=0.511\pm0.040$ &
  \\ \hline
\end{tabular}
\end{center}
\caption{\label{tab:NP-param-final} Values of $\chi^2$ and NP parameters obtained obtained in the global fit of DY and SIDIS data. The collinear distributions are NNPDF31 and DSS.}
\end{table}

\section{Comments on the extracted TMD distributions}
\label{sec:final}

\begin{figure}
\begin{center}
\includegraphics[width=0.44\textwidth]{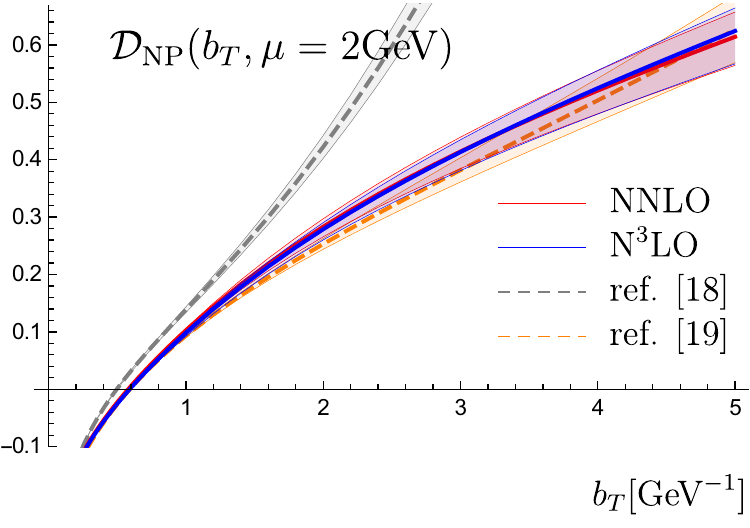}
\includegraphics[width=0.54\textwidth]{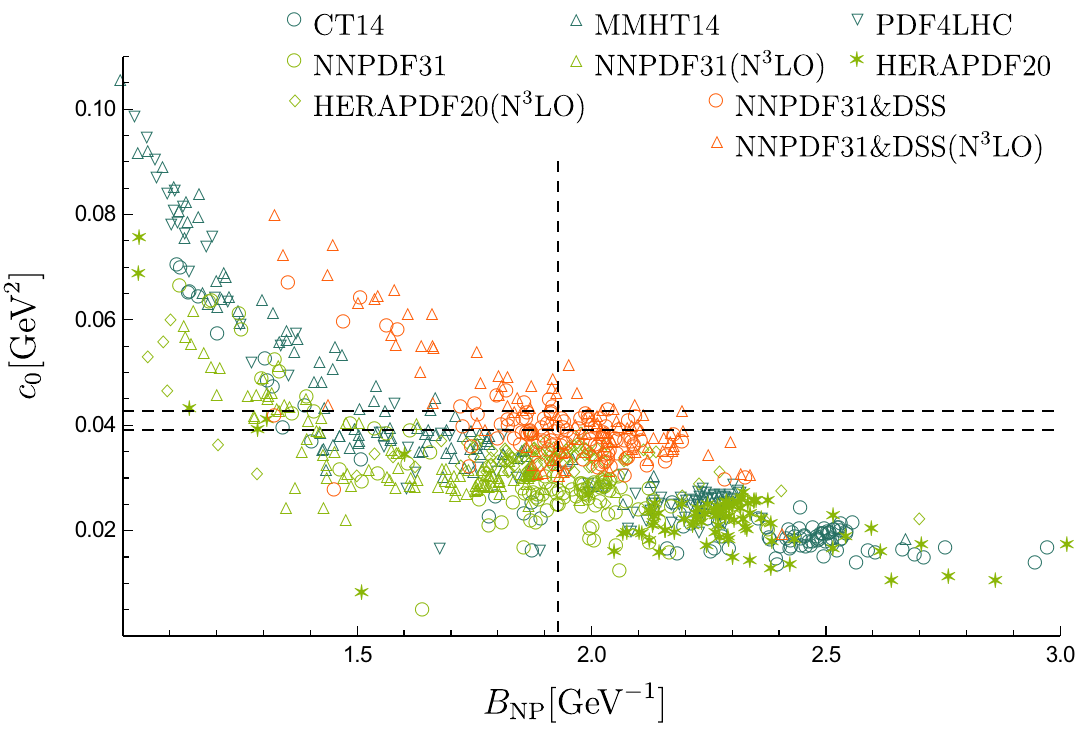}
\caption{\label{fig:RAD} (left) Comparison of NNLO RAD extracted in DY fit (NNPDF31), and global fit of DY and SIDIS (NNPDF31\& DSS). Shaded area shows the $1\sigma$-uncertainty band. The dashed lines show the extraction made in refs.\cite{Bacchetta:2017gcc} and \cite{Scimemi:2017etj} at LO and NNLO of RAD correspondingly. (right) Distribution of replica points in different fits of RAD. Dashed lines show the mean values of RAD extracted in the global fit of DY and SIDIS.}
\end{center}
\end{figure}

The non-perturbative distributions extracted in this work show  several features that are interesting for theory investigations. For instance, the RAD that measures the properties of the soft gluon exchanges and that is inclusively sensitive to the QCD vacuum structure. The factorization theorem ensures that the values of $B_{\text{NP}}$ and $c_0$ are totally uncorrelated from the rest of TMD parameters, because they are of complete different origin. As we have an extraction of these parameters from data we can expect that a certain correlation is re-introduced in the fitting process. In fig.~\ref{fig:correlation} (see also appendix~\ref{app:correlation}) we check this statement in the present global fit  and we find that it is qualitatively verified in our DY+SIDIS fit. In the figure the only non-perturbative parameters which show a higher (anti)correlation with the RAD are  $c_0$ and $\lambda_1$ in the TMDPDF.
Apart from this, the independence of the  RAD parameters from the  rest of TMD is certainly a success of the  $\zeta$-prescription, which allows a clear separation of all these effects.
In the rest of this section we report some specific comment for each of the functions that we have extracted.
\subsection{Non-perturbative RAD}
\label{sec:NP-RAD-final}

In fig.~\ref{fig:RAD} (left)  we plot the RAD as a function of $b$ with its uncertainty band. We present only the RAD extracted with NNPDF31 fits, but the picture does not change significantly for all other PDF sets.  In this figure we can test the universality of the RAD  looking at its extraction in DY and DY+SIDIS. At small $b$ the perturbative structure of the RAD dominates and we find practically no difference in its behavior as coming from different fits. The difference between these two cases  happens at large $b$ and it is at most of 10\%. The $1\sigma$-uncertainty bands of DY and global fit  do not  strictly overlap, which possibility indicates their underestimation. 

In the same fig.~\ref{fig:RAD} (left) we also compare our RAD with the one obtained in \cite{Bacchetta:2017gcc} and \cite{Scimemi:2017etj}. In refs.~\cite{Bacchetta:2017gcc,Scimemi:2017etj} a different shape of NP ansatz for RAD has been used, with a quadratic behavior at large-$b$. Such an ansatz has been used often, and (as we have also checked) it is able to describe the data. Nonetheless we disregard it because the global $\chi^2/N_{pt}$ is worse ($1.11$ and $1.34$ at NNLO and N$^3$LO, correspondingly), with much larger correlation among parameters. Additionally, the linear asymptotic behavior used in our ansatz is supported by non-perturbative models, e.g.~\cite{Vladimirov:2020umg}. Possibly, the uncertainty band is biased by this model, and the realistic band is larger by a factor two at most.

In fig.~\ref{fig:RAD} (right) we show the scattering of replicas in ($B_{\text{NP}},c_0$)-plane collected from all fits. It is clear that the parameters $B_{\text{NP}}$ and $c_0$ are strongly anti-correlated (see also fig.~\ref{fig:correlation}) and this is a consequence of the non-perturbative model, since the variation of  $c_0$ can be compensated by a variation of $B_{\text{NP}}$ up to $b^4$-corrections. The replicas of the global fit (orange points) are scattered in a much smaller area  and this provides a  $\sim 40\%$ smaller error-bands on parameters. Generally, the inclusion of the SIDIS data drastically constraints the values of $B_{\text{NP}}$, and for that reason they are very important for the determination of RAD. We conclude that the RAD extracted in the global fit is more reliable, in comparison to the one done using  DY data only.

The RAD that we have extracted is valid for all distributions and it has been used also to describe the pion-induced DY \cite{Vladimirov:2019bfa}. For further reduction of the uncertainty of the  RAD one should consider more precise low- and intermediate-energy processes, such as  up-coming JLab12 measurements, and the future EIC.


\subsection{TMD distributions}
\label{sec:final-TMD}

\begin{figure}
\begin{center}
\includegraphics[width=0.48\textwidth]{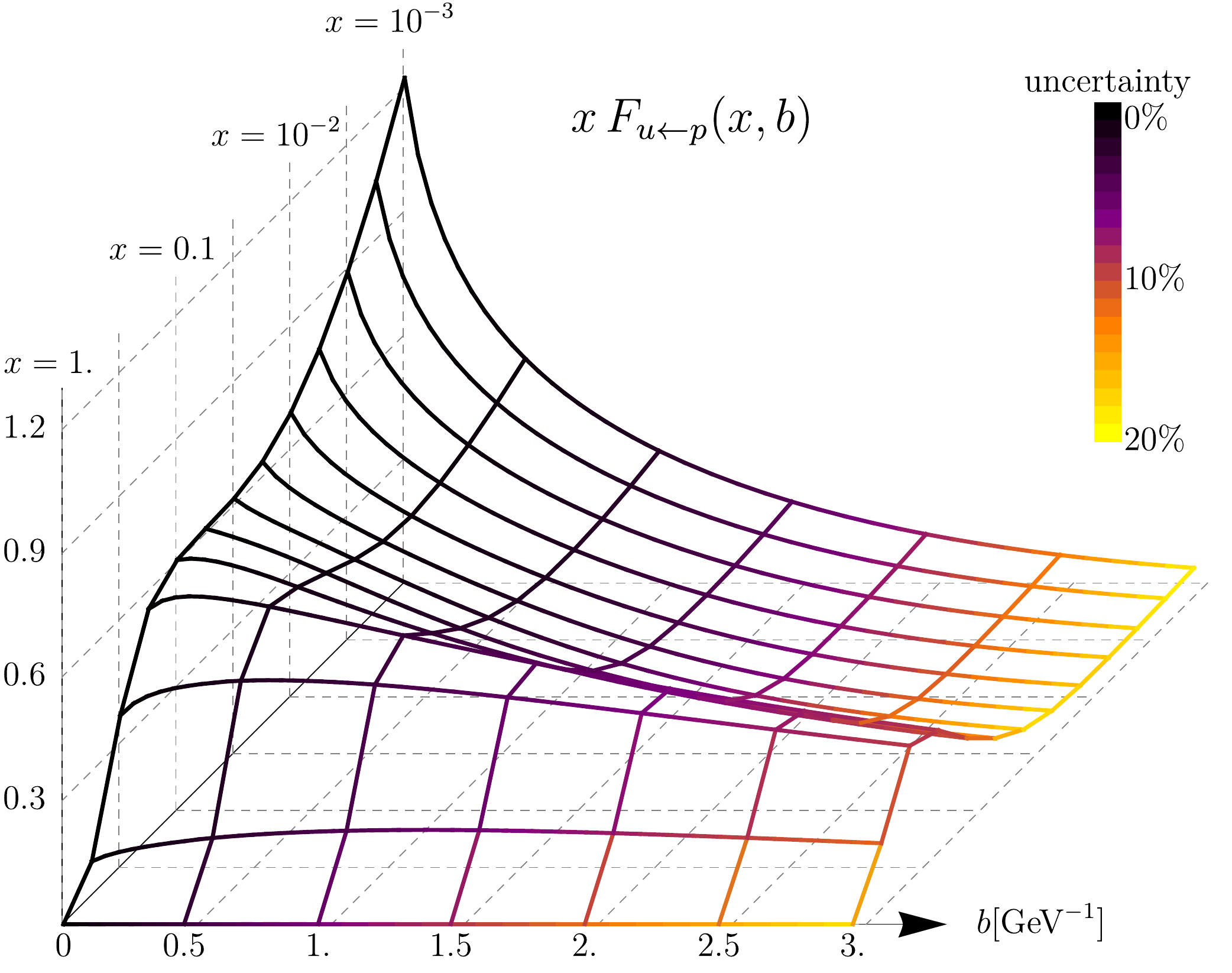}
\includegraphics[width=0.48\textwidth]{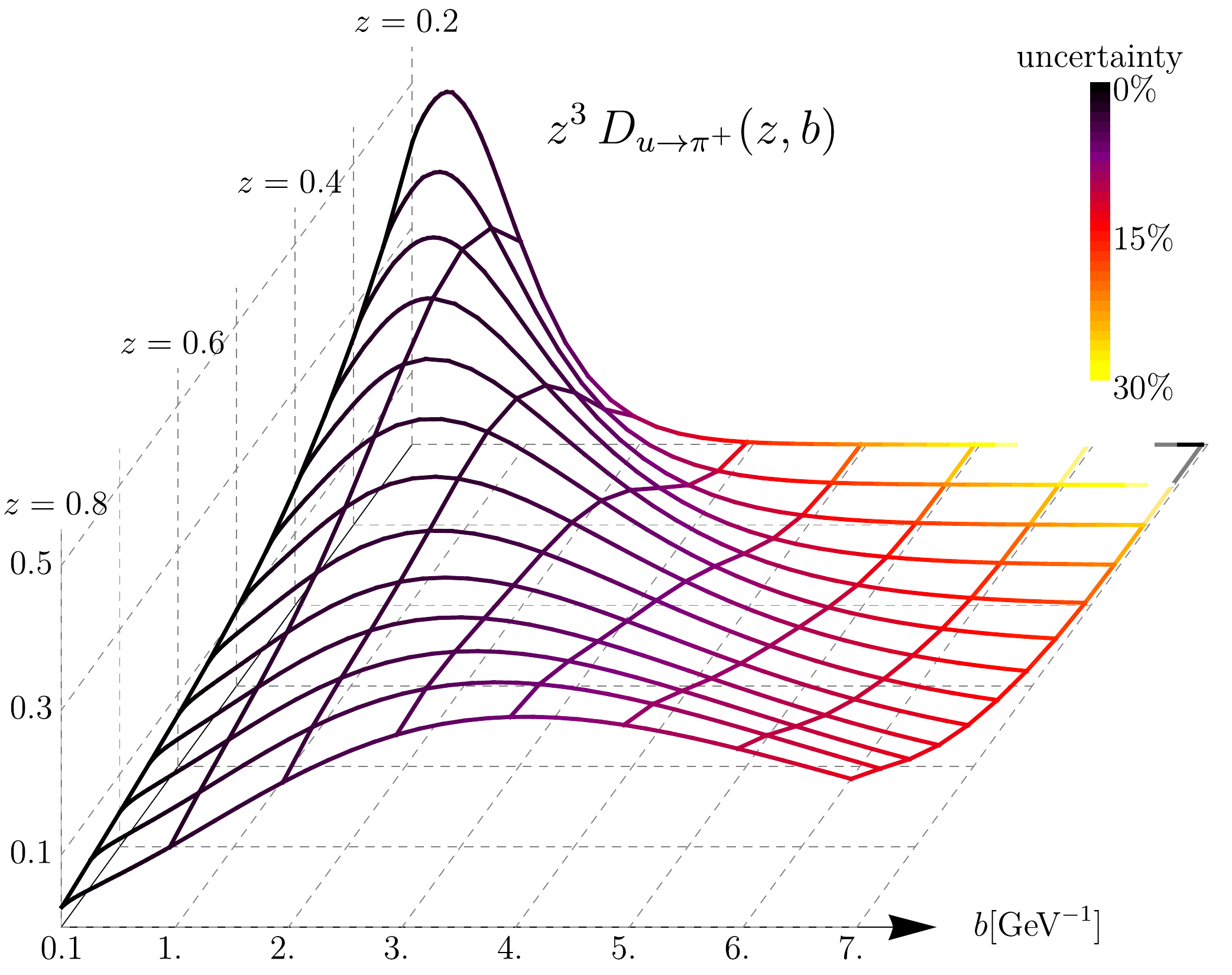}
\caption{\label{fig:TMDs} Example of extracted (optimal) unpolarized TMD distributions. The color indicates the relative size of the uncertainty band}
\end{center}
\end{figure}

The quark TMDPDF and TMDFF are extracted  simultaneously including  high QCD perturbative orders for the first time to our knowledge. The non-perturbative parameters  obtained using the PDF set NNPDF31 and the fragmentation set DSS are reported in table \ref{tab:NP-param-final}. 
Within one set of PDF the error induced from the PDF replicas dominates the experimental error of TMD. Thus, the error that we have reported on TMD parameters is certainly underestimated. To determine a realistic uncertainty band , one must invent a flexible ansatz for NP-part of TMD distributions that does not contradict the known theory. It appears to be a non-trivial task, which we leave for a future study.

The TMD distributions show a non-trivial intrinsic structure. An example of distributions in $(x,b)$-plane is presented in fig.~\ref{fig:TMDs}. Depending on $x$ the $b-$behavior apparently changes. We observe (the same observation has been made in ref. \cite{Bacchetta:2017gcc}) that the unpolarized TMDFF gains a large $b^2$-term in the NP part. It could indicate a non-trivial hadronisation physics, or a tension between colinear and TMD distributions. The study of its origin should be addressed by future studies.

\section{Conclusion}

Standing the TMD factorization of DY and SIDIS  cross-section, one identifies at least three non-perturbative QCD distributions in each cross-section -- two TMD parton distributions and a non-perturbative rapidity anomalous dimension (RAD). These functions should be extracted from the experimental data. Given such a large number of phenomenological functions, their universality plays a crucial role. In this work, we have shown that the TMD distributions and RAD are indeed universal functions. 

In order to confirm the universality statement, we have firstly extracted the RAD ($\mathcal{D}$) and the  unpolarized TMDPDF ($f_1$) from the DY data, and secondly  we have used them to describe the SIDIS data (extracting in addition the unpolarized TMDFF, $D_1$). To our best knowledge, this is the first clear-cut demonstration of the universality of the TMD non-perturbative components. This demonstration is the main result of this work. The subsidiary results are the values of extracted unpolarized TMD distributions and RAD, that could be used to predict and describe the low-$q_T$ spectrum of current (LHC, COMPASS, RHIC) and future (EIC, LHeC) experiments.

The sets of  data included in this analysis contain in total 1039 points (almost equally distributed between SIDIS, 582 points, and DY, 457 points). We have the data from fixed target DY measurements, Tevatron, RHIC, LHC, COMPASS, and HERMES. Unfortunately, only  low-energy measurements are available for SIDIS data.
 At the moment, we have not included any data from HERA multiplicities because they do not accomplish the kinematical requirements for the TMD factorization. Contrary to some observations in the literature~\cite{Anselmino:2013lza,Bacchetta:2017gcc}, we have not found any problem with the normalization of HERMES and  COMPASS data, although the systematic experimental errors quit precision to the final result.

The data analysis  is made with the current theory state-of-art, including all known perturbative QCD orders, i.e. N$^3$LO for the hard part and the evolution, and NNLO for the collinear matching. The NNLO and N$^3$LO predictions are very close to each other, which is a good signal indicating that the perturbative part of the cross-section is saturated. We have also collected all recent modifications and updates of the TMD factorization approach, such as target-mass corrections, frame-corrections, and exact evolution solution at large-$b$. Individually these aspects are subtle, however, cumulatively, they are sizable. In sec.~\ref{sec:2} we have presented a comprehensive collection of theory expressions used in this work. Let us also mention that the N$^3$LO evolution, as well as a non-trivial QCD matching for TMDFF (NNLO vs. LO) is used here for the first time. An open issue is represented by power corrections, which, given the current status of the art can be included only partially, as discussed in sec.~\ref{sec:2}.
More work concerning this problem should be addressed in the future.

The definition of matching scales and the evolution/modeling separation is done according to $\zeta$-prescription. The $\zeta$-prescription is equivalent to the popular CSS-scheme since it satisfies the same set of differential equations. Nonetheless, this equivalence is strict only within an all-order perturbation theory and it is numerically violated for any truncated series. The origin for this discrepancy is well-understood \cite{Scimemi:2018xaf} -- it comes from  spurious contributions in the CSS formalism that vanish in the exact perturbation theory. 
At LO and NLO, the numerical value of  spurious contributions is large, but it is tiny at N$^3$LO \cite{Scimemi:2018xaf}. Therefore, the $\zeta$-prescription provides a faster convergence and an improved stability of the perturbative series, as shown in fig.~\ref{fig:convergence}. Additionally, but not less importantly, the $\zeta$-prescription grants a strict separation of perturbative and non-perturbative pieces and thus it allows a stronger universality of the phenomenological functions, fig.~\ref{fig:correlation} and appendix~\ref{app:correlation}.
 In particular, the RAD extracted here can be used in the analysis of  jet-production \cite{Gutierrez-Reyes:2019msa,Gutierrez-Reyes:2019vbx,Gutierrez-Reyes:2018qez}. 
 The success of the present global fit confirms the reliability of the $\zeta$-prescription. 

Many points of the TMD phenomenology are discussed quantitatively for the first time (to our best knowledge). We critically consider each detail of the factorization  that have a disputable nature, f.i. power corrections to collinear variables. We demonstrated that the inclusion of these details improves the agreement between theory and the data. A particularly important check made here for the first time is the test of the limit of the TMD factorization approximation for SIDIS. In the DY case, the phenomenological limit of TMD factorization is $q_T\lesssim 0.25 Q$, as it has been shown in ref.~\cite{Scimemi:2017etj}. We have found that SIDIS also obeys this rule. This piece of information  is important because  it opens the door to reliable predictions of SIDIS experiments. 

The estimation of the uncertainty for extracted distributions is made by the replica method that gives a reliable error-propagation of experimental errors. On top of it one should include the uncertainty of other theoretical ingredients, and in particular  the collinear PDF error. We have checked that the prediction of the TMD factorization is crucially sensitive to the values of collinear PDFs. It indicates that our extraction has a considerable additional uncertainty due to the uncertainty of the collinear input. However, we were not able to accurately quantify the size of this uncertainty band, due to the high computational costs of such analysis. We leave this study for the future.

\section*{Acknowledgements}
We thank Valerio Bertone, Gunar Schnell, Pia Zurita and Elke Aschenauer for stimulation discussions and correspondence. I.S. is supported by the Spanish MECD grant FPA2016-75654-C2-2-P and PID2019-106080GB-C21.  This project has received funding from the European Union Horizon 2020 research and innovation program under grant agreement No 824093 (STRONG-2020). The research was supported by DFG grant N.430824754 as a part of the Research Unit FOR 2926.

\appendix
\section{Expression for $|C_V|$}

The hard coefficient for TMD factorization formula is the square of the hard matching coefficient for quark current $|C_V|^2$. At NLO hard matching coefficient reads
\begin{eqnarray}
C_V(q,\mu)=1+a_s C_F\(-\ln^2\(\frac{-q^2}{\mu^2}\)+3\ln\(\frac{-q^2}{\mu^2}\)-8+\frac{\pi^2}{6}\)+O(a_s^2).
\end{eqnarray}
The expression for NNLO can be found f.i. in ref.~\cite{Matsuura:1988sm,Kramer:1986sg}, and at N$^3$LO in \cite{Gehrmann:2010ue}. 

The hard coefficient for SIDIS and DY kinematics differ only by the sign of $q^2$ that is space-like (times-like) for DY (SIDIS). Since in our work $\mu^2=Q^2$ the expression simplifies. In particular, in the DY kinematics the logarithms turns to $\ln(-1)=\pm i\pi$, whereas in the case of SIDIS logarithm vanish. The NNLO expression for DY hard coefficient is
\begin{eqnarray}
|C_V(-Q^2,Q^2)|^2&=&1+a_s C_F\(-16+\frac{7\pi^2}{3}\)+a_s^2 C_F\Big[C_F\(\frac{511}{4}-\frac{83\pi^2}{3}-60\zeta_3+\frac{67\pi^4}{30}\)
\\\nn &&+C_A\(-\frac{51157}{324}+\frac{1061\pi^2}{54}+\frac{626\zeta_3}{9}-\frac{8\pi^4}{45}\)+N_f\(\frac{4085}{162}-\frac{91\pi^2}{27}+\frac{4\zeta_3}{9}\)\Big],
\end{eqnarray}
where $C_F(=4/3)$ and $C_A(=3)$ are quadratic Casimir eigenvalues of fundamental and adjoint representation of SU(3). $N_f$ is the number of active quark flavors. The NNLO expression for SIDIS hard coefficient is
\begin{eqnarray}
|C_V(Q^2,Q^2)|^2&=&1+a_s C_F\(-16+\frac{\pi^2}{3}\)+a_s^2 C_F\Big[C_F\(\frac{511}{4}-\frac{13\pi^2}{3}-60\zeta_3+\frac{13\pi^4}{30}\)
\\\nn &&+C_A\(-\frac{51157}{324}-\frac{337}{54}+\frac{626\zeta_3}{9}+\frac{22\pi^4}{45}\)+N_f\(\frac{4085}{162}+\frac{23\pi^2}{27}+\frac{4\zeta_3}{9}\)\Big].
\end{eqnarray}
One can see that the difference between SIDIS and DY hard coefficients is
\begin{eqnarray}
&&|C_V(Q^2,Q^2)|^2-|C_V(-Q^2,Q^2)|^2=-2\pi^2 a_sC_F
\\\nn&&\qquad +\pi^2a_s^2C_F\[C_F\(32-\frac{8\pi^2}{3}\)+C_A\(-\frac{233}{9}+\frac{2\pi^4}{3}\)+N_f\frac{38}{9}\]+O(a_s^3).
\end{eqnarray}
These corrections are known as $(\pi^2a_s)^n$-corrections. They could be resummed to all orders \cite{Ahrens:2008qu}. However, in the case of vector-boson production the correction coming from $(\pi^2a_s)^n$ is not significant (of order of next-to-given order correction \cite{Ahrens:2008qu}).

\section{Perturbative expression for $\mathcal{D}$}
\label{app:RAD}

The rapidity anomalous dimension (RAD) $\mathcal{D}(\mu,b)$ is generally non-perturbative function, which can be computed perturbatively only at small-b, see e.g.\cite{Vladimirov:2016dll,Echevarria:2015byo} for NNLO and N$^3$LO computations. RAD satisfies the integrability condition (\ref{def:integrability}) which can be seen as renormalization group equation,
\begin{eqnarray}\label{app:RAD:RGE}
\mu^2 \frac{d \mathcal{D}(\mu,b)}{d\mu^2}=\frac{\Gamma_{\text{cusp}}(\mu)}{2}.
\end{eqnarray}
Consequently, the $a_s^n$-order of RAD contains logarithms up to order $\mathbf{L}_\mu^{n}$. We define
\begin{eqnarray}\label{app:def:dnk}
\mathcal{D}_{\text{pert}}(\mu,b)=\sum_{n=1}^\infty a_s^n(\mu)\sum_{k=0}^n \mathbf{L}_\mu^k d^{(n,k)},
\end{eqnarray}
where $d^{(n,k)}$ are numbers, and
\begin{eqnarray}\label{app:Lmu}
\mathbf{L}_\mu=\ln\(\frac{\mu^2\vec b^2}{4e^{-2\gamma_E}}\).
\end{eqnarray}
The values for $d^{(n,k)}$ for $k>0$ can be computed from (\ref{app:RAD:RGE}) in the terms of $\Gamma_i$, $\beta_i$ and $d^{(n,0)}$. Here, and in the following we define beta-function, and coefficients of $\Gamma_{\text{cusp}}$ as
\begin{eqnarray}\label{app:def-beta-Gamma}
\mu^2 \frac{d a_s(\mu)}{d\mu^2}=-\beta(a_s)=-\sum_{n=0}^\infty a_s^{n+2}(\mu)\beta_n,\qquad \Gamma_{\text{cusp}}(\mu)=\sum_{n=0}^\infty a_s^{n+1}(\mu)\Gamma_n.
\end{eqnarray}
The leading terms are $\beta_0=\frac{11}{3}C_A-\frac{2}{3}N_f$, $\Gamma_0=4C_F$, and $d^{(1,0)}=0$. The values of $d^{(2,0)}$ and $d^{(3,0)}$ were computed in \cite{Becher:2010tm,Echevarria:2015byo} and \cite{Li:2016ctv,Vladimirov:2016dll} respectively. The $\beta$-function coefficients are known up to $\beta_4$ \cite{Baikov:2016tgj} and  the $\Gamma_i$ are known up to $\Gamma_3$-order for the quark case, see \cite{Moch:2017uml,Moch:2018wjh,Lee:2019zop} and references within.

The series (\ref{app:RAD:RGE}) has a small convergence radius since the expansion variable $(a_s\mathbf{L}_\mu)$ gets fastly bigger then 1 with the increase of $b$. To improve the convergence properties of RAD we use the resummed expression \cite{Echevarria:2012pw,Scimemi:2018xaf,Bizon:2018foh}. In this case we write
\begin{eqnarray}\label{app:RAD:resum}
\mathcal{D}_{\text{resum}}(\mu,b)=\sum_{n=0}^\infty a_s^n(\mu)d_n(X),\qquad X\equiv\beta_0a_s(\mu)\mathbf{L}_\mu.
\end{eqnarray}
The functions $d_n$ satisfy the set of equations
\begin{eqnarray}\label{app:RAD:eqn_dn}
\beta_0 d'_n(X)-\sum_{k=0}^n \beta_k ((n-k)d_{n-k}(X)+Xd'_{n-k}(X))=\frac{\Gamma_n}{2},
\end{eqnarray}
where $d_n'(X)=\partial d_n(X)/\partial X$. The boundary conditions are $d_n(X=0)=d^{(n,0)}$. These equations are to be solved recursively starting from the equation at $n=0$. The solutions of (\ref{app:RAD:eqn_dn}) are
\begin{eqnarray}\label{app:d0}
d_0(X)&=&-\frac{\Gamma_0}{2\beta_0}\ln(1-X),
\\
d_1(X)&=&\frac{1}{2\beta_0(1-X)}\Big[-\frac{\beta_1\Gamma_0}{\beta_0}(\ln(1-X)+X)+\Gamma_1X\Big],
\\
d_2(X)&=&\frac{1}{(1-X)^2}\Big[\frac{\Gamma_0\beta_1^2}{4\beta_0^3}\(\ln^2(1-X)-X^2\)
+\frac{\beta_1\Gamma_1}{4\beta_0^2}\(X^2-2X-2\ln(1-X)\)
\\\nn &&+\frac{\Gamma_0\beta_2}{4\beta_0^2}X^2-\frac{\Gamma_2}{4\beta_0}X(X-2)+d^{(2,0)}\Big],
\\\nn
d_3(X)&=&\frac{1}{(1-X)^3}\Big[
-\frac{\Gamma_0\beta_1^3}{6\beta_0^4}\(\ln^3(1-X)-\frac{3}{2}\ln^2(1-X)-3X\ln(1-X)+X^3-\frac{3}{2}X^2\)
\\ &&
+\frac{\beta_1^2\Gamma_1}{2\beta_0^3}\(\ln^2(1-X)+\frac{X^3}{3}-X^2\)-\frac{\beta_2\beta_1\Gamma_0}{2\beta_0^3}\(X\ln(1-X)+\frac{2}{3}X^3-X^2\)
\\\nn &&-\frac{\beta_1\Gamma_2}{2\beta_0^2}\(\ln(1-X)+\frac{X^3}{3}-X^2+X\)+\frac{X^2}{12\beta_0^2}\(\beta_3\Gamma_0(3-2X)+2\beta_2\Gamma_1(3-X)\)
\\\nn &&+\frac{\Gamma_3}{6\beta_0}X(3-3X+X^2)-\frac{2\beta_1 d^{(2,0)}}{\beta_0}\ln(1-X)+d^{(3,0)}\Big]
\end{eqnarray}
At $X\to1$ this expression has a pole, that is equivalent to Landau pole. This show the convergence radius of this expansion $\vec b\simeq 2e^{-\gamma_E}\Lambda_{QCD}^{-1}\approx 4$GeV$^{-1}$.

\section{Expression for $\zeta_\mu$}
\label{app:zeta-line}

The concept of the special null-evolution line plays a central role in $\zeta$-prescription. The $\zeta$-prescription, the double evolution and properties of TMD evolution have been elaborated in ref.~\cite{Scimemi:2018xaf}. In this appendix, we present the expressions for the special null-evolution line $\zeta_{\text{pert}}$ and $\zeta_{\text{NP}}$ that were used in the fit.

The definition of the special null-evolution line is discussed in the sec.~\ref{sec:evolution}. Parameterizing an equipotential line as $(\mu,\zeta_\mu(b))$, one finds the following equation
\begin{eqnarray}\label{app:sp-line}
\Gamma_{\text{cusp}}(\mu)\ln\(\frac{\mu^2}{\zeta_\mu(b)}\)-\gamma_V(\mu)=2\mathcal{D}(\mu,b) \frac{d \ln \zeta_\mu(b)}{d \ln \mu^2}.
\end{eqnarray}
The special null-evolution line is the line that passes thorough the saddle point $(\mu_0,\zeta_0)$ of the evolution field. The saddle point is defined as
\begin{eqnarray}\label{app:sp_boundary}
\mathcal{D}(\mu_0,b)=0,\qquad \gamma_F(\mu_0,\zeta_0)=0.
\end{eqnarray}
Note, that this boundary condition guarantees the finiteness of $\zeta_\mu(b)$ for all non-singular values of $\mu$.

Originally the $\zeta$-prescription has been implemented in the perturbative regime only \cite{Scimemi:2017etj}. This part is the most important since it gives the cancellation of double-logarithms in the matching coefficient. However, at large-$b$, non-perturbative corrections to RAD become large and can not be ignored (although they can be seen as a part of NP model, but it introduces an undesired correlation between $f_{NP}$ and $\mathcal{D}$). Therefore, one have to solve equation (\ref{app:sp-line}) with a generic non-perturbative RAD. Such solution can be found but its numerical implementation is problematic at very small-$b$. The problem is that it is very difficult to obtain the cancellation of \textit{perturbative} logarithms for the  \textit{exact} solution because at $b\to0$, because the numerical values of logarithms are huge. Therefore, a good practice is to use the perturbative solution at very small-$b$, (and hence cancel all logarithm exactly) and turn to the exact solution at larger $b$. This is implemented in the ansatz eq.~(\ref{NP:zeta}).

In the following sections we provide expressions for $\zeta_\mu^{\text{exact}}$ and $\zeta_\mu^{\text{pert}}$ that were used in the fit procedure.

\subsection{Perturbative expression}

The perturbative solution eq.~(\ref{app:sp-line}) is conveniently written as
\begin{eqnarray}
\zeta_\mu^{\text{pert}}(b)=\frac{\mu}{b}2e^{-\gamma_E}e^{v(\mu,b)},
\end{eqnarray}
where
\begin{eqnarray}\label{app:zeta-pert}
v(\mu,b)=\sum_{n=0}^\infty a_s^n(\mu) v_n(\mathbf{L}_\mu),
\end{eqnarray}
with $\mathbf{L}_\mu$ defined in eq.~(\ref{app:Lmu}). The general expression for $v_n$ can be found in \cite{Scimemi:2018xaf} (see eq.~(5.14)). We apply the boundary condition eq.~(\ref{app:sp_boundary}), which in the perturbative regime turns into requirement of finiteness of $v_n$ at $\mathbf{L}_\mu\to0$. The values of $v_n$ up to NNLO are
\begin{eqnarray}
v_0(\mathbf{L}_\mu)&=&\frac{\gamma_1}{\Gamma_0},
\\
v_1(\mathbf{L}_\mu)&=&\frac{\beta_0}{12}\mathbf{L}_\mu^2-\frac{\gamma_1\Gamma_1}{\Gamma_0^2}+\frac{\gamma_2+d^{(2,0)}}{\Gamma_0},
\\
v_2(\mathbf{L}_\mu)&=&\frac{\beta_0^2}{24}\mathbf{L}_\mu^3+\(\frac{\beta_1}{12}+\frac{\beta_0\Gamma_1}{\Gamma_0}\)\mathbf{L}_\mu^2+
\(\frac{\beta_0\gamma_2}{2\Gamma_0}+\frac{4\beta_0d^{(2,0)}}{3\Gamma_0}-\frac{\beta_0\gamma_1\Gamma_1}{2\Gamma_0^2}\)\mathbf{L}_\mu
\\\nn &&\qquad+\frac{\gamma_1\Gamma_1^2}{\Gamma_0^3}-\frac{\gamma_1\Gamma_2+\gamma_2\Gamma_1+d^{(2,0)}\Gamma_1}{\Gamma_0^2}+\frac{\gamma_3+d^{(3,0)}}{\Gamma_0}.
\end{eqnarray}
The definition of perturbative coefficients is given in eq.~(\ref{app:def:dnk}, \ref{app:def-beta-Gamma}) and
\begin{eqnarray}\label{app:gammaV}
\gamma_V(\mu)=\sum_{n=1}^\infty a_s^n(\mu) \gamma_n.
\end{eqnarray}
Similarly, to RAD the $\zeta_\mu^{\text{pert}}(b)$ can be resummed in terms of $a_s\mathbf{L}_\mu$ (see appendix A in \cite{Scimemi:2018xaf}). However, this is not necessary when using our ansatz eq.~(\ref{NP:zeta}), because the perturbative expression turns into exact much before the problems with convergence occur.

\subsection{Exact expression}

The evolution field, the equipotential line $\zeta_\mu$ and the position of the saddle point $(\mu_0,\zeta_0)$, depend on values of $b$, which is treated as a free parameter. 
This fact calls for some attention when implementating of the $\zeta$-prescription exactly. The lesser problem is that additional numerical computations are required to determine the position of saddle-point and the values of the line for different non-perturbative models of $\mathcal{D}$. The greater problem is that at larger $b$ the value of $\mu_0$ decreases and at some large value of $b$ (typically $b\sim 3$GeV$^{-1}$) $\mu_0$ is smaller than $\Lambda_{QCD}$.  Due to this behavior, it is impossible to determine the special null-evolution line at large-$b$ numerically. However, the special null-evolution line is still uniquely defined by the continuation from smaller values of $b$. 

In ref.~\cite{Vladimirov:2019bfa} a simple solution of this problem has been found. The central idea is to use the non-perturbative RAD as  generalized coordinates $(a_s,\mathcal{D})$ instead of  $(\mu,b)$. We introduce the function $g$ as
\begin{eqnarray}
\zeta_\mu^{\text{exact}}(b)=\mu^2 e^{-g(a_s(\mu),\mathcal{D}(\mu,b))/\mathcal{D}(\mu,b)}.
\end{eqnarray}
It satisfies the following linear equation in partial derivatives
\begin{eqnarray}\label{app:g-}
2\mathcal{D}+2\beta(a_s)\frac{\partial g(a_s,\mathcal{D})}{\partial a_s}-\Gamma_{\text{cusp}}(a_s)\frac{\partial g(a_s,\mathcal{D})}{\partial \mathcal{D}}+\gamma_V(a_s)=0.
\end{eqnarray}
The saddle point boundary condition eq.~(\ref{app:sp_boundary}) turns into
\begin{eqnarray}\label{app:g-boundary}
g(a_s,0)=0.
\end{eqnarray}
The equation (\ref{app:g-}) can be solved exactly, but the application of boundary condition eq.~(\ref{app:g-boundary}) requires the solution of functional equation with transcendental functions. 

On another hand, the values of $a_s$ used in the $\zeta$-prescription are always small, since they are evaluated at the hard scale $Q$, see (\ref{def:TMD-evolved}). Therefore, it is natural and numerically accurate to consider the expansion in $a_s$. Note, that such an expansion already incorporates the non-perturbative corrections exactly. Denoting
\begin{eqnarray}\label{app:g}
g(a_s,\mathcal{D})=\frac{1}{a_s}\frac{\Gamma_0}{2\beta_0^2}\sum_{n=0}^\infty a_s^n g_n(\mathcal{D}),
\end{eqnarray}
We find 
\begin{eqnarray}\label{app:g0}
g_0&=&e^{-p}-1+p,
\\
g_1&=&\frac{\beta_1}{\beta_0}\(e^{-p}-1+p-\frac{p^2}{2}\)-\frac{\Gamma_1}{\Gamma_0}(e^{-p}-1+p)+\frac{\beta_0\gamma_1}{\gamma_0}p,
\\\label{app:g2}
g_2&=&\(\frac{\Gamma_1^2}{\Gamma_0^2}-\frac{\Gamma_2}{\Gamma_0}\)(\cosh(p)-1)+\(\frac{\beta_1\Gamma_1}{\beta_0\Gamma_0}-\frac{\beta_2}{\beta_0}\)(\sinh(p)-p)
\\\nn && \qquad\qquad +\(\frac{\beta_0\gamma_2}{\Gamma_0}-\frac{\beta_0\gamma_1\Gamma_1}{\Gamma_0^2}\)(e^p-1),
\\\nn
g_3&=&\(\frac{\beta_1(\beta_1\Gamma_1-\beta_2\Gamma_0)}{12\beta_0^2\Gamma_0}-\frac{\beta_3\Gamma_0-2\beta_2\Gamma_1+\beta_1\Gamma_2}{12\beta_0\Gamma_0}-
\frac{\Gamma_1^3}{3\Gamma_0}+\frac{\Gamma_1\Gamma_2}{2\Gamma_0^2}-\frac{\Gamma_3}{6\Gamma_0}\)(e^{2p}+2e^{-p}-3)
\\\nn &&+\(\frac{\Gamma_1^3}{\Gamma_0^3}-\frac{\Gamma_1\Gamma_2}{\Gamma_0^2}-\frac{\beta_2\Gamma_1}{\beta_0\Gamma_0}\)(\cosh(p)-1)
+\frac{\beta_0}{\Gamma_0}\(\frac{\gamma_1\Gamma_1^2}{\Gamma_0^2}-\frac{\gamma_2\Gamma_1}{\Gamma_0}-\frac{\gamma_1\Gamma_2}{\Gamma_0}+\gamma_3\)e^p(e^p-1)
\\\label{app:g3} &&+\(\frac{\beta_1\gamma_2}{\Gamma_0}-\frac{\beta_1\gamma_1\Gamma_1}{\Gamma_0^2}-\frac{\beta_0\gamma_3}{\Gamma_0}+\frac{\beta_0\gamma_1\Gamma_2}{\Gamma_0^2}\)\frac{(e^p-1)^2}{2}+\frac{\beta_3}{2\beta_0}(e^{-p}+p-1)
\\\nn && + \frac{\beta_1}{2\beta_0}\(\frac{\beta_2}{\beta_0}-\frac{\beta_1\Gamma_1}{\beta_0\Gamma_0}+\frac{\Gamma_2}{\Gamma_0}\)(e^p-p-1).
\end{eqnarray}
where $p=2\beta_0 \mathcal{D}/\Gamma_0$. The expressions (\ref{app:g0}-\ref{app:g3}) provide a very accurate approximation, since $a_s$ is evaluated at $\mu=Q$ and typically $a_s=g^2/(4\pi)^2\sim 10^{-2}$. Most importantly  this expression is valid for all values of $b$, even when the saddle point is below $\Lambda_{QCD}$.

Let us mention that $g_2$ and $g_3$ exponentially grow at large-$\mathcal{D}$. It demonstrates that at large$-\mathcal{D}$ the series is an asymptotic series. However, this effect takes a place when $\mathcal{D}\sim 3-5$ which corresponds to  typical values for $b\sim 20-25$GeV$^{-1}$. It does not cause any problem but effectively cuts the Hankel integral (practically, the integration converges much earlier). 

\section{Correlation matrices in the numeric form}
\label{app:correlation}
\setcounter{MaxMatrixCols}{20}

Here we present the explicit expression for correlation matrices shown in fig.~\ref{fig:correlation}. The NNLO fit yields the following matrix
\begin{eqnarray}
\text{corr}_{\text{NNLO}}\!\!=\!\!
\left(
\begin{array}{rrrrrrrrrrr}
1~~ & -.632 & .179 & -.360 & .075 & .056 & .147 & -.553 & .259 & .197 & -.338
\\
-.632 & 1~~ & -.676 & .165 & .513 & .196 & -.075& .312  & -.104& .027 & -.159
\\ 
.179  & -.676& 1~~  &-.001 &-.825& .049& -.460& -.051 & .052 & .013 & .213
\\
-.360& .165 & -.001 & 1~~  & .097 & .163 &-.582 & .165  & .025 & .011 & .155
\\
.075 & .513& -.825 & .097 & 1~~   & .221 & .320 & -.047 & .118 & .066 & -.262
\\
.056 & .196& .049  & .163 & .221 & 1~~   & -.570& -.089 & .197 & .075 & -.101
\\
.147& -.075& -.460& -.582&  .320 & -.570& 1~~ & -.063 & -.100& -.179& .061
\\
-.553& .312& -.051& .165 & -.047& -.089& -.063& 1~~    & -.115 & -.024& .466
\\
.259& -.104&  .052& .025 & .118 &  .197& -.100& -.115& 1~~     & .292 & .252
\\
.197&  .027& .013 & .011 & .066 & .075& -.179 & -.024& .292 & 1~~   & -.590
\\
-.338& -.159& .213& .155 & -.262& -.101& .061 & .466 & .252 & -.590& 1~~
\end{array}
\right).\quad
\end{eqnarray}
The N$^3$LO fit yields the following matrix
\begin{eqnarray}
\text{corr}_{\text{N$^3$LO}}\!\!=\!\!
\left(
\begin{array}{rrrrrrrrrrr}
1~~     &  -.812 &  .606 &  -.038 &  -.374 &  .228 &  -.335 &  -.114 &  .387 &  .125 &  .381
\\
-.812 &  1~~     &  -.850&  -.016 &  .628  &  -.470&  .444  &  -.063 &  -.321&  -.086&  -.613
\\
.606  &  -.850 &  1~~    &  .023  &  -.781 &  .659 &  -.750 &  .182  &  .418 &  .087 & .726 
\\
-.038 &  -.016 &  .023 &  1~~     &  -.011 &  -.174&  -.256 &  .103  &  .266 &  .354 &  -.010
\\
-.374 &  .628  &  -.781&  -.011 &  1~~     &  -.343&  .625  &  -.128 &  -.374&  -.097&  -.610
\\ 
.228  &  -.470 &  .659 &  -.174 &  -.343 &  1~~    &  -.724 &  .282  &  .204 &  -.061&  .646
\\
-.335 &  .444  &  -.750&  -.256 &  .625  &  -.724&  1~~     &  -.261 &  -.511&  -.167&  -.638
\\
-.114 &  -.063 &  .182 &  .103  &  -.128 &  .282 &  -.261 &  1~~     &  .438 &  .487 &  .560
\\
 .387 &  -.321 &  .418 &  .266  &  -.374 &  .204 &  -.511 &  .438  &  1~~    &  .798 &  .574
\\ 
.125  &  -.086 &  .087 &  .354  &  -.097 &  -.061&  -.167 &  .487  &  .798 &  1~~    &  .159
\\
 .381 &  -.613 &  .726 &  -.010 &  -.610 &  .646 &  -.638 &  .560  &  .574 &  .159 &  1~~
\end{array}
\right).\quad
\end{eqnarray}
The enumeration of rows and columns in matrices corresponds to the NP parameters ordered as 
$\{B_{\text{NP}},c_0,\lambda_1,\lambda_2,\lambda_3,\lambda_4,\lambda_5,\eta_1,\eta_2,\eta_3,\eta_4\}$.

\bibliographystyle{JHEP}
\bibliography{TMD_ref}

\providecommand{\href}[2]{#2}\begingroup\raggedright\begin{thebibliography}{100}

\bibitem{Collins:1989gx}
J.~C. Collins, D.~E. Soper and G.~F. Sterman, \emph{{Factorization of Hard
  Processes in QCD}},
  \href{http://dx.doi.org/10.1142/9789814503266_0001}{\emph{Adv. Ser. Direct.
  High Energy Phys.} {\bfseries 5} (1989) 1--91},
  [\href{https://arxiv.org/abs/hep-ph/0409313}{{\ttfamily hep-ph/0409313}}].

\bibitem{Bacchetta:2006tn}
A.~Bacchetta, M.~Diehl, K.~Goeke, A.~Metz, P.~J. Mulders and M.~Schlegel,
  \emph{{Semi-inclusive deep inelastic scattering at small transverse
  momentum}},
  \href{http://dx.doi.org/10.1088/1126-6708/2007/02/093}{\emph{JHEP} {\bfseries
  02} (2007) 093}, [\href{https://arxiv.org/abs/hep-ph/0611265}{{\ttfamily
  hep-ph/0611265}}].

\bibitem{Bacchetta:2008xw}
A.~Bacchetta, D.~Boer, M.~Diehl and P.~J. Mulders, \emph{{Matches and
  mismatches in the descriptions of semi-inclusive processes at low and high
  transverse momentum}},
  \href{http://dx.doi.org/10.1088/1126-6708/2008/08/023}{\emph{JHEP} {\bfseries
  08} (2008) 023}, [\href{https://arxiv.org/abs/0803.0227}{{\ttfamily
  0803.0227}}].

\bibitem{Becher:2010tm}
T.~Becher and M.~Neubert, \emph{{{Drell-Yan} Production at Small $q_T$,
  Transverse Parton Distributions and the Collinear Anomaly}},
  \href{http://dx.doi.org/10.1140/epjc/s10052-011-1665-7}{\emph{Eur. Phys. J.}
  {\bfseries C71} (2011) 1665},
  [\href{https://arxiv.org/abs/1007.4005}{{\ttfamily 1007.4005}}].

\bibitem{Collins:2011zzd}
J.~Collins, \emph{{Foundations of perturbative QCD}}.
\newblock Cambridge University Press, 2013.

\bibitem{GarciaEchevarria:2011rb}
M.~G. Echevarria, A.~Idilbi and I.~Scimemi, \emph{{Factorization Theorem For
  Drell-Yan At Low $q_T$ And Transverse Momentum Distributions
  On-The-Light-Cone}},
  \href{http://dx.doi.org/10.1007/JHEP07(2012)002}{\emph{JHEP} {\bfseries 07}
  (2012) 002}, [\href{https://arxiv.org/abs/1111.4996}{{\ttfamily 1111.4996}}].

\bibitem{Echevarria:2012js}
M.~G. Echevarria, A.~Idilbi and I.~Scimemi, \emph{{Soft and Collinear
  Factorization and Transverse Momentum Dependent Parton Distribution
  Functions}},
  \href{http://dx.doi.org/10.1016/j.physletb.2013.09.003}{\emph{Phys. Lett.}
  {\bfseries B726} (2013) 795--801},
  [\href{https://arxiv.org/abs/1211.1947}{{\ttfamily 1211.1947}}].

\bibitem{Echevarria:2014rua}
M.~G. Echevarria, A.~Idilbi and I.~Scimemi, \emph{{Unified treatment of the QCD
  evolution of all (un-)polarized transverse momentum dependent functions:
  Collins function as a study case}},
  \href{http://dx.doi.org/10.1103/PhysRevD.90.014003}{\emph{Phys. Rev.}
  {\bfseries D90} (2014) 014003},
  [\href{https://arxiv.org/abs/1402.0869}{{\ttfamily 1402.0869}}].

\bibitem{Chiu:2012ir}
J.-Y. Chiu, A.~Jain, D.~Neill and I.~Z. Rothstein, \emph{{A Formalism for the
  Systematic Treatment of Rapidity Logarithms in Quantum Field Theory}},
  \href{http://dx.doi.org/10.1007/JHEP05(2012)084}{\emph{JHEP} {\bfseries 05}
  (2012) 084}, [\href{https://arxiv.org/abs/1202.0814}{{\ttfamily 1202.0814}}].

\bibitem{Vladimirov:2017ksc}
A.~Vladimirov, \emph{{Structure of rapidity divergences in multi-parton
  scattering soft factors}},
  \href{http://dx.doi.org/10.1007/JHEP04(2018)045}{\emph{JHEP} {\bfseries 04}
  (2018) 045}, [\href{https://arxiv.org/abs/1707.07606}{{\ttfamily
  1707.07606}}].

\bibitem{Scimemi:2018xaf}
I.~Scimemi and A.~Vladimirov, \emph{{Systematic analysis of double-scale
  evolution}}, \href{http://dx.doi.org/10.1007/JHEP08(2018)003}{\emph{JHEP}
  {\bfseries 08} (2018) 003},
  [\href{https://arxiv.org/abs/1803.11089}{{\ttfamily 1803.11089}}].

\bibitem{Angeles-Martinez:2015sea}
R.~Angeles-Martinez et~al., \emph{{Transverse Momentum Dependent (TMD) parton
  distribution functions: status and prospects}},
  \href{http://dx.doi.org/10.5506/APhysPolB.46.2501}{\emph{Acta Phys. Polon.}
  {\bfseries B46} (2015) 2501--2534},
  [\href{https://arxiv.org/abs/1507.05267}{{\ttfamily 1507.05267}}].

\bibitem{Sun:2013hua}
P.~Sun and F.~Yuan, \emph{{Transverse momentum dependent evolution: Matching
  semi-inclusive deep inelastic scattering processes to Drell-Yan and W/Z boson
  production}}, \href{http://dx.doi.org/10.1103/PhysRevD.88.114012}{\emph{Phys.
  Rev.} {\bfseries D88} (2013) 114012},
  [\href{https://arxiv.org/abs/1308.5003}{{\ttfamily 1308.5003}}].

\bibitem{Anselmino:2013lza}
M.~Anselmino, M.~Boglione, J.~O. Gonzalez~Hernandez, S.~Melis and A.~Prokudin,
  \emph{{Unpolarised Transverse Momentum Dependent Distribution and
  Fragmentation Functions from SIDIS Multiplicities}},
  \href{http://dx.doi.org/10.1007/JHEP04(2014)005}{\emph{JHEP} {\bfseries 04}
  (2014) 005}, [\href{https://arxiv.org/abs/1312.6261}{{\ttfamily 1312.6261}}].

\bibitem{Signori:2013mda}
A.~Signori, A.~Bacchetta, M.~Radici and G.~Schnell, \emph{{Investigations into
  the flavor dependence of partonic transverse momentum}},
  \href{http://dx.doi.org/10.1007/JHEP11(2013)194}{\emph{JHEP} {\bfseries 11}
  (2013) 194}, [\href{https://arxiv.org/abs/1309.3507}{{\ttfamily 1309.3507}}].

\bibitem{DAlesio:2014mrz}
U.~D'Alesio, M.~G. Echevarria, S.~Melis and I.~Scimemi, \emph{{Non-perturbative
  QCD effects in $q_{T}$ spectra of Drell-Yan and Z-boson production}},
  \href{http://dx.doi.org/10.1007/JHEP11(2014)098}{\emph{JHEP} {\bfseries 11}
  (2014) 098}, [\href{https://arxiv.org/abs/1407.3311}{{\ttfamily 1407.3311}}].

\bibitem{Aidala:2014hva}
C.~A. Aidala, B.~Field, L.~P. Gamberg and T.~C. Rogers, \emph{{Limits on
  transverse momentum dependent evolution from semi-inclusive deep inelastic
  scattering at moderate $Q$}},
  \href{http://dx.doi.org/10.1103/PhysRevD.89.094002}{\emph{Phys. Rev.}
  {\bfseries D89} (2014) 094002},
  [\href{https://arxiv.org/abs/1401.2654}{{\ttfamily 1401.2654}}].

\bibitem{Bacchetta:2017gcc}
A.~Bacchetta, F.~Delcarro, C.~Pisano, M.~Radici and A.~Signori,
  \emph{{Extraction of partonic transverse momentum distributions from
  semi-inclusive deep-inelastic scattering, Drell-Yan and Z-boson production}},
  \href{http://dx.doi.org/10.1007/JHEP06(2017)081}{\emph{JHEP} {\bfseries 06}
  (2017) 081}, [\href{https://arxiv.org/abs/1703.10157}{{\ttfamily
  1703.10157}}].

\bibitem{Scimemi:2017etj}
I.~Scimemi and A.~Vladimirov, \emph{{Analysis of vector boson production within
  TMD factorization}},
  \href{http://dx.doi.org/10.1140/epjc/s10052-018-5557-y}{\emph{Eur. Phys. J.}
  {\bfseries C78} (2018) 89},
  [\href{https://arxiv.org/abs/1706.01473}{{\ttfamily 1706.01473}}].

\bibitem{Bertone:2019nxa}
V.~Bertone, I.~Scimemi and A.~Vladimirov, \emph{{Extraction of unpolarized
  quark transverse momentum dependent parton distributions from
  Drell-Yan/Z-boson production}},
  \href{http://dx.doi.org/10.1007/JHEP06(2019)028}{\emph{JHEP} {\bfseries 06}
  (2019) 028}, [\href{https://arxiv.org/abs/1902.08474}{{\ttfamily
  1902.08474}}].

\bibitem{Vladimirov:2019bfa}
A.~Vladimirov, \emph{{Pion-induced Drell-Yan processes within TMD
  factorization}},  \href{https://arxiv.org/abs/1907.10356}{{\ttfamily
  1907.10356}}.

\bibitem{Bacchetta:2019sam}
A.~Bacchetta, V.~Bertone, C.~Bissolotti, G.~Bozzi, F.~Delcarro, F.~Piacenza
  et~al., \emph{{Transverse-momentum-dependent parton distributions up to
  N$^3$LL from Drell-Yan data}},
  \href{https://arxiv.org/abs/1912.07550}{{\ttfamily 1912.07550}}.

\bibitem{Gehrmann:2014yya}
T.~Gehrmann, T.~Luebbert and L.~L. Yang, \emph{{Calculation of the transverse
  parton distribution functions at next-to-next-to-leading order}},
  \href{http://dx.doi.org/10.1007/JHEP06(2014)155}{\emph{JHEP} {\bfseries 06}
  (2014) 155}, [\href{https://arxiv.org/abs/1403.6451}{{\ttfamily 1403.6451}}].

\bibitem{Echevarria:2015byo}
M.~G. Echevarria, I.~Scimemi and A.~Vladimirov, \emph{{Universal transverse
  momentum dependent soft function at NNLO}},
  \href{http://dx.doi.org/10.1103/PhysRevD.93.054004}{\emph{Phys. Rev.}
  {\bfseries D93} (2016) 054004},
  [\href{https://arxiv.org/abs/1511.05590}{{\ttfamily 1511.05590}}].

\bibitem{Echevarria:2015usa}
M.~G. Echevarria, I.~Scimemi and A.~Vladimirov, \emph{{Transverse momentum
  dependent fragmentation function at next-to-next-to leading order}},
  \href{http://dx.doi.org/10.1103/PhysRevD.93.011502,
  10.1103/PhysRevD.94.099904}{\emph{Phys. Rev.} {\bfseries D93} (2016) 011502},
  [\href{https://arxiv.org/abs/1509.06392}{{\ttfamily 1509.06392}}].

\bibitem{Echevarria:2016scs}
M.~G. Echevarria, I.~Scimemi and A.~Vladimirov, \emph{{Unpolarized Transverse
  Momentum Dependent Parton Distribution and Fragmentation Functions at
  next-to-next-to-leading order}},
  \href{http://dx.doi.org/10.1007/JHEP09(2016)004}{\emph{JHEP} {\bfseries 09}
  (2016) 004}, [\href{https://arxiv.org/abs/1604.07869}{{\ttfamily
  1604.07869}}].

\bibitem{Li:2016ctv}
Y.~Li and H.~X. Zhu, \emph{{Bootstrapping Rapidity Anomalous Dimensions for
  Transverse-Momentum Resummation}},
  \href{http://dx.doi.org/10.1103/PhysRevLett.118.022004}{\emph{Phys. Rev.
  Lett.} {\bfseries 118} (2017) 022004},
  [\href{https://arxiv.org/abs/1604.01404}{{\ttfamily 1604.01404}}].

\bibitem{Vladimirov:2016dll}
A.~A. Vladimirov, \emph{{Soft-/rapidity- anomalous dimensions correspondence}},
  \href{http://dx.doi.org/10.1103/PhysRevLett.118.062001}{\emph{Phys. Rev.
  Lett.} {\bfseries 118} (2017) 062001},
  [\href{https://arxiv.org/abs/1610.05791}{{\ttfamily 1610.05791}}].

\bibitem{Luo:2019hmp}
M.-X. Luo, X.~Wang, X.~Xu, L.~L. Yang, T.-Z. Yang and H.~X. Zhu,
  \emph{{Transverse Parton Distribution and Fragmentation Functions at NNLO:
  the Quark Case}},  \href{https://arxiv.org/abs/1908.03831}{{\ttfamily
  1908.03831}}.

\bibitem{Gehrmann:2010ue}
T.~Gehrmann, E.~W.~N. Glover, T.~Huber, N.~Ikizlerli and C.~Studerus,
  \emph{{Calculation of the quark and gluon form factors to three loops in
  QCD}}, \href{http://dx.doi.org/10.1007/JHEP06(2010)094}{\emph{JHEP}
  {\bfseries 06} (2010) 094},
  [\href{https://arxiv.org/abs/1004.3653}{{\ttfamily 1004.3653}}].

\bibitem{Baikov:2016tgj}
P.~A. Baikov, K.~G. Chetyrkin and J.~H. Kühn, \emph{{Five-Loop Running of the
  QCD coupling constant}},
  \href{http://dx.doi.org/10.1103/PhysRevLett.118.082002}{\emph{Phys. Rev.
  Lett.} {\bfseries 118} (2017) 082002},
  [\href{https://arxiv.org/abs/1606.08659}{{\ttfamily 1606.08659}}].

\bibitem{Moch:2017uml}
S.~Moch, B.~Ruijl, T.~Ueda, J.~A.~M. Vermaseren and A.~Vogt, \emph{{Four-Loop
  Non-Singlet Splitting Functions in the Planar Limit and Beyond}},
  \href{http://dx.doi.org/10.1007/JHEP10(2017)041}{\emph{JHEP} {\bfseries 10}
  (2017) 041}, [\href{https://arxiv.org/abs/1707.08315}{{\ttfamily
  1707.08315}}].

\bibitem{Moch:2018wjh}
S.~Moch, B.~Ruijl, T.~Ueda, J.~A.~M. Vermaseren and A.~Vogt, \emph{{On quartic
  colour factors in splitting functions and the gluon cusp anomalous
  dimension}},
  \href{http://dx.doi.org/10.1016/j.physletb.2018.06.017}{\emph{Phys. Lett.}
  {\bfseries B782} (2018) 627--632},
  [\href{https://arxiv.org/abs/1805.09638}{{\ttfamily 1805.09638}}].

\bibitem{Lee:2019zop}
R.~N. Lee, A.~V. Smirnov, V.~A. Smirnov and M.~Steinhauser, \emph{{Four-loop
  quark form factor with quartic fundamental colour factor}},
  \href{http://dx.doi.org/10.1007/JHEP02(2019)172}{\emph{JHEP} {\bfseries 02}
  (2019) 172}, [\href{https://arxiv.org/abs/1901.02898}{{\ttfamily
  1901.02898}}].

\bibitem{Landry:2002ix}
F.~Landry, R.~Brock, P.~M. Nadolsky and C.~P. Yuan, \emph{{Tevatron Run-1 $Z$
  boson data and Collins-Soper-Sterman resummation formalism}},
  \href{http://dx.doi.org/10.1103/PhysRevD.67.073016}{\emph{Phys. Rev.}
  {\bfseries D67} (2003) 073016},
  [\href{https://arxiv.org/abs/hep-ph/0212159}{{\ttfamily hep-ph/0212159}}].

\bibitem{Qiu:2000hf}
J.-w. Qiu and X.-f. Zhang, \emph{{Role of the nonperturbative input in QCD
  resummed Drell-Yan $Q_{T}$ distributions}},
  \href{http://dx.doi.org/10.1103/PhysRevD.63.114011}{\emph{Phys. Rev.}
  {\bfseries D63} (2001) 114011},
  [\href{https://arxiv.org/abs/hep-ph/0012348}{{\ttfamily hep-ph/0012348}}].

\bibitem{Bozzi:2008bb}
G.~Bozzi, S.~Catani, G.~Ferrera, D.~de~Florian and M.~Grazzini,
  \emph{{Transverse-momentum resummation: A Perturbative study of Z production
  at the Tevatron}},
  \href{http://dx.doi.org/10.1016/j.nuclphysb.2009.02.014}{\emph{Nucl. Phys.}
  {\bfseries B815} (2009) 174--197},
  [\href{https://arxiv.org/abs/0812.2862}{{\ttfamily 0812.2862}}].

\bibitem{Catani:2012qa}
S.~Catani, L.~Cieri, D.~de~Florian, G.~Ferrera and M.~Grazzini, \emph{{Vector
  boson production at hadron colliders: hard-collinear coefficients at the
  NNLO}}, \href{http://dx.doi.org/10.1140/epjc/s10052-012-2195-7}{\emph{Eur.
  Phys. J.} {\bfseries C72} (2012) 2195},
  [\href{https://arxiv.org/abs/1209.0158}{{\ttfamily 1209.0158}}].

\bibitem{Bizon:2018foh}
W.~Bizon, X.~Chen, A.~Gehrmann-De~Ridder, T.~Gehrmann, N.~Glover, A.~Huss
  et~al., \emph{{Fiducial distributions in Higgs and Drell-Yan production at
  N$^{3}$LL+NNLO}},
  \href{http://dx.doi.org/10.1007/JHEP12(2018)132}{\emph{JHEP} {\bfseries 12}
  (2018) 132}, [\href{https://arxiv.org/abs/1805.05916}{{\ttfamily
  1805.05916}}].

\bibitem{Bizon:2019zgf}
W.~Bizon, A.~Gehrmann-De~Ridder, T.~Gehrmann, N.~Glover, A.~Huss, P.~F. Monni
  et~al., \emph{{The transverse momentum spectrum of weak gauge bosons at N
  ${}^3$ LL + NNLO}},
  \href{http://dx.doi.org/10.1140/epjc/s10052-019-7324-0}{\emph{Eur. Phys. J.}
  {\bfseries C79} (2019) 868},
  [\href{https://arxiv.org/abs/1905.05171}{{\ttfamily 1905.05171}}].

\bibitem{Collins:1981va}
J.~C. Collins and D.~E. Soper, \emph{{Back-To-Back Jets: Fourier Transform from
  B to K-Transverse}},
  \href{http://dx.doi.org/10.1016/0550-3213(82)90453-9}{\emph{Nucl. Phys.}
  {\bfseries B197} (1982) 446--476}.

\bibitem{Aybat:2011zv}
S.~M. Aybat and T.~C. Rogers, \emph{{TMD Parton Distribution and Fragmentation
  Functions with QCD Evolution}},
  \href{http://dx.doi.org/10.1103/PhysRevD.83.114042}{\emph{Phys. Rev.}
  {\bfseries D83} (2011) 114042},
  [\href{https://arxiv.org/abs/1101.5057}{{\ttfamily 1101.5057}}].

\bibitem{web}
``\texttt{artemide} web-page, https://teorica.fis.ucm.es/artemide/ \\
  \texttt{artemide} repository,
  https://github.com/vladimirovalexey/artemide-public.''

\bibitem{Anselmino:2005nn}
M.~Anselmino, M.~Boglione, U.~D'Alesio, A.~Kotzinian, F.~Murgia and
  A.~Prokudin, \emph{{The Role of Cahn and sivers effects in deep inelastic
  scattering}}, \href{http://dx.doi.org/10.1103/PhysRevD.71.074006}{\emph{Phys.
  Rev.} {\bfseries D71} (2005) 074006},
  [\href{https://arxiv.org/abs/hep-ph/0501196}{{\ttfamily hep-ph/0501196}}].

\bibitem{Olive:2016xmw}
{\scshape Particle Data Group} collaboration, C.~Patrignani et~al.,
  \emph{{Review of Particle Physics}},
  \href{http://dx.doi.org/10.1088/1674-1137/40/10/100001}{\emph{Chin. Phys.}
  {\bfseries C40} (2016) 100001}.

\bibitem{Aad:2014xaa}
{\scshape ATLAS} collaboration, G.~Aad et~al., \emph{{Measurement of the
  $Z/\gamma^*$ boson transverse momentum distribution in $pp$ collisions at
  $\sqrt{s}$ = 7 TeV with the ATLAS detector}},
  \href{http://dx.doi.org/10.1007/JHEP09(2014)145}{\emph{JHEP} {\bfseries 09}
  (2014) 145}, [\href{https://arxiv.org/abs/1406.3660}{{\ttfamily 1406.3660}}].

\bibitem{Aad:2015auj}
{\scshape ATLAS} collaboration, G.~Aad et~al., \emph{{Measurement of the
  transverse momentum and $\phi ^*_{\eta }$ distributions of Drell-Yan lepton
  pairs in proton-proton collisions at $\sqrt{s}=8$ TeV with the ATLAS
  detector}},
  \href{http://dx.doi.org/10.1140/epjc/s10052-016-4070-4}{\emph{Eur. Phys. J.}
  {\bfseries C76} (2016) 291},
  [\href{https://arxiv.org/abs/1512.02192}{{\ttfamily 1512.02192}}].

\bibitem{Chatrchyan:2011wt}
{\scshape CMS} collaboration, S.~Chatrchyan et~al., \emph{{Measurement of the
  Rapidity and Transverse Momentum Distributions of $Z$ Bosons in $pp$
  Collisions at $\sqrt{s}=7$ TeV}},
  \href{http://dx.doi.org/10.1103/PhysRevD.85.032002}{\emph{Phys. Rev.}
  {\bfseries D85} (2012) 032002},
  [\href{https://arxiv.org/abs/1110.4973}{{\ttfamily 1110.4973}}].

\bibitem{Khachatryan:2016nbe}
{\scshape CMS} collaboration, V.~Khachatryan et~al., \emph{{Measurement of the
  transverse momentum spectra of weak vector bosons produced in proton-proton
  collisions at $ \sqrt{s}=8 $ TeV}},
  \href{http://dx.doi.org/10.1007/JHEP02(2017)096}{\emph{JHEP} {\bfseries 02}
  (2017) 096}, [\href{https://arxiv.org/abs/1606.05864}{{\ttfamily
  1606.05864}}].

\bibitem{Bastami:2018xqd}
S.~Bastami et~al., \emph{{Semi-Inclusive Deep Inelastic Scattering in
  Wandzura-Wilczek-type approximation}},
  \href{http://dx.doi.org/10.1007/JHEP06(2019)007}{\emph{JHEP} {\bfseries 06}
  (2019) 007}, [\href{https://arxiv.org/abs/1807.10606}{{\ttfamily
  1807.10606}}].

\bibitem{Balitsky:2017gis}
I.~Balitsky and A.~Tarasov, \emph{{Power corrections to TMD factorization for
  Z-boson production}},
  \href{http://dx.doi.org/10.1007/JHEP05(2018)150}{\emph{JHEP} {\bfseries 05}
  (2018) 150}, [\href{https://arxiv.org/abs/1712.09389}{{\ttfamily
  1712.09389}}].

\bibitem{Boer:2006eq}
D.~Boer and W.~Vogelsang, \emph{{Drell-Yan lepton angular distribution at small
  transverse momentum}},
  \href{http://dx.doi.org/10.1103/PhysRevD.74.014004}{\emph{Phys. Rev.}
  {\bfseries D74} (2006) 014004},
  [\href{https://arxiv.org/abs/hep-ph/0604177}{{\ttfamily hep-ph/0604177}}].

\bibitem{Mulders:1995dh}
P.~J. Mulders and R.~D. Tangerman, \emph{{The Complete tree level result up to
  order 1/Q for polarized deep inelastic leptoproduction}},
  \href{http://dx.doi.org/10.1016/S0550-3213(96)00648-7,
  10.1016/0550-3213(95)00632-X}{\emph{Nucl. Phys.} {\bfseries B461} (1996)
  197--237}, [\href{https://arxiv.org/abs/hep-ph/9510301}{{\ttfamily
  hep-ph/9510301}}].

\bibitem{Bacchetta:2004jz}
A.~Bacchetta, U.~D'Alesio, M.~Diehl and C.~A. Miller, \emph{{Single-spin
  asymmetries: The Trento conventions}},
  \href{http://dx.doi.org/10.1103/PhysRevD.70.117504}{\emph{Phys. Rev.}
  {\bfseries D70} (2004) 117504},
  [\href{https://arxiv.org/abs/hep-ph/0410050}{{\ttfamily hep-ph/0410050}}].

\bibitem{Arnold:2008kf}
S.~Arnold, A.~Metz and M.~Schlegel, \emph{{Dilepton production from polarized
  hadron hadron collisions}},
  \href{http://dx.doi.org/10.1103/PhysRevD.79.034005}{\emph{Phys. Rev.}
  {\bfseries D79} (2009) 034005},
  [\href{https://arxiv.org/abs/0809.2262}{{\ttfamily 0809.2262}}].

\bibitem{Nefedov:2018vyt}
M.~Nefedov and V.~Saleev, \emph{{Off-shell initial state effects, gauge
  invariance and angular distributions in the Drell–Yan process}},
  \href{http://dx.doi.org/10.1016/j.physletb.2018.12.071}{\emph{Phys. Lett.}
  {\bfseries B790} (2019) 551--556},
  [\href{https://arxiv.org/abs/1810.04061}{{\ttfamily 1810.04061}}].

\bibitem{Echevarria:2012pw}
M.~G. Echevarria, A.~Idilbi, A.~Schafer and I.~Scimemi,
  \emph{{Model-Independent Evolution of Transverse Momentum Dependent
  Distribution Functions (TMDs) at NNLL}},
  \href{http://dx.doi.org/10.1140/epjc/s10052-013-2636-y}{\emph{Eur. Phys. J.}
  {\bfseries C73} (2013) 2636},
  [\href{https://arxiv.org/abs/1208.1281}{{\ttfamily 1208.1281}}].

\bibitem{Scimemi:2019gge}
I.~Scimemi, A.~Tarasov and A.~Vladimirov, \emph{{Collinear matching for Sivers
  function at next-to-leading order}},
  \href{https://arxiv.org/abs/1901.04519}{{\ttfamily 1901.04519}}.

\bibitem{Moch:1999eb}
S.~Moch and J.~A.~M. Vermaseren, \emph{{Deep inelastic structure functions at
  two loops}},
  \href{http://dx.doi.org/10.1016/S0550-3213(00)00045-6}{\emph{Nucl. Phys.}
  {\bfseries B573} (2000) 853--907},
  [\href{https://arxiv.org/abs/hep-ph/9912355}{{\ttfamily hep-ph/9912355}}].

\bibitem{Stratmann:1996hn}
M.~Stratmann and W.~Vogelsang, \emph{{Next-to-leading order evolution of
  polarized and unpolarized fragmentation functions}},
  \href{http://dx.doi.org/10.1016/S0550-3213(97)00182-X}{\emph{Nucl. Phys.}
  {\bfseries B496} (1997) 41--65},
  [\href{https://arxiv.org/abs/hep-ph/9612250}{{\ttfamily hep-ph/9612250}}].

\bibitem{Scimemi:2016ffw}
I.~Scimemi and A.~Vladimirov, \emph{{Power corrections and renormalons in
  Transverse Momentum Distributions}},
  \href{http://dx.doi.org/10.1007/JHEP03(2017)002}{\emph{JHEP} {\bfseries 03}
  (2017) 002}, [\href{https://arxiv.org/abs/1609.06047}{{\ttfamily
  1609.06047}}].

\bibitem{Tafat:2001in}
S.~Tafat, \emph{{Nonperturbative corrections to the Drell-Yan transverse
  momentum distribution}},
  \href{http://dx.doi.org/10.1088/1126-6708/2001/05/004}{\emph{JHEP} {\bfseries
  05} (2001) 004}, [\href{https://arxiv.org/abs/hep-ph/0102237}{{\ttfamily
  hep-ph/0102237}}].

\bibitem{Vladimirov:2020umg}
A.~A. Vladimirov, \emph{{Self-contained definition of Collins-Soper kernel}},
  \href{https://arxiv.org/abs/2003.02288}{{\ttfamily 2003.02288}}.

\bibitem{Hautmann:2020cyp}
F.~Hautmann, I.~Scimemi and A.~Vladimirov, \emph{{Non-perturbative
  contributions to vector-boson transverse momentum spectra in hadronic
  collisions}},  \href{https://arxiv.org/abs/2002.12810}{{\ttfamily
  2002.12810}}.

\bibitem{Collins:2014jpa}
J.~Collins and T.~Rogers, \emph{{Understanding the large-distance behavior of
  transverse-momentum-dependent parton densities and the Collins-Soper
  evolution kernel}},
  \href{http://dx.doi.org/10.1103/PhysRevD.91.074020}{\emph{Phys. Rev.}
  {\bfseries D91} (2015) 074020},
  [\href{https://arxiv.org/abs/1412.3820}{{\ttfamily 1412.3820}}].

\bibitem{Airapetian:2012ki}
{\scshape HERMES} collaboration, A.~Airapetian et~al., \emph{{Multiplicities of
  charged pions and kaons from semi-inclusive deep-inelastic scattering by the
  proton and the deuteron}},
  \href{http://dx.doi.org/10.1103/PhysRevD.87.074029}{\emph{Phys. Rev.}
  {\bfseries D87} (2013) 074029},
  [\href{https://arxiv.org/abs/1212.5407}{{\ttfamily 1212.5407}}].

\bibitem{Aghasyan:2017ctw}
{\scshape COMPASS} collaboration, M.~Aghasyan et~al.,
  \emph{{Transverse-momentum-dependent Multiplicities of Charged Hadrons in
  Muon-Deuteron Deep Inelastic Scattering}},
  \href{http://dx.doi.org/10.1103/PhysRevD.97.032006}{\emph{Phys. Rev.}
  {\bfseries D97} (2018) 032006},
  [\href{https://arxiv.org/abs/1709.07374}{{\ttfamily 1709.07374}}].

\bibitem{Derrick:1995xg}
{\scshape ZEUS} collaboration, M.~Derrick et~al., \emph{{Inclusive charged
  particle distributions in deep inelastic scattering events at HERA}},
  \href{http://dx.doi.org/10.1007/s002880050075}{\emph{Z. Phys.} {\bfseries
  C70} (1996) 1--16}, [\href{https://arxiv.org/abs/hep-ex/9511010}{{\ttfamily
  hep-ex/9511010}}].

\bibitem{Adloff:1996dy}
{\scshape H1} collaboration, C.~Adloff et~al., \emph{{Measurement of charged
  particle transverse momentum spectra in deep inelastic scattering}},
  \href{http://dx.doi.org/10.1016/S0550-3213(96)00675-X}{\emph{Nucl. Phys.}
  {\bfseries B485} (1997) 3--24},
  [\href{https://arxiv.org/abs/hep-ex/9610006}{{\ttfamily hep-ex/9610006}}].

\bibitem{Asaturyan:2011mq}
R.~Asaturyan et~al., \emph{{Semi-Inclusive Charged-Pion Electroproduction off
  Protons and Deuterons: Cross Sections, Ratios and Access to the Quark-Parton
  Model at Low Energies}},
  \href{http://dx.doi.org/10.1103/PhysRevC.85.015202}{\emph{Phys. Rev.}
  {\bfseries C85} (2012) 015202},
  [\href{https://arxiv.org/abs/1103.1649}{{\ttfamily 1103.1649}}].

\bibitem{Adolph:2013stb}
{\scshape COMPASS} collaboration, C.~Adolph et~al., \emph{{Hadron Transverse
  Momentum Distributions in Muon Deep Inelastic Scattering at 160 GeV/$c$}},
  \href{http://dx.doi.org/10.1140/epjc/s10052-013-2531-6,
  10.1140/epjc/s10052-014-3255-y}{\emph{Eur. Phys. J.} {\bfseries C73} (2013)
  2531}, [\href{https://arxiv.org/abs/1305.7317}{{\ttfamily 1305.7317}}].

\bibitem{Ito:1980ev}
A.~S. Ito et~al., \emph{{Measurement of the Continuum of Dimuons Produced in
  High-Energy Proton - Nucleus Collisions}},
  \href{http://dx.doi.org/10.1103/PhysRevD.23.604}{\emph{Phys. Rev.} {\bfseries
  D23} (1981) 604--633}.

\bibitem{Moreno:1990sf}
G.~Moreno et~al., \emph{{Dimuon production in proton - copper collisions at
  $\sqrt{s}$ = 38.8-GeV}},
  \href{http://dx.doi.org/10.1103/PhysRevD.43.2815}{\emph{Phys. Rev.}
  {\bfseries D43} (1991) 2815--2836}.

\bibitem{McGaughey:1994dx}
{\scshape E772} collaboration, P.~L. McGaughey et~al., \emph{{Cross-sections
  for the production of high mass muon pairs from 800-GeV proton bombardment of
  H-2}}, \href{http://dx.doi.org/10.1103/PhysRevD.50.3038,
  10.1103/PhysRevD.60.119903}{\emph{Phys. Rev.} {\bfseries D50} (1994)
  3038--3045}.

\bibitem{Aidala:2018ajl}
{\scshape PHENIX} collaboration, C.~Aidala et~al., \emph{{Measurements of
  $\mu\mu$ pairs from open heavy flavor and Drell-Yan in $p+p$ collisions at
  $\sqrt{s}=200$ GeV}}, {\emph{Submitted to: Phys. Rev. D} (2018) },
  [\href{https://arxiv.org/abs/1805.02448}{{\ttfamily 1805.02448}}].

\bibitem{Affolder:1999jh}
{\scshape CDF} collaboration, T.~Affolder et~al., \emph{{The transverse
  momentum and total cross section of $e^+e^-$ pairs in the $Z$ boson region
  from $p\bar{p}$ collisions at $\sqrt{s} = 1.8$ TeV}},
  \href{http://dx.doi.org/10.1103/PhysRevLett.84.845}{\emph{Phys. Rev. Lett.}
  {\bfseries 84} (2000) 845--850},
  [\href{https://arxiv.org/abs/hep-ex/0001021}{{\ttfamily hep-ex/0001021}}].

\bibitem{Aaltonen:2012fi}
{\scshape CDF} collaboration, T.~Aaltonen et~al., \emph{{Transverse momentum
  cross section of $e^+e^-$ pairs in the $Z$-boson region from $p\bar{p}$
  collisions at $\sqrt{s}=1.96$ TeV}},
  \href{http://dx.doi.org/10.1103/PhysRevD.86.052010}{\emph{Phys. Rev.}
  {\bfseries D86} (2012) 052010},
  [\href{https://arxiv.org/abs/1207.7138}{{\ttfamily 1207.7138}}].

\bibitem{Abbott:1999wk}
{\scshape D0} collaboration, B.~Abbott et~al., \emph{{Measurement of the
  inclusive differential cross section for $Z$ bosons as a function of
  transverse momentum in $\bar{p}p$ collisions at $\sqrt{s} = 1.8$ TeV}},
  \href{http://dx.doi.org/10.1103/PhysRevD.61.032004}{\emph{Phys. Rev.}
  {\bfseries D61} (2000) 032004},
  [\href{https://arxiv.org/abs/hep-ex/9907009}{{\ttfamily hep-ex/9907009}}].

\bibitem{Abazov:2007ac}
{\scshape D0} collaboration, V.~M. Abazov et~al., \emph{{Measurement of the
  shape of the boson transverse momentum distribution in $p \bar{p} \to Z /
  \gamma^{*} \to e^+ e^- + X$ events produced at $\sqrt{s}$=1.96-TeV}},
  \href{http://dx.doi.org/10.1103/PhysRevLett.100.102002}{\emph{Phys. Rev.
  Lett.} {\bfseries 100} (2008) 102002},
  [\href{https://arxiv.org/abs/0712.0803}{{\ttfamily 0712.0803}}].

\bibitem{Abazov:2010kn}
{\scshape D0} collaboration, V.~M. Abazov et~al., \emph{{Measurement of the
  normalized $Z/\gamma^* -> \mu^+\mu^-$ transverse momentum distribution in
  $p\bar{p}$ collisions at $\sqrt{s}=1.96$ TeV}},
  \href{http://dx.doi.org/10.1016/j.physletb.2010.09.012}{\emph{Phys. Lett.}
  {\bfseries B693} (2010) 522--530},
  [\href{https://arxiv.org/abs/1006.0618}{{\ttfamily 1006.0618}}].

\bibitem{Aaij:2015gna}
{\scshape LHCb} collaboration, R.~Aaij et~al., \emph{{Measurement of the
  forward $Z$ boson production cross-section in $pp$ collisions at $\sqrt{s}=7$
  TeV}}, \href{http://dx.doi.org/10.1007/JHEP08(2015)039}{\emph{JHEP}
  {\bfseries 08} (2015) 039},
  [\href{https://arxiv.org/abs/1505.07024}{{\ttfamily 1505.07024}}].

\bibitem{Aaij:2015zlq}
{\scshape LHCb} collaboration, R.~Aaij et~al., \emph{{Measurement of forward W
  and Z boson production in $pp$ collisions at $ \sqrt{s}=8 $ TeV}},
  \href{http://dx.doi.org/10.1007/JHEP01(2016)155}{\emph{JHEP} {\bfseries 01}
  (2016) 155}, [\href{https://arxiv.org/abs/1511.08039}{{\ttfamily
  1511.08039}}].

\bibitem{Aaij:2016mgv}
{\scshape LHCb} collaboration, R.~Aaij et~al., \emph{{Measurement of the
  forward Z boson production cross-section in pp collisions at $\sqrt{s} = 13$
  TeV}}, \href{http://dx.doi.org/10.1007/JHEP09(2016)136}{\emph{JHEP}
  {\bfseries 09} (2016) 136},
  [\href{https://arxiv.org/abs/1607.06495}{{\ttfamily 1607.06495}}].

\bibitem{Bacchetta:2019tcu}
A.~Bacchetta, G.~Bozzi, M.~Lambertsen, F.~Piacenza, J.~Steiglechner and
  W.~Vogelsang, \emph{{Difficulties in the description of Drell-Yan processes
  at moderate invariant mass and high transverse momentum}},
  \href{https://arxiv.org/abs/1901.06916}{{\ttfamily 1901.06916}}.

\bibitem{Bertone:2018ecm}
{\scshape NNPDF} collaboration, V.~Bertone, N.~P. Hartland, E.~R. Nocera,
  J.~Rojo and L.~Rottoli, \emph{{Charged hadron fragmentation functions from
  collider data}},
  \href{http://dx.doi.org/10.1140/epjc/s10052-018-6130-4}{\emph{Eur. Phys. J.}
  {\bfseries C78} (2018) 651},
  [\href{https://arxiv.org/abs/1807.03310}{{\ttfamily 1807.03310}}].

\bibitem{Bertone:2017tyb}
{\scshape NNPDF} collaboration, V.~Bertone, S.~Carrazza, N.~P. Hartland, E.~R.
  Nocera and J.~Rojo, \emph{{A determination of the fragmentation functions of
  pions, kaons, and protons with faithful uncertainties}},
  \href{http://dx.doi.org/10.1140/epjc/s10052-017-5088-y}{\emph{Eur. Phys. J.}
  {\bfseries C77} (2017) 516},
  [\href{https://arxiv.org/abs/1706.07049}{{\ttfamily 1706.07049}}].

\bibitem{Bertone:2013vaa}
V.~Bertone, S.~Carrazza and J.~Rojo, \emph{{APFEL: A PDF Evolution Library with
  QED corrections}},
  \href{http://dx.doi.org/10.1016/j.cpc.2014.03.007}{\emph{Comput. Phys.
  Commun.} {\bfseries 185} (2014) 1647--1668},
  [\href{https://arxiv.org/abs/1310.1394}{{\ttfamily 1310.1394}}].

\bibitem{Ball:2008by}
{\scshape NNPDF} collaboration, R.~D. Ball, L.~Del~Debbio, S.~Forte,
  A.~Guffanti, J.~I. Latorre, A.~Piccione et~al., \emph{{A Determination of
  parton distributions with faithful uncertainty estimation}},
  \href{http://dx.doi.org/10.1016/j.nuclphysb.2008.09.037,
  10.1016/j.nuclphysb.2009.02.027}{\emph{Nucl. Phys.} {\bfseries B809} (2009)
  1--63}, [\href{https://arxiv.org/abs/0808.1231}{{\ttfamily 0808.1231}}].

\bibitem{Ball:2012wy}
R.~D. Ball et~al., \emph{{Parton Distribution Benchmarking with LHC Data}},
  \href{http://dx.doi.org/10.1007/JHEP04(2013)125}{\emph{JHEP} {\bfseries 04}
  (2013) 125}, [\href{https://arxiv.org/abs/1211.5142}{{\ttfamily 1211.5142}}].

\bibitem{Gutierrez-Reyes:2019rug}
D.~Gutierrez-Reyes, S.~Leal-Gomez, I.~Scimemi and A.~Vladimirov,
  \emph{{Linearly polarized gluons at next-to-next-to leading order and the
  Higgs transverse momentum distribution}},
  \href{https://arxiv.org/abs/1907.03780}{{\ttfamily 1907.03780}}.

\bibitem{Ball:2017nwa}
{\scshape NNPDF} collaboration, R.~D. Ball et~al., \emph{{Parton distributions
  from high-precision collider data}},
  \href{http://dx.doi.org/10.1140/epjc/s10052-017-5199-5}{\emph{Eur. Phys. J.}
  {\bfseries C77} (2017) 663},
  [\href{https://arxiv.org/abs/1706.00428}{{\ttfamily 1706.00428}}].

\bibitem{Abramowicz:2015mha}
{\scshape H1, ZEUS} collaboration, H.~Abramowicz et~al., \emph{{Combination of
  measurements of inclusive deep inelastic ${e^{\pm }p}$ scattering cross
  sections and QCD analysis of HERA data}},
  \href{http://dx.doi.org/10.1140/epjc/s10052-015-3710-4}{\emph{Eur. Phys. J.}
  {\bfseries C75} (2015) 580},
  [\href{https://arxiv.org/abs/1506.06042}{{\ttfamily 1506.06042}}].

\bibitem{Harland-Lang:2014zoa}
L.~A. Harland-Lang, A.~D. Martin, P.~Motylinski and R.~S. Thorne, \emph{{Parton
  distributions in the LHC era: MMHT 2014 PDFs}},
  \href{http://dx.doi.org/10.1140/epjc/s10052-015-3397-6}{\emph{Eur. Phys. J.}
  {\bfseries C75} (2015) 204},
  [\href{https://arxiv.org/abs/1412.3989}{{\ttfamily 1412.3989}}].

\bibitem{Dulat:2015mca}
S.~Dulat, T.-J. Hou, J.~Gao, M.~Guzzi, J.~Huston, P.~Nadolsky et~al.,
  \emph{{New parton distribution functions from a global analysis of quantum
  chromodynamics}},
  \href{http://dx.doi.org/10.1103/PhysRevD.93.033006}{\emph{Phys. Rev.}
  {\bfseries D93} (2016) 033006},
  [\href{https://arxiv.org/abs/1506.07443}{{\ttfamily 1506.07443}}].

\bibitem{Butterworth:2015oua}
J.~Butterworth et~al., \emph{{PDF4LHC recommendations for LHC Run II}},
  \href{http://dx.doi.org/10.1088/0954-3899/43/2/023001}{\emph{J. Phys.}
  {\bfseries G43} (2016) 023001},
  [\href{https://arxiv.org/abs/1510.03865}{{\ttfamily 1510.03865}}].

\bibitem{Buckley:2014ana}
A.~Buckley, J.~Ferrando, S.~Lloyd, K.~Noerdstrom, B.~Page, M.~Ruefenacht
  et~al., \emph{{LHAPDF6: parton density access in the LHC precision era}},
  \href{http://dx.doi.org/10.1140/epjc/s10052-015-3318-8}{\emph{Eur. Phys. J.}
  {\bfseries C75} (2015) 132},
  [\href{https://arxiv.org/abs/1412.7420}{{\ttfamily 1412.7420}}].

\bibitem{Ball:2010gb}
{\scshape NNPDF} collaboration, R.~D. Ball, V.~Bertone, F.~Cerutti,
  L.~Del~Debbio, S.~Forte, A.~Guffanti et~al., \emph{{Reweighting NNPDFs: the W
  lepton asymmetry}}, \href{http://dx.doi.org/10.1016/j.nuclphysb.2011.03.017,
  10.1016/j.nuclphysb.2011.10.024, 10.1016/j.nuclphysb.2011.09.011}{\emph{Nucl.
  Phys.} {\bfseries B849} (2011) 112--143},
  [\href{https://arxiv.org/abs/1012.0836}{{\ttfamily 1012.0836}}].

\bibitem{Gutierrez-Reyes:2019msa}
D.~Gutierrez-Reyes, Y.~Makris, V.~Vaidya, I.~Scimemi and L.~Zoppi,
  \emph{{Probing Transverse-Momentum Distributions With Groomed Jets}},
  \href{http://dx.doi.org/10.1007/JHEP08(2019)161}{\emph{JHEP} {\bfseries 08}
  (2019) 161}, [\href{https://arxiv.org/abs/1907.05896}{{\ttfamily
  1907.05896}}].

\bibitem{Gutierrez-Reyes:2019vbx}
D.~Gutierrez-Reyes, I.~Scimemi, W.~J. Waalewijn and L.~Zoppi, \emph{{Transverse
  momentum dependent distributions in $e^+e^-$ and semi-inclusive
  deep-inelastic scattering using jets}},
  \href{http://dx.doi.org/10.1007/JHEP10(2019)031}{\emph{JHEP} {\bfseries 10}
  (2019) 031}, [\href{https://arxiv.org/abs/1904.04259}{{\ttfamily
  1904.04259}}].

\bibitem{Gutierrez-Reyes:2018qez}
D.~Gutierrez-Reyes, I.~Scimemi, W.~J. Waalewijn and L.~Zoppi, \emph{{Transverse
  momentum dependent distributions with jets}},
  \href{http://dx.doi.org/10.1103/PhysRevLett.121.162001}{\emph{Phys. Rev.
  Lett.} {\bfseries 121} (2018) 162001},
  [\href{https://arxiv.org/abs/1807.07573}{{\ttfamily 1807.07573}}].

\bibitem{deFlorian:2014xna}
D.~de~Florian, R.~Sassot, M.~Epele, R.~J. Hernández-Pinto and M.~Stratmann,
  \emph{{Parton-to-Pion Fragmentation Reloaded}},
  \href{http://dx.doi.org/10.1103/PhysRevD.91.014035}{\emph{Phys. Rev.}
  {\bfseries D91} (2015) 014035},
  [\href{https://arxiv.org/abs/1410.6027}{{\ttfamily 1410.6027}}].

\bibitem{deFlorian:2017lwf}
D.~de~Florian, M.~Epele, R.~J. Hernandez-Pinto, R.~Sassot and M.~Stratmann,
  \emph{{Parton-to-Kaon Fragmentation Revisited}},
  \href{http://dx.doi.org/10.1103/PhysRevD.95.094019}{\emph{Phys. Rev.}
  {\bfseries D95} (2017) 094019},
  [\href{https://arxiv.org/abs/1702.06353}{{\ttfamily 1702.06353}}].

\bibitem{Sato:2019yez}
{\scshape JAM} collaboration, N.~Sato, C.~Andres, J.~J. Ethier and
  W.~Melnitchouk, \emph{{Strange quark suppression from a simultaneous Monte
  Carlo analysis of parton distributions and fragmentation functions}},
  \href{https://arxiv.org/abs/1905.03788}{{\ttfamily 1905.03788}}.

\bibitem{Gonzalez-Hernandez:2018ipj}
J.~O. Gonzalez-Hernandez, T.~C. Rogers, N.~Sato and B.~Wang, \emph{{Challenges
  with Large Transverse Momentum in Semi-Inclusive Deeply Inelastic
  Scattering}}, \href{http://dx.doi.org/10.1103/PhysRevD.98.114005}{\emph{Phys.
  Rev.} {\bfseries D98} (2018) 114005},
  [\href{https://arxiv.org/abs/1808.04396}{{\ttfamily 1808.04396}}].

\bibitem{Matsuura:1988sm}
T.~Matsuura, S.~C. van~der Marck and W.~L. van Neerven, \emph{{The Calculation
  of the Second Order Soft and Virtual Contributions to the Drell-Yan
  Cross-Section}},
  \href{http://dx.doi.org/10.1016/0550-3213(89)90620-2}{\emph{Nucl. Phys.}
  {\bfseries B319} (1989) 570--622}.

\bibitem{Kramer:1986sg}
G.~Kramer and B.~Lampe, \emph{{Two Jet Cross-Section in e+ e- Annihilation}},
  \href{http://dx.doi.org/10.1007/BF01679868}{\emph{Z. Phys.} {\bfseries C34}
  (1987) 497}.

\bibitem{Ahrens:2008qu}
V.~Ahrens, T.~Becher, M.~Neubert and L.~L. Yang, \emph{{Origin of the Large
  Perturbative Corrections to Higgs Production at Hadron Colliders}},
  \href{http://dx.doi.org/10.1103/PhysRevD.79.033013}{\emph{Phys. Rev.}
  {\bfseries D79} (2009) 033013},
  [\href{https://arxiv.org/abs/0808.3008}{{\ttfamily 0808.3008}}].

\end{thebibliography}\endgroup
\end{document}